\documentclass[usenatbib]{mnras}
\usepackage{hyperref}
\usepackage{amsmath}
\usepackage{mathptmx}
\usepackage{txfonts}
\usepackage[T1]{fontenc}
\usepackage{ae,aecompl}

\usepackage{graphicx}
\usepackage{natbib}
\usepackage{txfonts}
\usepackage{amssymb}
\usepackage{rotating}
\usepackage{pdflscape}

\newcommand{\myarcsec}{\hbox{$.\!\!^{\prime\prime}$}}
\newcommand{\myarcmin}{\hbox{$.\!\!^{\prime}$}}

\newcommand{\angstrom}{\AA}
\newcommand{\angstromblank}{\AA$\;$}
\newcommand{\Lya}{Ly\,$\alpha$ }
\newcommand{\Lyan}{Ly\,$\alpha$}
\newcommand{\Ha}{H\,$\alpha$ }

\newcommand{\Han}{H\,$\alpha$}
\newcommand{\Hbn}{H\,$\beta$}
\newcommand{\Hgn}{H\,$\gamma$}
\newcommand{\llya}{L_{\rm Ly\,\alpha}}
\newcommand{\kms}{km\,s$^{-1}$}
\newcommand{\cmss}{cm$^{-2}$}
\newcommand{\ergs}{erg\,s$^{-1}$}
\newcommand{\ergscm}{erg\,s$^{-1}$\,cm$^{-2}$}

\DeclareRobustCommand{\ion}[2]{\textup{#1\,\textsc{\lowercase{#2}}}}

\title[AGN flickering and low-redshift Lyman-$\alpha$ blobs]
      {About AGN ionization echoes, thermal echoes, and ionization deficits 
        in low redshift Lyman-$\alpha$ blobs
}
\author[M. Schirmer et al.]{
Mischa Schirmer,$^{1}$\thanks{E-mail: mschirme@gemini.edu}
Sangeeta Malhotra,$^{2}$
Nancy A. Levenson,$^{1}$
Hai Fu,$^{3}$
\newauthor
Rebecca L. Davies,$^{4}$
William C. Keel,$^{5}$
Paul Torrey,$^{6,7}$
Vardha N. Bennert,$^{8}$
\newauthor
Anna Pancoast$^{6,\dagger}$
and James E. H. Turner$^{1}$
\\
$^{1}$Gemini Observatory, Casilla 603, La Serena, Chile\\
$^{2}$School of Earth and Space Exploration, Arizona State University, Tempe, AZ 85287, USA\\
$^{3}$Department of Physics and Astronomy, University of Iowa, Iowa City, IA 52245, USA\\
$^{4}$Research School of Astronomy and Astrophysics, Australian National University, Cotter Road, Weston, ACT 2611, Australia\\
$^{5}$Department of Physics and Astronomy, University of Alabama, Box 870324, Tuscaloosa, AL 35487, USA\\
$^{6}$Harvard-Smithsonian Center for Astrophysics, 60 Garden Street, Cambridge, MA 02138, USA\\
$^{7}$MIT Kavli Institute for Astrophysics \& Space Research, Cambridge, MA 02139, USA\\
$^{8}$Physics Department, California Polytechnic State University, San Luis Obispo, CA 93407, USA\\
$^{\dagger}$Einstein Fellow\\}

\date{Accepted XXX. Received YYY; in original form ZZZ}

\pubyear{2016}

\begin{document}
\label{firstpage}
\pagerange{\pageref{firstpage}--\pageref{lastpage}}
\maketitle

\begin{abstract}
We report the discovery of 14 Lyman-$\alpha$ blobs (LABs) at $z\sim0.3$, existing at least $4-7$ 
billion years later in the Universe than all other LABs known. Their optical diameters are $20-70$\,kpc, and
\textit{GALEX} data imply \Lya luminosities of $(0.4-6.3)\times10^{43}$\,\ergs. Contrary to high-z 
LABs, they live in low-density areas. They are ionized by AGN, suggesting that cold accretion 
streams as a power source must deplete between $z=2$ and $z=0.3$. We also show that transient 
AGN naturally explain the ionization deficits observed in many LABs: Their \Lya and X-ray fluxes 
decorrelate below $\lesssim10^6$ years because of the delayed escape of resonantly scattering \Lya 
photons. High \Lya luminosities do not require \textit{currently} powerful AGN, independent 
of obscuration. \textit{Chandra} X-ray data reveal intrinsically weak AGN, confirming the luminous 
optical nebulae as impressive \textit{ionization echoes}. For the first time, we also report mid-infrared 
\textit{thermal echoes} from the dusty tori. We conclude that the AGN have faded by $3-4$ orders of 
magnitude within the last $10^{4-5}$ years, leaving fossil UV, optical and thermal radiation behind. The 
host galaxies belong to the group of previously discovered Green Bean galaxies (GBs). \textit{Gemini} 
optical imaging reveals smooth spheres, mergers, spectacular outflows and ionization cones. Because 
of their proximity and high flux densities, GBs are perfect targets to study AGN feedback, mode 
switching and the \Lya escape. The fully calibrated, coadded optical FITS images are publicly available.
\end{abstract}



\defcitealias{sdh13}{S13}
\defcitealias{bbb13}{B13}

\section{Introduction}{\label{intro}
Lyman-$\alpha$ blobs (LABs) are extended \Lya nebulae with luminosities of 
$\llya=10^{42-44}$\,\ergs, populating the Universe at $z\gtrsim2$. They are often
selected using optical narrow-band filters that isolate redshifted \Lya emission
\citep[e.g.][]{ktk03,myh04,myh11,dbs05,ooe09,yzt09,yze10}. LABs are $20-200$\,kpc in 
size and show a bewildering range of properties. They can be associated with Lyman 
Break Galaxies \citep{sph95}, visible and obscured AGN, starburst sub-mm galaxies 
and passively evolving red galaxies \citep[e.g.][]{fwc01,csw04,myh04,gal09,wyh09}. 

LABs are landmarks of ongoing massive galaxy formation \citep{myh06,pkd08}, yet the
ionizing sources in many of them remain mysterious. Our understanding of these processes 
would greatly benefit from studying the physical conditions in LABs. However, this is 
difficult as (1) cosmological surface brightness dimming reduces the flux densities by 
factors $\gtrsim100$, (2) the \Lya line is resonant, and (3) non-resonant optical 
lines are redshifted into and beyond the near-infrared atmospheric passbands.
The resonant character of \Lya causes two main problems with the interpretation of 
\Lya data.

First, \Lya photons scatter efficiently in space and frequency when propagating 
through a moving medium. Three-dimensional radiative transfer calculations 
\citep[e.g.][]{mer02,vsm06,kzd10} reveal a great variety of double-peaked \Lya line 
profiles emerging for various static and kinematic source/halo configurations 
\citep[for a one-dimensional analytic description in a static medium see][]{neu90}. 
It is difficult at best to infer the gas kinematics and the \Lya production sites from 
\Lya imaging and spectroscopy alone. A multi-wavelength perspective is required, 
including optically thin lines such as [\ion{O}{III}] and \Ha
\citep[e.g.][]{sso08,wbg10,yzj11,yzj14,mcm14,znf15,svs15}. 

The second problem is that we need to understand the processes that govern how many
\Lya photons manage to escape, so that they become observable at all. Dust, neutral 
hydrogen, metallicity and gas outflows control this escape fraction. The latter can 
range from less than 1 per cent to more than 50 per cent \citep[][and references 
therein]{ymg15}. Intrinsically, the \Han/\Lya line ratio is fixed for a photo-ionized 
nebula in equilibrium; the \Ha line can then be used to estimate the escape fraction 
and the total amount of \Lya produced. However, it is only for a small redshift window 
of $z=1.9-2.4$ that \textit{both} lines are observable from the ground. Worse, 
AGN variability may change the escape fraction further, by orders of magnitude, due 
to delayed \Lya escape \citep{rsf10,xwf11}. 

Probably the largest mystery with LABs is their frequent lack of ionizing sources; some LABs 
show no continuum counterparts at all. The lack of accessible diagnostic lines has prevented 
consistent conclusions on many occasions, and various processes have been suggested that 
could power LABs. For example, LABs are preferentially found in denser areas and filaments 
\citep{sso08,yze10,ebs11,myh11}, where LABs easily accrete cold neutral hydrogen from the 
cosmic web \citep{hsq00,dil09,gds10}. This is a requirement by $\Lambda$CDM galaxy formation 
models. The \Lya emission arises because of collisional excitation of hydrogen (virial 
temperature of $10^{4-5}$\,K) when it sinks into the dark matter haloes. It can contribute to 
an LAB's ionization over $10-30$ per cent of the Hubble time \citep{dil09}. Some calculations 
including self-shielding and realistic gas phases indicate that cold accretion alone could be 
insufficient to explain the \Lya fluxes of luminous LABs \citep{fkd10}. On the other hand, 
\citet{rob12} and \citet{cez13} find cold streams to be rather powerful, reproducing the 
size--luminosity function of observed LABs. The challenges in modelling the cold streams are 
mirrored on the observational side. One such gravitationally powered LAB \citep{nfm06} is
questioned by \citet{pmb15}, who discovered an embedded obscured AGN and argue that the original 
data speak \textit{against} cold accretion. 

Alternatively, LABs can be shock-ionized by starburst-driven superwinds \citep{tas00}, and 
photoionized by obscured AGN or starbursts \citep[e.g.][]{cls2001,gal09}. Starbursts 
alone cannot explain \Lya equivalent widths (EW) higher than about 240\,\angstromblank 
\citep{mar02,sso08}; LABs often exceed this value. Another possibility is centrally produced 
\Lyan, resonantly scattered by neutral hydrogen in the circum-galactic medium 
\citep{las07,sbs11}. This leads to a characteristic polarization signal and can thus be 
distinguished from photo-ionization and shock heating which produce \Lya \textit{in situ} 
\citep{hss11,hvv13}. However, \citet{tvb14} find that similar polarization signals may arise 
in cold streams as well.

Evidence for obscured AGN in some LABs has been found in infrared and sub-mm data 
\citep{bzs04,gal09,myh11,ond13,pmb15}, and for other LABs they have been postulated. For example, 
\citet{myh04} have found 35 LABs, some of which likely powered by superwinds and others by 
cooling flows \citep{myh06}. About 30 per cent lack UV continuum counterparts. Associated visible 
AGN are uncommon in this sample, and obscured star formation has been ruled out \citep{tmi13}. 
\citet{gal09} have shown that 24 out of 29 of these LABs remain undetected in a 400\,ks 
\textit{Chandra} exposure even after statistical stacking. They have suggested buried AGN and 
star-bursts as power sources instead of cold accretion. This, however, requires particular 
combinations of geometrical and radiative transfer effects to explain the substantial escape of 
\Lyan, while simultaneously preserving the thick obscuration along the line-of-sight 
\citep[see also][]{sas00}. While this certainly holds for some of these LABs, it seems unlikely
to be the case for all of them.

In this paper we investigate the effects of episodic AGN duty cycles (\textit{flickering}) on the 
UV, optical and mid-infrared (MIR) properties of LABs, forming optical \textit{ionization echoes}
and MIR \textit{thermal echoes}. Ionization echoes have been reported before, mostly at lower redshift
and in smaller and less luminous nebulae \citep{sev10,kls12,kcb12,sdh13,ssk13,kmb15}. We show that 
transient AGN naturally explain the ionization deficits in LABs. Our analysis is based on the 
\textit{Green Bean} galaxies \citep[GBs;][hereafter S13]{sdh13} at $z\sim0.3$, hosting luminous 
extended emission line regions (EELRs). We show that these EELRs are indeed LABs, and that they also 
host recently faded AGN. GBs form the most impressive ionization and thermal echoes currently known. 

This paper is structured as follows. In Section 2 we present an overview of the GBs, our 
\textit{Chandra} X-ray data, archival \textit{GALEX} data, and our Gemini/GMOS optical 
observations. The data are analysed in Sections 3, 4 and 5, respectively. In Section 
6 we discuss the evidence for AGN flickering, and its effect on the UV, MIR and optical 
properties of LABs. In Section 7 we discuss the LAB size--luminosity function and the evolution 
of the LAB comoving density. Our summary and conclusions are presented in Section 8. Details 
about individual targets are given in Appendix \ref{targetnotes}. We assume a flat cosmology 
with $\Omega_m=0.27$, $\Omega_\Lambda=0.73$ and $H_0=70\,$\kms$\,{\rm Mpc}^{-1}$.


\section{\label{sectionobs}Sample selection, observations and data reduction}
\subsection{Identifying Green Bean galaxies}
\subsubsection{\label{GPvsGB}Differences between Green Peas and Green Beans}
Green Peas \citep[GPs,][]{css09} are compact galaxies with strong [\ion{O}{III}] emission lines
that are redshifted into $r$-band. GPs have been discovered in SDSS $gri$ images because 
of their green colour. Amongst the 112 spectroscopically confirmed GPs (out of $\sim40000$ known) 
are 80 star-forming galaxies with high specific star formation rates, 13 composite objects 
revealing both AGN and star formation, 9 Seyfert-1s and 10 Seyfert-2s. Whilst the fraction of star 
forming GPs has been studied in detail \citep{apv10,apv12,igt11,pvm12,haw12,jao13,hsm15,ymg15}, 
the AGN fraction has remained largely unexplored.

Green Beans \citepalias[GBs,][]{sdh13} are much larger and more luminous than GPs. Their spectra 
are dominated by narrow lines with high EWs (e.g. $950$\,\angstromblank for [\ion{O}{III}] in 
J2240$-$0927), and they are (or were) powered by radio-quiet/weak type-2 quasars. Some of the GBs 
are probably extreme versions of Seyfert-2 GPs, whereas other GBs have different formation 
histories and/or ionization sources. We investigate these aspects in Section \ref{opticalresults}. 
Exploring further links between GPs and GBs is beyond the scope of this paper, also because very 
few observations exist for the AGN fraction amongst the GPs.

\subsubsection{GB sample selection and completeness}\label{contamination}
GBs are highly unusual, yet they were not identified earlier despite their brightness, size and 
colour. This is because (1) GBs are extremely rare, and (2) the SDSS $ugriz$ colour space 
occupied by the GBs is contaminated to 95 per cent by artefacts. The 17 GBs known to date were
found by an automatic SQL query mining the SDSS-DR8 photometric data base (14500 deg$^2$). The 
query consisted of broad-band colour criteria and a lower size threshold close to the resolution 
limit of SDSS. The genuine GBs were isolated from the artefacts by visual inspection of SDSS post 
stamp images. Details about the selection, the original SQL filter and the spectroscopic 
verification can be found in \citetalias{sdh13}.

Wide and deep imaging surveys with higher resolution than SDSS are ideal to find GBs. 
In Fig. \ref{GB_colorspace} we plot the $g-r$ and $r-i$ colours of GBs against those of galaxies in 
the Canada-France-Hawaii Telescope Lens Survey \citep[CFHTLenS, 158\,deg$^2$;][]{hek12,ehm13}. In the CFHTLenS 
catalogue we retained only bright ($r<20.5$) and large (half flux diameter more than 1\myarcsec1, like GBs) 
galaxies. We also rejected objects near bright stars and within filter ghosts (their {\tt MASK} parameter). 
The GBs are separated by a large margin from all other galaxies in this colour space. Only one object 
(CFHTLenS ID \#W4m0m0\_29728) initially remained in the space occupied by GBs. We removed it as its 
$r$-band photometry was falsified by a close pass of minor planet 704 \textit{Interamnia} (mag 11.3) 
on 2006 June 07. 

One can select GBs using 
\begin{equation}
g-r\,>0.8
\end{equation}
and
\begin{equation}
g-r\,>1.5\,(r-i\,)+0.9\,.
\end{equation}
These criteria are based on \textit{SExtractor} \citep{ber06} {\tt MAG\_AUTO} parameters. They do not 
change when using aperture magnitudes ({\tt MAG\_APER}) with diameters of 1\myarcsec5 
(isolating the nuclei) and 4\myarcsec5 (including most of the EELR flux). 

Figure \ref{GB_colorspace} also shows a significant redshift dependence because \Ha moves from 
$i$-band into $z$-band for $z\gtrsim0.295$. This decreases $r-i$ and increases $i-z$. A further 
discriminator is $u-g$ because of [\ion{O}{II}]$\lambda$3726,29 falling into $g$-band for 
$0.06\lesssim z\lesssim0.48$. Our original SDSS selection criteria therefore also included $u$- and 
$z$-band photometry \citepalias{sdh13}.

\begin{figure}
  \includegraphics[width=1.0\hsize]{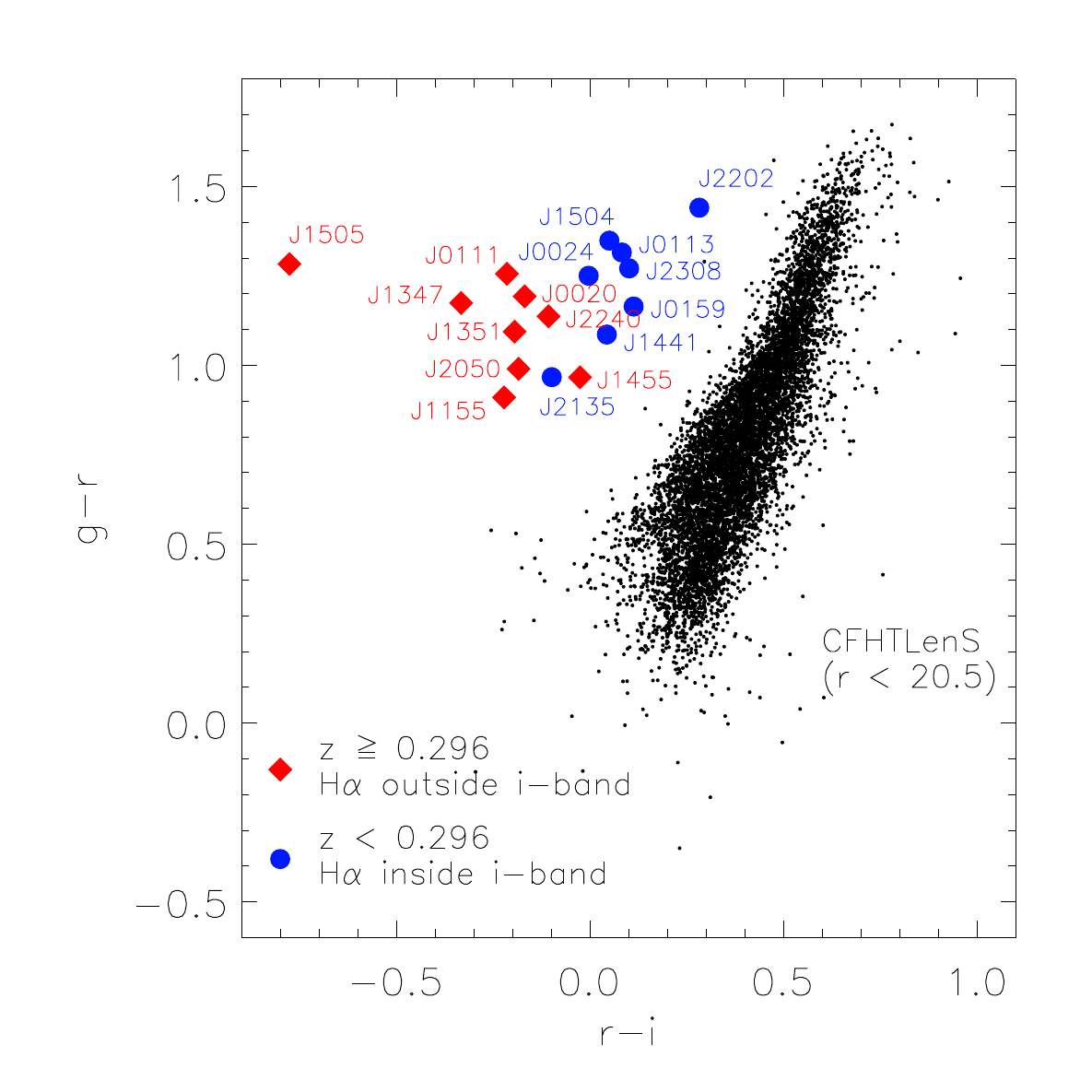}
  \caption{\label{GB_colorspace}{Gemini/GMOS broad-band colours of GBs compared with bright galaxies 
      from the 158 square degrees of CFHTLenS.}}
\end{figure}

The GB sample is fairly complete over the SDSS-DR8 footprint and the $z=0.12-0.36$ redshift 
range, with three caveats. First, some SDSS data have lower quality leading to an excess 
of ``green'' artefacts, seemingly related to poor seeing. Any GBs in these unusable survey areas 
were missed. This does not bias the sample as it is simply a matter of slightly lower sky coverage. 
Second, due to our lower size threshold, smaller GBs might not be recovered from some areas due to
seeing variations. Most likely there is a smooth transition between large GPs and small GBs. We are 
not concerned by this incompleteness as we are interested in the most extended sources, only.
Third, if an EELR coincides with a luminous elliptical galaxy ($M_i\gtrsim-23$ mag), then the 
[\ion{O}{III}] EW might not be high enough to distinguish the object and it would be overlooked
(see also Section \ref{densityevolutionclusters}). Lastly, we mention that our search entirely 
misses low-z LABs that do not emit in [\ion{O}{III}] (should they exist at $z\sim0.3$). This, however, 
does not count against the completeness of our initial goal of identifying strong [\ion{O}{III}] 
emitters.

Not all GBs have been discovered yet as SDSS covers a third of the sky, only. We expect 
$20-30$ more GBs that are still awaiting their discovery at southern declinations.

\subsubsection{\label{previousdiscoveries}Previous discoveries of Green Bean galaxies}
GBs have a surface density of $1.1\times10^{-3}$\,deg$^{-2}$, falling off the grid of smaller 
surveys and random observations. There are three exceptions, though. First, J2240$-$0927 
($z=0.326$) was a chance discovery in a CFHT 
wide-field data set \citep[program ID: 2008BO01;][]{shk11}. Only because of this 
coincidence did we learn about the existence of GBs and initiate our survey. Second, 
J0113+0106 ($z=0.281$) was selected automatically for SDSS spectroscopic follow-up to construct 
a flux-limited $u$-band sample (SDSS3 target flags {\tt U\_EXTRA2} and {\tt U\_PRIORITY}).
Third, J1155$-$0147 ($z=0.306$; the brightest and largest GB), was picked 
up independently by the Quasar Equatorial Survey Team \citep[QUEST;][]{sny98,rma04}. It
is the only GB in the QUEST survey area, a 2.4\,deg wide equatorial strip drift scanned 
for emission line objects. \textit{Chandra} images were taken in 2003 
(PI: Coppi; \textit{Chandra} Proposal Number 03700891; title: \textit{The X-ray Emission Of High 
Luminosity Emission Line Galaxies: Quasar-2s And The Starburst-AGN Connection}). We did not find
any publications of these data, nor about J1155$-$0147 itself.

\begin{table*}
\caption{\label{genprops}
  General properties of the GBs and their optical data. Column 1 lists the full names, which we abbreviate 
  in the main text to the first four digits in RA and Dec. Columns 2 and 3 contain the decimal sky coordinates. 
  The spectroscopic redshift is given in column 4. Columns 5, 6 and 7 contain the SDSS $r$-band AB magnitude 
  and the $g-r$ and $r-i$ colours, respectively. Column 8 lists the $r$-band image seeing, translated to a physical 
  resolution (at the respective source redshift) in column 9.}
  \begin{tabular}{lrrccrrcc}
    \noalign{\smallskip}
    \hline 
    \noalign{\smallskip}
    Name & $\alpha_{2000.0}$ & $\delta_{2000.0}$ & z & $r$ & $g-r$ & $r-i$ & Seeing & Resolution\\
    &  [deg] & [deg] & & [mag] & [mag] & [mag] & & [kpc]\\
    \noalign{\smallskip}
    \hline 
    \noalign{\smallskip}
    SDSS J002016.44$-$053126.6 &   5.06852 & -5.52405 & 0.334 & 18.3 & $1.19$ & $-0.17$ & 0\myarcsec59 & 2.8\\
    SDSS J002434.90$+$325842.7 &   6.14543 & 32.97852 & 0.293 & 18.2 & $1.25$ & $-0.00$ & 0\myarcsec69 & 3.0\\
    SDSS J011133.31$+$225359.1 &  17.88879 & 22.89976 & 0.319 & 19.1 & $1.26$ & $-0.21$ & 0\myarcsec59 & 2.8\\
    SDSS J011341.11$+$010608.5 &  18.42129 &  1.10237 & 0.281 & 18.5 & $1.32$ & $0.08$  & 0\myarcsec77 & 3.3\\
    SDSS J015930.84$+$270302.2 &  29.87851 & 27.05062 & 0.278 & 18.9 & $1.16$ & $0.11$  & 0\myarcsec57 & 2.4\\
    SDSS J115544.59$-$014739.9 & 178.93580 & -1.79443 & 0.306 & 17.9 & $0.91$ & $-0.22$ & 0\myarcsec70 & 3.2\\
    SDSS J134709.12$+$545310.9 & 206.78802 & 54.88637 & 0.332 & 18.7 & $1.17$ & $-0.33$ & 0\myarcsec37 & 1.8\\
    SDSS J135155.48$+$081608.4 & 207.98117 &  8.26900 & 0.306 & 19.0 & $1.09$ & $-0.19$ & 0\myarcsec71 & 3.2\\
    SDSS J144110.95$+$251700.1 & 220.29561 & 25.28337 & 0.192 & 18.5 & $1.09$ & $0.04$  & 0\myarcsec52 & 1.7\\
    SDSS J145533.69$+$044643.2 & 223.89036 &  4.77866 & 0.334 & 18.5 & $0.97$ & $-0.03$ & 0\myarcsec55 & 2.7\\
    SDSS J150420.68$+$343958.2 & 226.08615 & 34.66618 & 0.294 & 18.7 & $1.35$ & $0.05$  & 0\myarcsec37 & 1.6\\
    SDSS J150517.63$+$194444.8 & 226.32347 & 19.74578 & 0.341 & 17.9 & $1.28$ & $-0.78$ & 0\myarcsec52 & 2.5\\
    SDSS J205058.08$+$055012.8 & 312.74198 &  5.83688 & 0.301 & 18.6 & $0.99$ & $-0.18$ & 0\myarcsec77 & 3.5\\
    SDSS J213542.85$-$031408.8 & 323.92855 & -3.23577 & 0.246 & 19.2 & $0.97$ & $-0.10$ & 0\myarcsec72 & 2.8\\
    SDSS J220216.71$+$230903.1 & 330.56961 & 23.15086 & 0.258 & 18.9 & $1.44$ & $0.28$  & 0\myarcsec53 & 2.1\\
    SDSS J224024.11$-$092748.1 & 340.10044 & -9.46335 & 0.326 & 18.3 & $1.14$ & $-0.11$ & 0\myarcsec69 & 3.3\\
    SDSS J230829.37$+$330310.5 & 347.12239 & 33.05291 & 0.284 & 19.1 & $1.27$ & $0.10$  & 0\myarcsec67 & 2.9\\
    \hline
  \end{tabular}
\end{table*}

\subsection{\label{GMOSobs}Optical imaging with Gemini/GMOS}
We obtained $gri$ broad-band imaging of all 17 GBs. J2240$-$0927 was discovered earlier in deep 
CFHT data (Section \ref{previousdiscoveries}). The remaining GBs 
were observed with GMOS-N and GMOS-S at the 8-m Gemini Telescopes in Hawaii and Chile, respectively 
(program IDs GS-2013A-Q-48, GN-2014B-Q-78, GN-2015A-DD-3, GN-2015A-FT-23). $20$ minutes 
exposure time per filter were sufficient as the targets are bright ($r\sim18$ mag). 
The GMOS data were obtained in grey time, under clear and thin cirrus conditions with good seeing. 

Several of our targets were re-observed in $griz$ bands at the 4-m Southern Astrophysical Research 
Telescope (SOAR, Chile) during the science verification runs of the SOAR Adaptive Module (SAM) 
and the SAM Imager \citep[SAMI;][]{tts10,tts12}. SAMI has a 3\myarcmin1 field of view. SAM corrects 
for ground layer turbulence using three natural guide stars and one UV laser guide star, achieving 
a homogeneous PSF across the field and at optical wavelengths. Unfortunately, the seeing during 
these nights was dominated by turbulence in the upper atmosphere and SAM's ground layer correction 
could not yield an improvement over the GMOS data. The only exception are the $r$-band data of 
J0113+0106 for which the corrected seeing is 0\myarcsec62 while the DIMM seeing was stable between 
1\myarcsec0$-$1\myarcsec1; the GMOS image seeing for this object is 0\myarcsec80. We use the SAMI 
images to characterize the morphology of J0113+0106.

Image processing was done with \textsc{THELI} \citep{sch13,esd05} using standard procedures. 
Photometric zeropoints were tied to SDSS field magnitudes, correcting for non-photometric 
conditions. The physical resolution of the coadded images is between $1.6-3.5$\,kpc, depending on
seeing and source redshift. Table \ref{genprops} summarizes the optical characteristics. The fully 
calibrated, coadded $gri$ {\tt FITS} images are publicly 
available\footnote{{\tt https://zenodo.org/record/56059}}.

\subsection{\label{GMOSobsspec}Optical spectroscopy with Gemini/GMOS}
We conducted a sparse redshift survey around 13 GBs to study their environment, using GMOS
poor weather programs (GN-2015A-Q-99, GS-2015A-Q-99; thin cirrus, seeing $\gtrsim$1\myarcsec2, 
bright moon). We used the 1\myarcsec5 long-slit with the B600 grating, tuning the central 
wavelength to the 4000\,\angstromblank break at the GBs' redshifts. Main redshift 
indicators are \ion{Ca}{H+K}, [\ion{O}{II}], [\ion{O}{III}] and the Balmer series. $3\times600$\,s 
exposures were used with $4\times4$ detector binning. Target selection was heterogeneous and 
incomplete. We aimed at galaxies whose angular diameters, magnitudes and colours suggest similar 
redshifts as the GBs. Red sequence galaxies were preferred if present. High priority was given 
to galaxies in the immediate vicinity of the GBs, in particular if merger signatures such as tidal 
tails, extended haloes, and warps are visible. Position angles were chosen to maximize the number 
of objects (up to 7) on the slit. Up to three slit positions were observed per target area, and 
a total of 52 redshifts were obtained. Results are presented in Section \ref{longslit} and
Table \ref{redshiftlist}.

\subsection{\label{Lickobsspec}Optical spectroscopy with Lick/Kast}
Four of the GBs from \citetalias{sdh13} were observed with the Kast 
double-beam spectrograph at the 3-m Shane telescope of Lick Observatory to determine 
their redshifts. A dichroic beamsplitter divided the beams at 4600\,\AA. The blue arm 
used a grism setting spanning the $3390-4720$\,\AA\;range with a dispersion of 
0.65\,\AA\;pixel$^{-1}$ and a resolution of 3.3\,\AA\;FWHM. The red spectra covered 
$5010-7840$\,\AA\;at 2.4\,\AA\;pixel$^{-1}$ and resolution of 6.1\,\AA\;FWHM. The
$2^{\prime\prime}$ slit was oriented on a position angle chosen for each object to maximize 
the line flux included (and the angular span for any kinematic information). 

Individual 30-minute exposures were obtained on 12 and 13 March 2013 UT. Reduction used the
{\tt IRAF} long-slit tasks. A flux calibration was provided by observation of the standard stars 
G191B2B and BD +26 2606 with the same grism and grating settings each night.

Two of the targets, J1347+5453 ($z=0.332$) and J1504+3439 ($z=0.294$),
belong to the sample studied in this paper because (1) their redshifted [\ion{O}{III}] 
line falls into the $r$-band, and (2) with log([\ion{O}{III}]/H$\beta$)=1.00 and 0.94
they are also highly ionized as all other GBs. The other two, J1721+6322 and 
J1913+6211, are also highly ionized, yet their redshifts of $z=0.544$ and $z=0.552$ are 
above our upper redshift cut-off.

\subsection{\label{Chandraobs}Chandra X-ray imaging of 9 GBs}
We selected 9 GBs for follow-up with \textit{Chandra}, adding to the archival data of J1155$-$0147. 
The target sample is comprised of GBs with different morphology, [\ion{O}{III}] 
line structure, and [\ion{O}{III}] vs. mid-IR excess. The latter criterion was chosen to 
include AGN at different stages of the fading process.

We used the aim point on ACIS/S3 for greater soft response and spectral resolution. As our 
sources are faint we used the VFAINT mode, and pileup is well below 1 per cent according to 
{\tt PIMMS}. The full emission region of each galaxy fit on the single CCD, and on-chip 
background measurements were sufficient. No other bright X-ray sources are present in the
fields. The setup for the archival observations of J1155$-$0147 was similar.

\begin{table*}
\caption{\label{targetlist2}
  MIR and X-ray properties. We list the names in columns 1, the X-ray fluxes, count rates and
  fractional difference hardness ratios ($HR$) in columns $2-4$. Columns 5 and 6 contain the exposure time
  and \textit{Chandra} dataset IDs, respectively.}
\begin{tabular}{lrrrrrr}
  \noalign{\smallskip}
  \hline 
  \noalign{\smallskip}
  Name & $F_{22\mu{\rm m}}$ & $F_{0.3-8\,{\rm keV}}^{X}$ & $R_{\rm obs}$ & 
  $HR$ & $T_{\rm exp}^{\rm Chandra}$ & \textit{Chandra}\\
  & [mJy] & [\ergscm] & [s$^{-1}$] & & [ks] & dataset ID\\
  \noalign{\smallskip}
  \hline 
  \noalign{\smallskip}
SDSS J002016.44$-$053126.6 & 11.7 & $2.43\times10^{-14}$ & $2.57\times10^{-3}$  & $0.137\pm0.015$ & 30  & 16100\\
SDSS J002434.90$+$325842.7 & 25.4 & $1.19\times10^{-14}$ & $1.25\times10^{-3}$  & $-0.083\pm0.040$ & 20  & 16101\\
SDSS J011133.31$+$225359.1 & 23.1 & $-$                 & $-$                 & $-$ & $-$ & $-$ \\
SDSS J011341.11$+$010608.5 & 39.7 & $8.29\times10^{-14}$ & $8.98\times10^{-3}$  & $0.715\pm0.010$ & 15  & 16102\\
SDSS J015930.84$+$270302.2 & 18.1 & $1.95\times10^{-15}$ & $2.20\times10^{-4}$  & low & 30  & 16107\\
SDSS J115544.59$-$014739.9 & 16.9 & $1.18\times10^{-13}$ & $2.01\times10^{-2}$  & $0.351\pm0.002$ & 30  & 3140\\
SDSS J134709.12$+$545310.9 &  5.3 & $-$                 & $-$                 & $-$ & $-$ & $-$ \\
SDSS J135155.48$+$081608.4 & 25.7 & $-$                 & $-$                 & $-$ & $-$ & $-$ \\
SDSS J144110.95$+$251700.1 & 19.6 & $1.75\times10^{-14}$ & $1.83\times10^{-3}$  & $0.082\pm0.019$ & 30  & 16108\\
SDSS J145533.69$+$044643.2 & 20.4 & $7.51\times10^{-15}$ & $7.95\times10^{-4}$  & $-0.866\pm0.088$ & 20  & 16103\\
SDSS J150420.68$+$343958.2 &  7.6 & $-$                 & $-$                 & $-$  & $-$ & $-$ \\
SDSS J150517.63$+$194444.8 & 49.9 & $3.62\times10^{-14}$ & $3.86\times10^{-3}$  & $0.316\pm0.020$ & 15  & 16104\\
SDSS J205058.08$+$055012.8 & 49.5 & $4.70\times10^{-14}$ & $5.00\times10^{-3}$  & $0.719\pm0.016$ & 15  & 16106\\
SDSS J213542.85$-$031408.8 &  3.2 & $-$                 & $-$                 & $-$ & $-$ & $-$ \\
SDSS J220216.71$+$230903.1 & 24.8 & $-$                 & $-$                 & $-$ & $-$ & $-$ \\
SDSS J224024.11$-$092748.1 & 37.4 & $9.76\times10^{-15}$ & $1.04\times10^{-3}$  & $-0.077\pm0.067$ & 15  & 16105\\
SDSS J230829.37$+$330310.5 & 13.9 & $-$                 & $-$                 & $-$ & $-$ & $-$ \\
  \hline
\end{tabular}
\end{table*}

Count rates were estimated with the AGN MIR X-ray correlation of 
\citet{iut12} using SWIFT/BAT and AKARI. By using the hard $14-195$\,keV band, \citet{iut12} 
avoid complications by absorption at softer energies. We used their offset between the 
WISE and AKARI bandpasses, and estimated the GBs' intrinsic $14-195$\,keV luminosities from WISE 
22\,$\mu$m data. We then inferred the \textit{Chandra}/ACIS-S $0.3-8$\,keV count rate using {\tt PIMMS}, 
assuming a power law with photon index $\Gamma=1.9$ for the unabsorbed AGN spectrum, and a 
column density of $N_{\rm H}=10^{23}$\,\cmss. Our choices for $\Gamma$ and $N_{\rm H}$ 
were based on our analysis of the J1155$-$0147 data (Section \ref{j1155chandra}).

We chose \textit{Chandra}/ACIS exposure times of $15-30$\,ks, aiming at a total of $\sim1000$
counts per target. This would secure a successful spectral analysis for the entire sample. 
In the limiting case of Compton thick absorption ($N_{\rm H} = 1.5\times10^{24}$\,\cmss),
without reflection the continuum count rate would be $\sim0.002$\,s$^{-1}$. In these instances 
we would still detect an absorbed AGN in the strong Fe K$\alpha$ line. Note that even a low
column density of $N_{\rm H} = 10^{22}$\,\cmss\; is sufficient to account for the non-detection 
of all targets by ROSAT (which is sensitive at soft energies only).

Observations were carried out in \textit{Chandra} cycle 15, and the event files were processed in 
CIAO following standard procedures. We corrected the WCS of the final X-ray maps by about half a 
\textit{Chandra} pixel. The offset was calculated from the mean displacement observed between other 
X-ray detected AGN in the field and their counterparts in the optical GMOS images. The GBs were 
excluded from this calculation to avoid biasing by any true offset of their AGN with respect to 
the peak of the optical emission. Our X-ray measurements are summarized in Table \ref{targetlist2}. 



\begin{figure}
  \includegraphics[width=1.0\hsize]{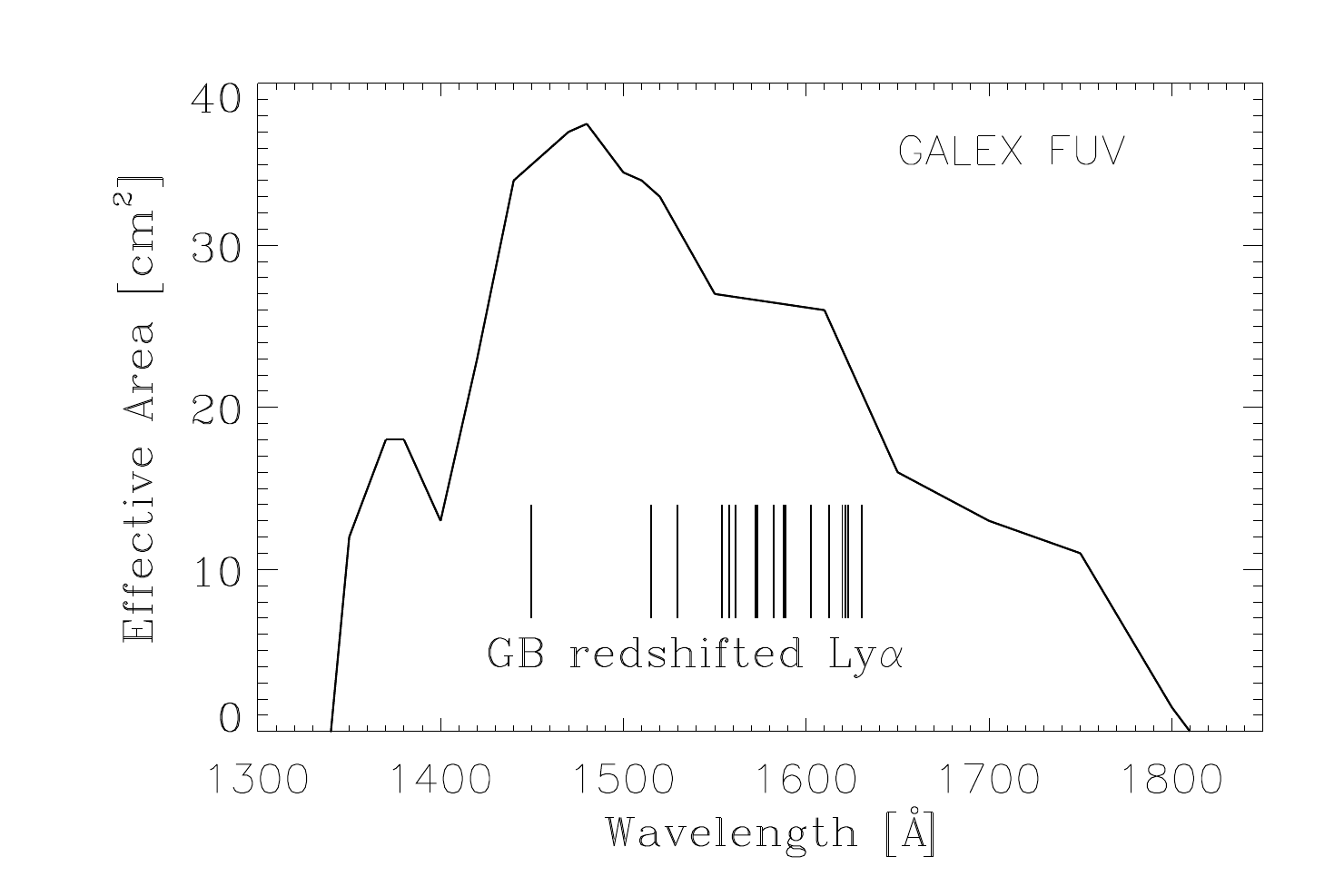}
  \caption{\label{galex_effarea}{The \textit{GALEX} FUV effective area, reproduced from the 
      \textit{GALEX} online documentation. The GBs' redshifted \Lya wavelengths fall into the 
      region of highest sensitivity.}}
\end{figure}

\subsection{\label{galexobs}GALEX observations}
\textit{GALEX} data are perfectly suited to detect redshifted \Lya from the GBs in the FUV 
channel ($1340-1800$\,\AA; Fig. \ref{galex_effarea}). In the course of the various \textit{GALEX} surveys,
5 GBs were observed in the FUV with exposure times of $1200-2800$\,s, and 10 GBs with exposure 
times of $60-300$\,s. Spectroscopic data were not taken. 14 out of 15 GBs are detected in the FUV, 
with ${\rm S/N}=2-21$. For two GBs no FUV data are available, but they are visible in the NUV 
($1750-2800$\,\AA). Only J0111+2253 is not detected in the FUV nor the NUV.

Flux measurements were taken from the \textit{GALEX} DR6 catalogue query page. 
If multiple measurements of the same source were available, then we used the one with
the longest exposure time. The only exception is J1455$+$0446, which is marginally blended 
in the \textit{GALEX} data with a large foreground galaxy and not available as a separate catalogue entry. 
We downloaded the calibrated FUV and NUV images and measured the fluxes in a 10\myarcsec5 wide 
circular aperture, cleanly separating J1455$+$0446 from its neighbour. The background signal and 
measurement errors were estimated by placing the same aperture at 10 randomly chosen 
nearby blank positions. The FUV and NUV spectral flux densities, exposure times,
galactic reddening and estimated \Lya luminosities are listed in Table \ref{targetUVprop}.

\section{Analysis of the X-ray data}
\subsection{Absence of kpc-scale AGN binaries}
Many GBs are interacting and/or merging (Sects. \ref{mergerrates} and 
\ref{morphologies}), and could perhaps host binary AGN. Mergers boost the accretion rates of 
supermassive black holes (SMBHs) by funnelling more gas toward the centres. This also holds 
for binary AGN as shown by \citet{lss12}, who find that the log([\ion{O}{III}]) luminosity 
increases by $0.7\pm0.1$ in AGN binaries when their separation decreases from 100 to 5\,kpc. 
Therefore, the GBs' high [\ion{O}{III}] luminosities make binary AGN at least plausible. The 
fraction of binaries with separations of tens of kpc amongst optically selected AGN is small 
\citep[3.6 per cent;][]{lss11}, yet it could be enhanced in GBs. The GBs' complex line profiles 
\citetext{\citetalias{sdh13}, \citealp{dst15}}, though, are much more likely caused by gas 
kinematics \citep[e.g.][]{slg11,cgs12,cog14,ass15}. 

We find that the nuclei in GBs are X-ray point sources. If binary AGN are present, then their 
separations must be smaller than $1.7-1.9$\,kpc (about one \textit{Chandra} ACIS pixel), 
and/or the secondary AGN is below our detection limit. Nonetheless, the advanced merger states 
make it worthwhile to search for sub-kpc binaries at other wavelengths.

\subsection{Offsets between X-ray and [OIII] peaks}{\label{xrayoffsets}}
The positions of the X-ray peaks are fully consistent with the positions of the optical 
peaks in the $r$-band images, i.e. the location of highest [\ion{O}{III}] brightness. The only 
exception is J1505+1944, where the X-ray peak is offset by 0\myarcsec5 (2.4\,kpc) to the 
West. Interestingly, the [\ion{O}{III}] nebula in J1505+1944 fragments in East-West direction. 
The brightest [\ion{O}{III}] part could be powered by a shock or be part of an outflow. 
Alternatively, it could harbour a second SMBH that is either deeply buried, or faded from our 
view recently while its ionizing radiation is still propagating outwards; dynamic data are not 
yet available for this system.

\subsection{Diffuse X-ray emission}
The X-ray contours are extended for 60 per cent of the targets
(J0024, J0159, J1155, J1455, J1055, J2240). While this is weakly significant for most 
targets individually (caused by just $1-3$ extra counts), in all cases the extended 
X-ray flux traces the most luminous parts of the [\ion{O}{III}] gas. We think this is 
caused by photoionized emission from the gas. For the remaining 40 per cent, any diffuse 
emission is below our detection threshold.

\subsection{\label{fadingagn}Compton-thick or intrinsically weak?}
Our observations yielded much lower count rates than anticipated, to the point where spectral 
fitting became meaningless ($7-80$ total counts). Therefore, we did not obtain power law 
indices, column densities and model fluxes apart from J1155$-$0147.

Figure \ref{ichikawa12} shows the MIR X-ray relation of \citet{iut12}. Based on the observed
\textit{Chandra} $0.3-8.0$\,keV count rates and a power law index of $\Gamma=1.9$, we calculate 
the expected X-ray fluxes for four different intrinsic column densities, 
${\rm log}\,(N_{\rm H}\,\rm{cm}^{2})=22.0$, 23.0, 24.0 and 24.3; error bars account for an 
uncertainty of 0.1 in $\Gamma$. The result can be
interpreted in two ways: Either, most GBs are nearly or fully Compton-thick (Section \ref{argcthick}), 
or they have faded recently and quickly, quicker than the typical response time of the
dusty tori's MIR emission (Section \ref{argcfading}).

\begin{figure}
  \includegraphics[width=1.0\hsize]{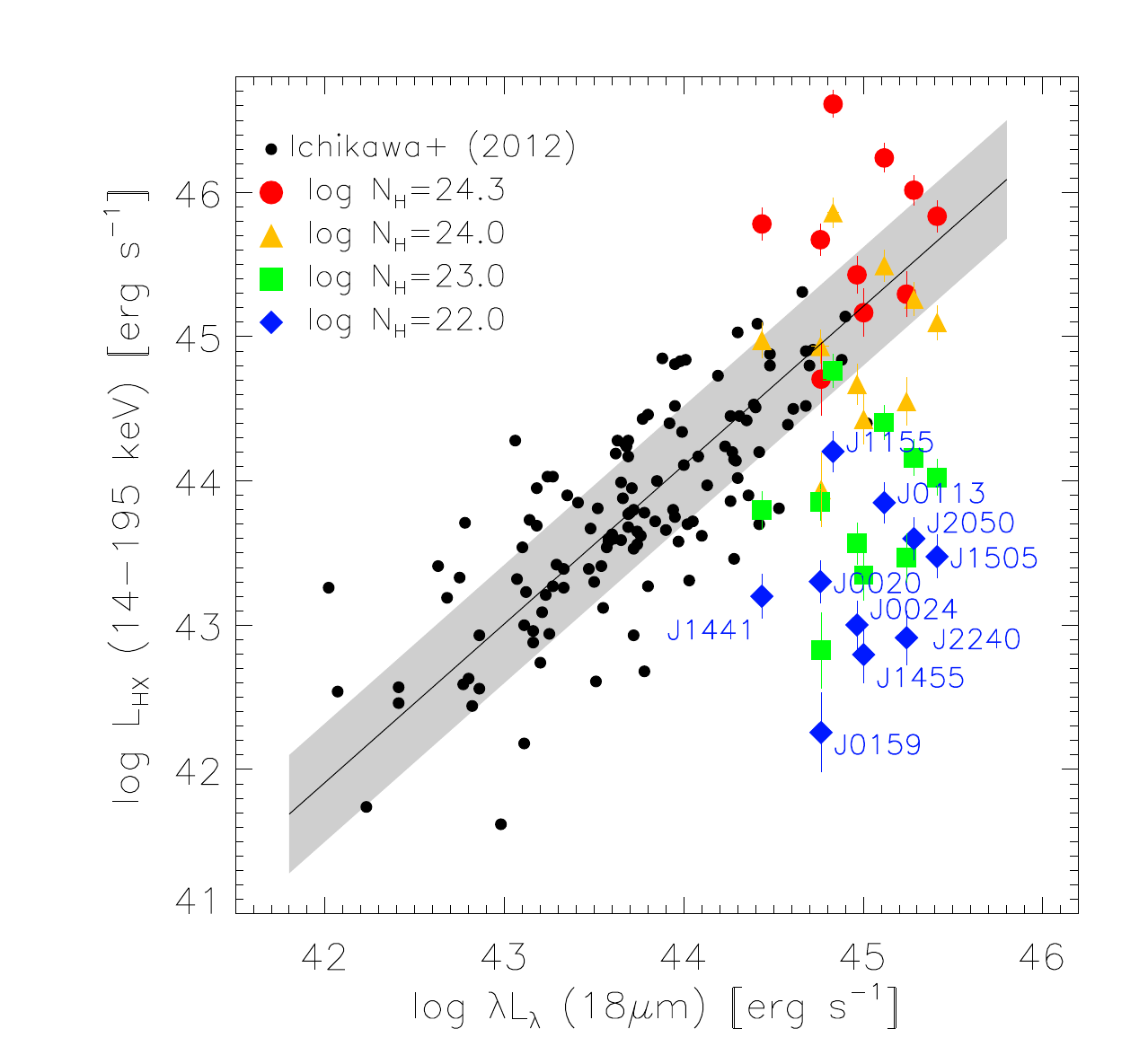}
  \caption{\label{ichikawa12}{AGN thermal echoes. The black dots show the 
      AGN MIR X-ray relation studied by \citet{iut12}, the shaded area 
      represents its intrinsic scattering. Overlaid are the expected
      X-ray luminosities of 10 GBs, based on their observed \textit{Chandra}
      $0.3-8.0$\,keV fluxes, a power law and 4 different values of intrinsic 
      absorption. Horizontal error bars are smaller than the symbol size, vertical 
      error bars include the \textit{Chandra} measurement error and a fiducial uncertainty 
      of $\Delta\Gamma=0.1$ for the photon index. The GBs follow the observed MIR X-ray 
      relation only if Compton-thick; since we can exclude Compton-thick obscuration
      for the sample, the GBs' X-ray fluxes must have faded quicker than their MIR 
      fluxes, causing thermal echoes. The typical MIR response time-scale to 
      X-ray fading is $10-1000$ years.}}
\end{figure}

\subsubsection{\label{argcthick}GBs cannot be Compton-thick as a sample}
The low count rates could be explained if, on average, GBs are obscured with 
log$\,(N_{\rm H}\,\rm{cm}^{2})=24.1$, implying a Compton-thick fraction $f_{\rm CTK}\sim0.9$.
The latter is unusually high; for comparison, \citet{rms99}, \citet{gmp05} and \cite{msb09} 
find $f_{\rm CTK}=0.4-0.5$ for optically and X-ray selected Seyfert-2s at $z\leq0.035$, 
and \citet{bdm99} report $f_{\rm CTK}=0.23-0.30$.

It is well-known that the fraction of absorbed AGN with log$\,(N_{\rm H}\,\rm{cm}^{2})>22$ decreases with 
increasing X-ray luminosity \citep{has08,uah14,mbb14}. How luminous are the AGN in GBs, and what 
fraction of Compton-thick sources should we expect? Compared to the type-2 samples of \citet{rzs08} 
and \citet{maf13}, GBs rank amongst the most [\ion{O}{III}] luminous type-2 AGN known, and should 
harbour AGN of high bolometric luminosity \citep{bdm99,hkb04,lbm09}. For e.g. J2240$-$0927, 
\citetalias{sdh13} measure an extinction corrected $L_{[\ion{O}{III}]}=(5.7\pm0.9)\times10^{43}$\,\ergs, 
translating to $L_{\rm bol}\sim2.3\times10^{46}$\,\ergs\; following \citet{lbm09}. We should therefore 
expect low values for $f_{\rm CTK}$. 

Just how low can be estimated from \citet{uah14} and their fig. 13, showing the fractions of 
moderately absorbed AGN (log$\,(N_{\rm H}\,\rm{cm}^{2})=22-24$) and Compton-thick AGN 
(log$\,(N_{\rm H}\,\rm{cm}^{2})=24-26$). 
Amongst type-2 AGN, $f_{\rm CTK}=0.3$ and 0.05 for the lower and higher X-ray luminous sources, 
respectively. Both values are in strong disagreement with $f_{\rm CTK}=0.9$ for the GBs.

However, these statistical arguments alone are insufficient to reject the hypothesis that 
nearly all GBs are Compton-thick. After all, GBs were discovered only recently and have not 
been studied before. The selection function of GBs (essentially, $r$-band excess caused by 
[\ion{O}{III}]) favours the selection of optically absorbed AGN: if unabsorbed type-1 AGN were 
present amongst the GBs, then their continuum contribution to the broad-band photometry would 
reduce the $r$-band excess and they would not be selected. Therefore, some obscuration amongst 
GBs is expected, but they do not have to be exclusively Compton-thick.

\subsubsection{\label{argcfading}GBs are intrinsically X-ray weak}
If GBs were indeed Compton-thick, then we would still detect the fluorescent K$\alpha$ line 
\citep{krk87}. However, this line is largely absent in our sample, favouring intrinsically weak 
AGN over heavy obscuration. The only GB for which we detect K$\alpha$ is J1155$-$0147, which is 
sufficiently bright to allow for spectral modelling. This is a moderately obscured source 
(Section \ref{j1155chandra}).

Another indicator for intrinsically weak AGN comes from the fractional difference hardness ratio, 
\begin{equation}
HR = \frac{H-S}{H+S}\;.
\end{equation}
Here, $S$ and $H$ are the counts in the soft ($0.3-2.0$\,keV) and hard ($2.0-8.0$\,keV) bands, 
respectively. We observe a moderate sample mean of $\langle HR\,\rangle=0.14\pm0.48$ (Table 
\ref{targetlist2}), meaning that the AGN cannot be deeply buried as a group.

\begin{table*}
\caption{\label{targetUVprop}
  UV properties of the GBs, including the observed monochromatic FUV and NUV spectral flux 
  densities (uncorrected for galactic extinction), the colour excess for the respective 
  line of sight, an extinction-corrected estimate of the \Lya luminosity assuming no continuum, 
  and the \textit{GALEX} integration times. See text for details.}
\begin{tabular}{lrrrrrr}
  \noalign{\smallskip}
  \hline 
  \noalign{\smallskip}
  Name & $f_{\nu}^{\;\rm FUV}$ & $f_{\nu}^{\;\rm NUV}$ & $E(B-V)$ & $L_{\rm Ly\alpha}$ & $T_{\rm exp}^{\;\rm FUV}$ & $T_{\rm exp}^{\;\rm NUV}$ \\
  & [$\mu$Jy] & [$\mu$Jy] & [mag] & [$10^{43}$\,\ergs] & [s] & [s]\\
  \noalign{\smallskip}
  \hline 
  \noalign{\smallskip}
SDSS J002016.44$-$053126.6 & $ 9.5\pm2.3$ & $ 9.0\pm2.4$  & 0.030 & $2.42\pm0.59$  &   206 &  206 \\
SDSS J002434.90$+$325842.7 & $ 5.9\pm2.2$ & $15.1\pm1.8$  & 0.051 & $1.12\pm0.42$  &   247 &  501 \\
SDSS J011133.31$+$225359.1 & undetected   & undetected    & 0.034 & $\lesssim0.31$ &   110 &  110 \\
SDSS J011341.11$+$010608.5 & $15.86\pm0.50$ & $15.44\pm0.50$  & 0.028 & $2.29\pm0.07$  &  2743 & 7999 \\
SDSS J015930.84$+$270302.2 & $ 7.1\pm2.4$ & $ 9.9\pm2.2$  & 0.056 & $1.23\pm0.42$  &   186 &  186 \\
SDSS J115544.59$-$014739.9 & $40.8\pm1.9$ & $41.2\pm1.1$  & 0.019 & $6.64\pm0.31$  &  2768 & 2768 \\
SDSS J134709.12$+$545310.9 & $31.0\pm5.3$ & $14.3\pm2.4$  & 0.010 & $6.47\pm1.11$  &   190 &  190 \\
SDSS J135155.48$+$081608.4 & $18.1\pm5.2$ & $ 7.6\pm2.5$  & 0.020 & $2.97\pm0.86$  &   106 &  106 \\
SDSS J144110.95$+$251700.1 & $29.5\pm8.5$ & $16.6\pm0.9$  & 0.023 & $1.29\pm0.37$  &    61 & 1690 \\
SDSS J145533.69$+$044643.2 & $19.8\pm1.6$ & $ 7.4\pm1.5$  & 0.033 & $5.14\pm0.42$  &  1650 & 1650 \\
SDSS J150420.68$+$343958.2 & $ 2.7\pm1.1$ & $10.1\pm0.8$  & 0.012 & $0.38\pm0.16$  &   306 & 2275 \\
SDSS J150517.63$+$194444.8 & $22.7\pm3.5$ & $25.4\pm2.6$  & 0.033 & $6.84\pm1.06$  &   234 &  234 \\
SDSS J205058.08$+$055012.8 & $ 6.6\pm2.5$ & $12.2\pm1.3$  & 0.088 & $1.77\pm0.68$  &   169 & 1616 \\
SDSS J213542.85$-$031408.8 & $18.0\pm1.3$ & $15.9\pm1.3$  & 0.033 & $1.58\pm0.11$  &  1561 & 1561 \\
SDSS J220216.71$+$230903.1 &  no data     & $11.4\pm2.8$  & 0.072 & $-$            &   $-$ &  173 \\
SDSS J224024.11$-$092748.1 & $21.2\pm1.6$ & $14.3\pm1.0$  & 0.052 & $5.43\pm0.41$  &  1578 & 1578 \\
SDSS J230829.37$+$330310.5 &  no data     & $ 5.7\pm2.5$  & 0.073 & $-$            &   $-$ &  158 \\
  \hline
\end{tabular}
\end{table*}

\section{Analysis of the GALEX data}\label{GALEXresults}
In this Section we estimate the \Lya luminosities of the GBs using \textit{GALEX} FUV and NUV broad-band 
imaging data (Table \ref{targetUVprop}). In the absence of \textit{GALEX} spectra, we must estimate continuum
contributions to the FUV, which could be mistaken for \Lya emission. We do not have sufficient ancillary 
data available to perform this for all GBs in our sample. Nonetheless, in four cases we can do this,
and we show that continuum emission contributes at most a few 10 per cent to the FUV flux. As the 
properties of the GBs are similar, we argue that our conclusions hold for the sample as a whole.

If the continuum contribution was 25 per cent, then 85 per cent (53 per cent) of the GBs have 
\Lya luminosities in excess of $1\times10^{43}$ ($2\times10^{43}$)\,\ergs\;with \Lya EWs 
of up to $1000$\,\angstrom. We conclude that we have indeed found LABs at low redshift, 
17 years after their initial discovery at high redshift (see also Fig. \ref{labstats}).

\subsection{\label{estlya}Estimating the Ly\,$\alpha$ luminosities}
We correct the FUV spectral flux densities for galactic extinction using the \citet{scf11} 
tables and a $R_v=3.1$ dust model. The correction factors range between 1.08 (J1347+5453) and 
1.97 (J2050+0550), and are calculated for the redshifted \Lya wavelengths assuming that most 
of the FUV flux is caused by this line. Other bright lines such as CIV$\lambda$1549 are redshifted 
beyond the \textit{GALEX} FUV bandpass even for the lowest redshift in our sample ($z=0.192$, J1441+2517).

We must account for the relative response function of \textit{GALEX} when estimating the total 
\Lya flux from the FUV broad-band data. The bandpass-averaged observed monochromatic spectral 
flux density, $f_\nu^{\;\rm obs}$, is calculated from the redshifted source spectrum, $f_\nu(\nu)$, 
as
\begin{equation}
\label{fnuobs}
f_\nu^{\;\rm obs} = \frac{\int{f_\nu(\nu)\,T(\nu)\,{\rm d}\nu}}{\int{T(\nu)\,{\rm d}\nu}}\;.
\end{equation}
Here, $T(\nu)$ is the unnormalized relative system throughput which we interpolate from the
FUV effective area (Fig. \ref{galex_effarea}). 

We approximate the spectrum as the sum of a constant continuum and some line profile. The 
continuum is parametrized as a fraction $c$ of the observed spectral flux density, 
$f_\nu^{\;\rm obs}$, and the spectrum is written as
\begin{equation}
f_\nu(\nu) = c\,f_\nu^{\;\rm obs}\,+\,f_\nu^{\;\rm line}(\nu)\,.
\end{equation}
Insert this into equation (\ref{fnuobs}) and we have
\begin{equation}
(1-c)\,f_\nu^{\;\rm obs} = \frac{\int{f_\nu^{\;\rm line}(\nu)\,T(\nu)\,{\rm d}\nu}}{\int{T(\nu)\,{\rm d}\nu}}\;.
\end{equation}

The \Lya line width is just a few \angstromblank even for a velocity dispersion of 1000\,\kms. 
The \textit{GALEX} FUV response can be considered constant over such small a wavelength range. 
We model the line profile as a Dirac delta function, normalized to yield the total line flux 
density, $F_{{\rm Ly}\alpha}$, when integrated over frequency:
\begin{equation}
(1-c)\,f_\nu^{\;\rm obs} = \frac{\int{F_{{\rm Ly}\alpha}\;\delta_{\rm D}(\nu-\nu_{{\rm Ly}\alpha})\,T(\nu)\,{\rm d}\nu}}{\int{T(\nu)\,{\rm d}\nu}}=
F_{{\rm Ly}\alpha}\;\frac{T(\nu_{{\rm Ly}\alpha})}{\int{T(\nu)\,{\rm d}\nu}}\;.
\end{equation}
Here, $\nu_{{\rm Ly}\alpha}$ is the frequency of the redshifted \Lya line. We solve for $F_{{\rm Ly}\alpha}$
and derive the \Lya luminosity using the luminosity distance. In Table \ref{targetUVprop} we list the 
\Lya luminosities assuming no continuum ($c=0$). If a fraction $c$ of the FUV flux is in 
the continuum, then the true line flux will be $(1-c)$ times the tabulated value.

Possible continuum sources are stars, the nebular continuum, and scattered AGN light:
\begin{equation}
c=c^{\,\rm stellar}+c^{\,\rm nebular}+c^{\,\rm scatter}\,.
\end{equation}
We discuss each of these terms below.

\subsection{Young and old stars must be considered}
Young hot stars contribute to the UV continuum. GBs are gas rich and often found in 
mergers (Section \ref{mergerrates}), a combination known to boost star formation. AGN feedback 
may also trigger star formation by shock-inducing cloud collapse \citep[e.g.][]{sdd13,sil13}.
However, our images also bear evidence for strong AGN-driven outflows, which may quench star 
formation by removing gas \citep[e.g.][]{gdg14}. High values of log([\ion{O}{III}]/H\,$\beta)\sim1$ 
show that star formation plays a minor role at least for the optical line emission \citepalias{sdh13}.

Old stars with high surface temperatures also contribute to the UV continuum. This includes 
binary stars \citep{hpl07}, low-mass helium-burning stars in the horizontal branch 
\citep[e.g.][]{cyl11,rjo12} and evolved post-AGB stars \citep[e.g.][]{cog10}. These types are 
thought to cause the UV excess (UVX) observed in elliptical galaxies, in particular bluewards of 
2000\,\angstromblank \citep[UV upturn; for a review see][]{crw99}. 

Which stellar populations are present in GBs? The red colours of the host galaxies (e.g. J1347+5453, 
J1504+3439) and of tidal stellar debris (J0024+3258, J0111+2253) are consistent with the prevalence 
of older stars. In most other cases the host galaxies are too compact to determine reliable colours 
in the presence of the nebular emission. Four GBs (J1155$-$0147, J1505+1944, J2050+0550, J2202+2309) 
are in groups or clusters with masses of at most a few $10^{13}\;M_{\odot}$ (e.g. Section 
\ref{j11550147target}). Dynamical friction \citep{cha43,nus99} and merging is efficient in such
low velocity environments, as witnessed by the presence of red sequence galaxies. It is plausible 
that these four GBs are also red sequence galaxies as they share the same environment with 
their neighbours.

We have to assume that both young and old stars contribute to the FUV continuum. We can estimate this 
for J2240$-$0927, for which we have useful spectra (Section \ref{starlight}), and for three other GBs 
where sufficient red sequence galaxies and \textit{GALEX} data are available (Section \ref{fuvJ1155}).

\begin{figure*}
  \includegraphics[width=1.0\hsize]{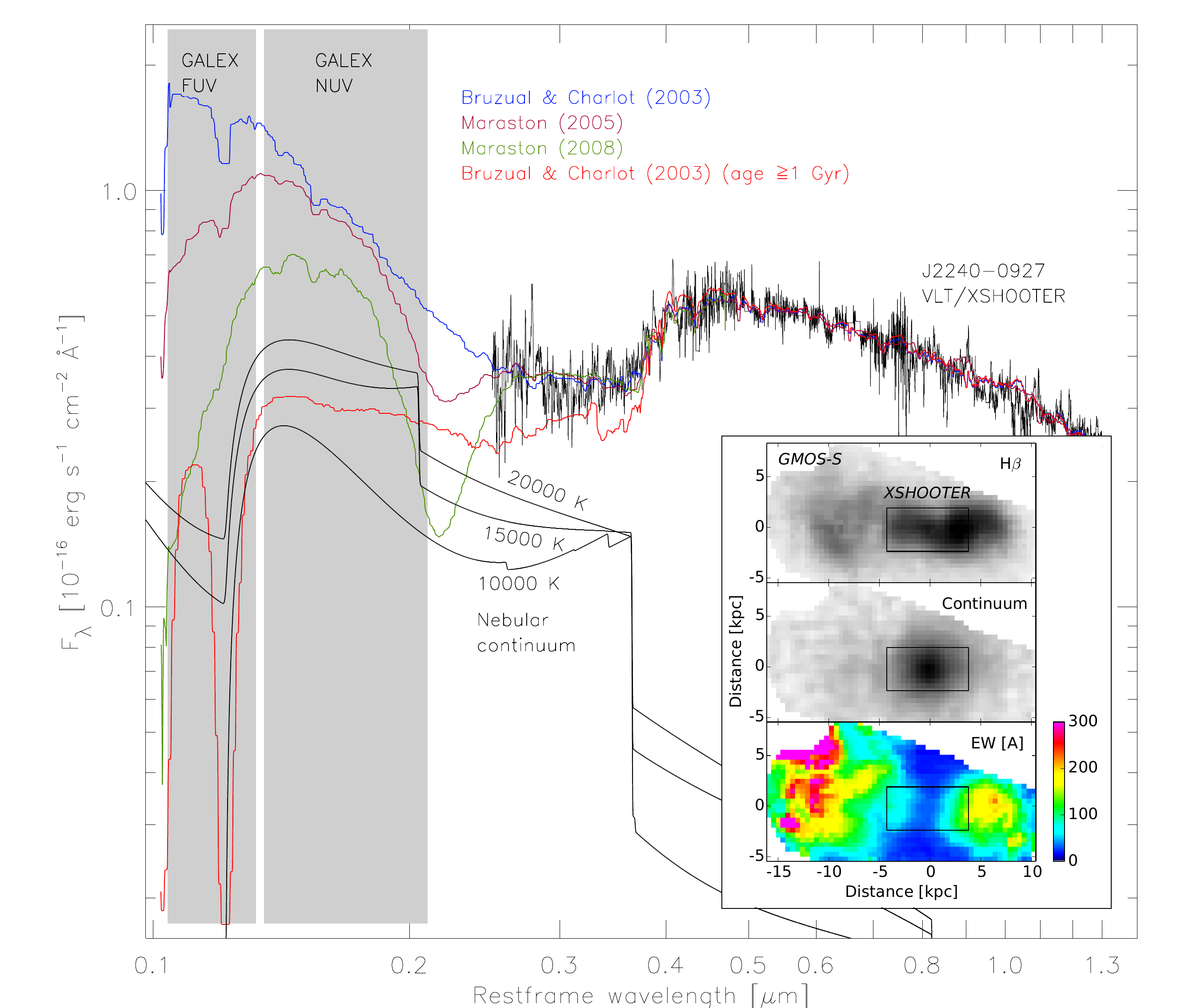}
  \caption{\label{j2241-0927_starlight}{Continuum restframe VLT/XSHOOTER spectrum of J2240$-$0927 
      (black, emission lines and nebular continuum removed). Overplotted are three representative 
      composite SSP models, used to predict the luminosities in the (blueshifted) \textit{GALEX} FUV 
      and NUV bandpasses (shaded areas). Comparison with an old stellar base population (red line) reveals 
      excess flux below 3800\,\AA, requiring additional young stars. The SSP models are divergent below 
      2500\,\angstrom$\;$for which we do not have observational data. Also shown is the nebular continuum
      for three electron temperatures and $n_{\rm e}=100$\,cm$^{-3}$. The XSHOOTER spectrum was centred on 
      the nucleus, and the extraction area is shown in the lower right inset. The latter displays the 
      reconstructed 2D images (nonlinear stretch) from the GMOS-S data cube, near and on the H$\beta$ 
      line, and the H$\beta$ equivalent width.}}
\end{figure*}

\subsection{\label{starlight}Constraining the stellar FUV/NUV flux for J2240$-$0927 using SED fitting}
\citet{dst15} have presented a 3D spectroscopic study of J2240$-$0927. The continuum maps reveal a 
compact spherical galaxy with $\lesssim9$\,kpc diameter. Our VLT/XSHOOTER spectrum \citepalias{sdh13} 
of the nucleus covers the $2500-18800$\,\angstromblank restframe range. It is corrected for galactic 
extinction using a $R_V=3.1$ dust model and the extinction maps of \citet{scf11}, de-redshifted, 
and we subtract the nebular continuum for a 15000\,K hydrogen helium gas mix in ionization equilibrium 
(see also Section \ref{nebcontinuum}). We then fit a combination of simple stellar populations (SSPs) 
with the {\textsc STARLIGHT} code \citep{msc06}, and determine the FUV and NUV fluxes permitted by the 
models.

We use the evolutionary models of \citet{brc03}, the SSPs of \citet{mar05}, which include the effects 
of thermally pulsating AGB stars (TP-AGB), \citet{mnb09}, the binary SSPs of \citet{hpl07}, and 
lastly the SSPs of \citet{cog10}, which better describe the UV properties of the horizontal branch 
and post-AGB stars. A comparison can be found in \citet{cog10}. 

We run {\textsc STARLIGHT} on a grid with 90 different configurations (SSPs, dust and extinction 
models, hard and soft convergence criteria, wavelength ranges). Figure \ref{j2241-0927_starlight} 
shows the spectrum together with a choice of three composite SSP models, displaying the 
full range of predicted FUV/NUV fluxes including extreme cases. We summarize our findings:

First, none of the fits is superior. The continuum levels and slopes are reproduced 
by all fits for $\lambda\gtrsim2800$\,\angstrom, whereas absorption features (CaH+K, G-band, Mg, 
NaD, \ion{Ca}{II} triplet, etc) are recovered with lesser accuracy. Inclusion of different dust 
models and extinction laws do not change the fits significantly.

Second, we observe absorption bands near 8500\,\angstromblank and 9500\,\angstrom, attributed to 
TP-AGB stars, CN and CO bands, and they are better described by the MA05 models. All fits reproduce 
the spectrum well between $13000-19000$\,\angstromblank (not shown in Fig. \ref{j2241-0927_starlight}).

Third, a reference model consisting of old populations with ages $\geq1$ Gyr only (red line), 
reveals excess flux below 3800\,\AA. Consequently, all models require
the presence of a younger population with age of a few to a few ten Myrs. Depending on the 
{\textsc STARLIGHT} setup, $10-30$ per cent of the bolometric luminosity are caused by young 
stars. This contribution drops to 2 per cent when we exclude wavelengths shorter than 
3800\,\angstromblank from the fit.

Fourth, all models diverge below 2500\,\angstrom. This is caused by the strong metallicity and age 
dependence of the UV upturn which is unconstrained by our data.

Finally, we shift the various composite SEDs to the redshift of J2240$-$0927, add back the
reddening, and calculate bandpass-averaged FUV/NUV spectral flux densities for comparison with 
the \textit{GALEX} data. We apply aperture correction factors, as \textit{GALEX} integrated the entire 
galaxy light, whereas XSHOOTER observed through a 0\myarcsec9 wide slit. To this end we use the 3D 
GMOS-S spectroscopic cube of \citet{dst15}, which covers the full spatial extent of J2240$-$0927. We 
integrate the light over the reconstructed IFU image, once over the full field, and once within the XSHOOTER 
aperture. Seeing corrections are unnecessary because both data sets were taken with a seeing of 
0\myarcsec5$-$0\myarcsec6. For the stellar continuum (taken near H$\beta$) and the nebular emission (taken 
on H$\beta$) we determine aperture correction factors of $2.10\pm0.04$ and $2.65\pm0.05$, respectively. 
The nebular correction factor is larger because the continuum light is much more concentrated (see the 
inset in Fig. \ref{j2241-0927_starlight}).

\subsubsection{Results of the SED fitting}
The observed \textit{GALEX} FUV flux density of J2240$-$0927 is $21.2\pm1.6$\,$\mu$Jy (Table \ref{targetUVprop}). 
We find the stellar model fluxes to range between $2.7-12.1$\,$\mu$Jy (most extreme values, $13-57$ per cent 
contribution). Deeper observations below restframe wavelengths of 2500\,\AA$\;$(observed 3400\,\AA) are 
required to better discriminate between the models. The flux calibration and S/N of our XSHOOTER observations 
(${\rm S/N}\sim1$ after $32\times$ spectral binning) are too poor for this purpose. Given these data alone, 
the most plausible contribution is $\sim5-8$\,$\mu$Jy, i.e.
\begin{equation}
c^{\,{\rm stellar}} = 0.2-0.4\;\; ({\rm J2240-0927}).
\end{equation}

For the NUV, we find stellar model flux densities between $10.1-16.7$\,$\mu$Jy, compared to an observed 
value of $14.28\pm0.95$\,$\mu$Jy. Consequently, the model that produces the lowest FUV contribution of 
13 per cent accounts for 70 per cent of the observed NUV flux, ruling out models that contribute
more than about $20$ per cent to the FUV. Including this constraint from the NUV data, we update
\begin{equation}
c^{\,{\rm stellar}} \lesssim 0.1-0.2\;\; ({\rm J2240-0927}).
\end{equation}


\subsection{\label{fuvJ1155}Constraining the stellar FUV flux in 3 GBs from red sequence galaxies}
J1155$-$0147, J1505$+$1944 and J2050$+$0550 are in spectroscopically confirmed galaxy groups with 
a red sequence. We derive mean stellar FUV to $i$-band flux ratios, 
$\langle f_\nu^{\,\rm FUV} / f_\nu^{\,i}\rangle$, for the red sequence members. Assuming that the GBs 
are also red sequence members, we use their $i$-band magnitude to estimate their stellar FUV 
flux (note that \Ha is redshifted beyond $i$-band in all three cases).

We place apertures over the red sequence members in the $i$-band image and measure their $i$-band 
spectral flux density. The apertures are then transformed to the \textit{GALEX} FUV image accounting 
for the larger plate scale and PSF, and the measurement is repeated. The red sequence members are not 
detected individually by \textit{GALEX}. Integrating over all apertures, we derive 
$\langle f_\nu^{\,\rm FUV} / f_\nu^{\,i}\rangle = (5.0\pm6.5)\times10^{-3}$, $(-2.7\pm2.0)\times10^{-3}$, 
and $(1.07\pm0.62)\times10^{-2}$, respectively, for these three systems. This describes the total FUV 
contribution from young and old stars. Comparison with the observed FUV flux densities yields
\begin{eqnarray}
c^{\,{\rm stellar}} & = & 0.027\pm0.035\;\; ({\rm J1155-0147})\\
c^{\,{\rm stellar}} & < & 0.018\;\; ({\rm J1505+1944;\;95\,\%\;confidence)}\\
c^{\,{\rm stellar}} & = & 0.11\pm0.07\;\; ({\rm J2050+0550}).
\end{eqnarray}
The value for J1505$+$1944 is an upper limit. These contributions are lower than or equal to 
what we have found for J2240$-$0927 using SED fitting. All GB host galaxies have similar 
$i$-band magnitudes ($18.8\pm0.4$ mag). Stars, therefore, cannot explain their high FUV fluxes 
(accounting for a few per cent, at most a few 10 per cent of the FUV flux).

\subsection{Nebular continuum for J2240-0927}\label{nebcontinuum}
The nebular continuum also contributes to the UV. We model it using our custom-made software
{\tt NEBULAR} (Schirmer 2016, submitted), which is publicly 
available\footnote{{\tt https://zenodo.org/record/55843}}.
In particular, we use a mixed hydrogen helium plasma in ionization equilibrium, with a helium 
abundance (by parts) of 0.1. 
The continuum of the nebular spectrum is comprised of free-bound recombination emission from 
\ion{H}{I}, \ion{He}{I} and \ion{He}{II}, free-free emission from electrons scattering at charged ions,
and the two-photon continuum.

The two-photon continuum far exceeds the free-bound emissivity below 2000\,\AA. It arises in 
hydrogenic ions from the decay of the 2$^2S$ level to the 1$^2S$ level by simultaneous emission 
of two photons (a single photon decay is prohibited by the dipole selection rules). The energy of 
the two photons adds up to the \Lya energy. The two-photon spectrum has a natural upper cut-off at 
the Ly$\alpha$ frequency, and peaks at half the Ly$\alpha$ frequency when expressed in frequency units. 
We approximate the two-photon spectrum following \citet{nus84}. The 2$^2S$ level is 
increasingly de-populated by collisions for electron densities $n_{\rm e}>1000$\,cm$^{-3}$ \citep{pes64}, 
reducing the two-photon continuum. This process can be ignored in the low-density gas 
\citep{dst15} of the GBs.

At optical wavelengths, the nebular continuum is faint and dominated by the stellar continuum.
Fotunately, its amplitude is fixed with respect to the intensity of the Balmer lines at a 
given electron temperature and density. Using {\tt NEBULAR}, we derive H$\beta$ equivalent 
widths of 1370, 770 and 650\,\AA$\;$over the nebular continuum, for electron temperatures of 
10000, 15000 and 20000\,K, respectively (and $n_{\rm e}=100$\,cm$^{-3}$). Using the total observed 
H$\beta$ flux from our GMOS-S data 
cube ($7.5\times10^{-15}$\,\ergs), we find the following: In the FUV, the nebular continuum 
contributes 0.5, 1.6 and 2.1\,$\mu$Jy for $T_{\rm e}=10000$, 15000 and 20000\,K, respectively, i.e. 
2, 8 and 10 per cent of the observed total FUV flux. \cite{dst15} have shown that the typical 
gas temperature in J2240$-$0927 is around 13800\,K, and 15500\,K if the hotter nuclear region 
is included as well.

As can be seen in Fig. \ref{j2241-0927_starlight}, the (redshifted) nebular continuum peaks in the 
\textit{GALEX} NUV channel (mostly because of the strong two-photon spectrum). Consequently, we determine
much higher NUV flux densities of 5.4, 9.0 and 10.3\,$\mu$Jy for $T_{\rm e}=10000$, 15000 and 20000\,K, 
respectively (38, 63 and 73 per cent of the observed total NUV flux).

\subsubsection{Results for the nebular continuum}
The nebular continuum contributes $2-10$ per cent to the FUV flux of J2240$-$0927, and $38-73$ per cent 
of the observed NUV flux. It is much better constrained than the stellar contribution 
from SED fitting, because the H$\beta$ line is detected with high S/N and the nebular continuum is
fixed to the H$\beta$ flux. The stellar FUV/NUV continuum is much more uncertain as it is mostly 
unconstrained by observational data below 2500\AA. For J2240$-$0927, we have

\begin{eqnarray}
c^{\,{\rm nebular}} & = & 0.02\;\; ({\rm T=10000\,K})\\
c^{\,{\rm nebular}} & = & 0.08\;\; ({\rm T=15000\,K})\\
c^{\,{\rm nebular}} & = & 0.10\;\; ({\rm T=20000\,K}).
\end{eqnarray}

The most conservative estimates from the nebular continuum and the stellar continuum easily account 
for the entire \textit{GALEX} NUV flux. In the FUV, on the other hand, the largest conceivable combination
yields about 30 per cent, and the remainder must be attributed to \Lya.

\subsection{Scattered AGN light}
Another source of UV continuum is light from the AGN accretion disk, scattered in areas that have an 
unobscured view of the nucleus \citep[see][for examples of scattering in the near UV]{pdr93,zss05}. 
Without FUV polarization measurements we cannot constrain this effect directly. The absence of 
scattered broad lines in the shallow optical spectra of \citetalias{sdh13} indicates that this effect 
is insignificant for the sample as a whole. We have also shown above for J2240$-$0927 that the stellar
and the nebular continuum fully account for the NUV observations, leaving little to no headroom
for additional scattered light (Sects. \ref{starlight} and \ref{nebcontinuum}). Therefore, 
\begin{equation}
c^{\,\rm scatter}\approx 0\,.
\end{equation}

\section{Analysis of the optical data}\label{opticalresults}
In this Section we describe global characteristics of the GBs. Notes about individual targets 
are given in Appendix \ref{targetnotes}.

\subsection{\label{longslit}Environment}
We obtained 52 spectroscopic redshifts (Table \ref{redshiftlist}) of selected field galaxies 
to determine the local environment of 13 GBs. The selection function is described in Section 
\ref{GMOSobsspec}.

The majority of the GBs live in low-density areas. 35 per cent are isolated, and for 
25 per cent we can currently not say whether they are isolated as well, or have $1-3$ possible 
companion galaxies. $15-20$ per cent are located in sparse groups with low concentration and 
perhaps $3-5$ members. The remaining 25 per cent are found in richer groups of galaxies 
with $M_{200}=(1-6)\times10^{13}\; M_{\odot}$ and well-defined red sequences (see Table 
\ref{targetlist} and Appendix \ref{targetnotes}).

This is in stark contrast with LABs at high redshift, which are preferentially found in filaments 
and clusters. Possibly, at $z=0.3$ the cold accretion streams have been exhausted, and low-z LABs 
are mostly formed and ionized by AGN. If a GB is located in an apparent
group or cluster, then it is found near the centre of the distribution of galaxies. Particularly 
noteworthy is J1155$-$0147, dominating the group with its size and luminosity. This is the only 
GB whose morphology could match a cold accretion stream.

\subsection{\label{mergerrates}Merger rates}
$50-65$ per cent of the GBs interact or merge as evidenced by extended warped stellar haloes, 
tidal stellar streams and close companions. Some of the companions show tidal distortions 
(e.g. J2240$-$0927), others appear spherically compact, undisturbed and just embedded in 
the gas (e.g. J0020$-$0531). Only 15 per cent of the GBs reveal seemingly tidally undisturbed host 
galaxies (J0159+2703, J1347+5453, J1504+3439; it is possible that some signs of tidal tails and 
interactions have been missed due to their low surface brightness). In all other cases, the bright 
EELR prevents a clear view of the hosts, or the hosts are obviously interacting with their 
companions. \cite{ylz13} have shown with hydrodynamical simulations and radiative transfer 
calculations that binary galaxy mergers will produce LABs with \Lya luminosities of $10^{42-44}$\,\ergs\; 
and typical sizes of $10-50$\,kpc (like GBs), albeit at $z=3-7$. The \Lya emission 
in these model mergers is mostly produced by intense star formation and gravitational 
cooling, whereas in GBs the main power sources are AGN. This is a another indication 
of a strong redshift evolution of LABs. We discuss this in Section \ref{discussion}.

\subsection{\label{morphologies}Morphologies of the host galaxies}
The host galaxies of five GBs are easy to classify because of the EELRs' low EWs and the hosts' 
large diameters. In J0159+2703 we find a large, 46\,kpc face-on barred spiral galaxy. J1347+5453 
is an edge-on disk with 21\,kpc diameter and an axis-ratio of at least 5:1 (the minor axis is 
not spatially resolved). The $i$-band data reveal a bright bulge or unresolved nucleus. 
J1504+3439 is an elliptical galaxy with a major axis of 37\,kpc. J2202+2309 is a luminous 
elliptical near the centre of a galaxy cluster. It can be traced over at least $40\times25$ 
kpc and is embedded in a common halo with two other ellipticals of similar size and luminosity. 
The system could form the future brightest cluster galaxy (BCG) of this structure. J2308+3303 
is comprised of a 6\,kpc bright nucleus surrounded by a face-on featureless disk with 
$22\times20$\,kpc diameter.

The classification of most other hosts is hampered by the low spatial resolution and
strong line emission in the $gri$ filters. They appear to be compact with major axes of 
$8-18$\,kpc (Table \ref{targetlist}). Colours of tidal stellar streams suggest older stellar 
populations, but that does not exclude ongoing star formation. Perhaps the most bizarre object 
is J1455+0446, consisting of a 40\,kpc large jumbled mix of ionized gas and stars as judged 
by its large colour variations. The bright nucleus is found at the edge of the system.
Continuum images from 3D spectroscopy, and $K$-band images of relatively line-free regions 
of the spectrum would greatly help the classification.

\subsection{Morphologies of the emission line regions}
The emission line regions in the GBs extend over several 10\,kpc. In the absence of kinematic data,
the spatial image resolution of $2-3$\,kpc allows for some constraints on the formation of the GBs. 
Most compelling is the bewildering range of morphologies arising from the combination of various 
intrinsic shapes and viewing angles.

\citet{hhg13} have spectroscopically determined the size of [\ion{O}{III}] narrow-line regions (NLRs) 
around luminous type-2 quasars, measuring within an $1\times10^{-15}$\,\ergscm\,arcsec$^{-2}$ 
isophote. They have found typical sizes of $6-8$ kpc, with an upper limit of $10-20$\,kpc. 
We do not have spectroscopic data at hand for a direct comparison with their results; however, within 
the same $r$-band surface brightness, and along the minor axis, we find typical sizes of $12-25$\,kpc.
At these radii the flux is dominated by [\ion{O}{III}] emission, not by $r$-band continuum, thus a 
comparison of our measurements and those of \citet{hhg13} are still meaningful. \citet{hhg13} have 
argued that their size limit is caused by the unavailability of gas at larger radii to be ionized. 
Likely, this is the reason why our sample differs so much: it was selected because of its extreme 
broad-band colours, caused by very gas-rich systems.

\citet{lzg13} and \citet{ham14} have also studied the properties of [\ion{O}{III}] NLRs around luminous 
radio-quiet type-2 quasars. The sizes of the [\ion{O}{III}] nebulae in GBs are 
consistent or somewhat larger compared to their results; no corrections are made for methodology. 
Both authors find mostly circular or moderately elliptical, smooth morphologies for the outflows. Irregular
morphologies are commonly coupled with radio excess. This is in stark contrast with the nebulae observed 
in GBs. Most are highly asymmetric, irregular and patchy, apart from J1351+0816 and J2050+0550, which
reveal rather smooth spheres. Again, this could be a selection effect: we found 17 objects in SDSS with 
extreme broad-band photometry, whereas \citet{ham14} chose 16 AGN out of a parent sample of 24000 SDSS
AGN. It is entirely possible that some GBs could have ended up in the sample of \citet{ham14};
however, as we have mentioned in Section \ref{contamination}, the SDSS colour space occupied by GBs is 
too contaminated for automatic source extraction.

For some targets we give simple estimates about the duration of an AGN burst, and/or the time it 
must have occurred in the past. For simplicity, we assume a single, average outflow velocity of 
$v=1000$\,\kms\;and an inclination angle of $\theta=90$ degrees, i.e. the outflows are moving 
perpendicular to the line of sight. Therefore, time estimates must be scaled by 
$1000\;{\rm km\;s^{-1}}\;v_{\rm obs}^{\;-1}\;{\rm sin}^{-1}\,\theta_{\rm obs}$
to obtain the true values.

\subsubsection{\label{agnoutflows}AGN driven outflows}
$65-75$ per cent of the EELRs have AGN outflow signatures, such as collimated beams or symmetric ejecta 
in opposite directions (Table \ref{targetlist}). The outflows 
are usually launched by the injection of thermal energy into the surrounding gas during an AGN 
burst. The heated gas expands and sweeps up (and shocks) colder material along its path. Such 
outflows have been well studied both observationally as well as theoretically; a detailed account 
of these efforts is beyond the scope of our work. We compare our findings to the simulations of 
\cite{gab14}, who typically find unipolar outflows with wide opening angles. Accordingly, denser 
gas on one side of the nucleus may fully stop an outflow and reradiate its energy, while the 
outflow may escape on the other side through a thinner interstellar medium. One object in our sample, 
J0111+2253, fits this picture well. It displays a strong unipolar outflow emerging on one side of 
the nucleus where it is also wide; a weaker second outflow (or ionized material) is seen at a 
30 deg angle, and no outflow is found on the opposite side.

However, J0111+2253 appears to be the exception. For example, we observe bipolar outflows in J0024+3258, 
J0113+0106 and J1347+5453 that are well focused near the geometric centre of the host galaxy or their 
nuclei. The southern outflow in J0024+3258 even appears collimated over 15\,kpc. J1347+5453 is a posterchild 
bipolar outflow, launched from the nucleus of a spiral galaxy perpendicular to the edge-on disk.

Some of the outflows must have been sustained over a prolonged time because their gas is
continuously distributed all the way to the nucleus. Differential velocities in the outflow 
will enhance this effect. In case of J0024+3258, the burst would have lasted $8-11$ Myr assuming no 
velocity dispersion within the stream. Such long (and Eddington-limited) accretion phases are also found 
by \citet{gab14}. Higher resolution images are needed to detect discontinuities in that 
outflow. J0113+0106, on the other hand, appears to have experienced a powerful event $\sim5$ 
Myr ago producing two superbubbles $5-8$\,kpc in size on either side of the nucleus. The bubbles are 
$1.5-2$ times larger than the seeing disk and therefore not well resolved. If the observed distances 
of the gas from the nucleus are caused by differential gas velocities, then this event could have been 
much shorter than 1 Myr. Ionized material at larger distances shows that this recent burst was preceded 
by another one, perhaps $20-30$ Myr ago. Recurrent events likely occurred as well in J1347+5453, J1441+2517, 
and J1504+3439.

Other systems have a more multipolar character with outflows in different directions.
This could be caused by variable gas densities near the nucleus which may partially stop an outflow or 
divide it (as in the northwestern outflow in J1347+5453, and in J0111+2253).

\subsubsection{Cloud systems}
Another typical feature are single or multiple regions of gas, apparently detached from the nucleus.
We refer to them as \textit{clouds}. This could be tidally stripped gas contributed by gas-rich mergers 
and now passing through the AGN's ionization cones \citep[like in \textit{Hanny's Voorwerp}, 
see][]{lsk09,rgj10,kls12}. Typically, these clouds have a relatively smooth appearance and a physical 
size of $5-15$\,kpc. Examples are J1441+2517, J1504+3439, J1505+1944, J2050+0550, and most spectacular 
in J2240$-$0927 \citep[][]{dst15}.

Alternatively, the clouds were ejected during one or more previous bursts, and then disconnected from 
the nucleus and now reside in the galaxies' haloes. Currently, this disconnection could be happening in 
J0024+3258, whose northern outflow appears to be still feeding such a cloud, and in J1347+5453, whose 
southeastern outflow has a similar structure. In both cases the clouds are significantly misaligned with 
the feeding stream, as if they experienced tidal dragging or other interactions with the intergalactic 
medium (see also Section \ref{warps}).

\subsubsection{\label{warps}Warps}
Several EELRs show warps and other symmetric and asymmetric deformations that could be caused by 
various mechanisms. The gas in J0020$-$0531 resembles a spiral with two widely opened arms that become 
thinner with increasing nuclear distance. This could be differential orbital motion, tidal 
interaction with two embedded compact ellipticals, or a continuous change in outflow direction. In 
J0024+3258, J0113+0106, and J1347+5453 it appears that the ejection direction has changed during or 
between bursts, or that the gas has been shaped by interactions with the surrounding halo and/or a 
radio jet. GBs are mostly radio quiet or radio weak \citepalias{sdh13}, and thus jet interaction is 
unlikely. Radio data from the VLA FIRST survey \citep{wbh97} have insufficient resolution 
and depth for further investigation.

Alternatively, the warps could be caused by a change in the ionization cone's opening angle and 
strength because of intervening or sublimating dust. Spin precession of the SMBH and its 
accretion disk could also play a role. Typical precession periods of $10^{3-7}$ years fully overlap 
with the duration of AGN bursts (Section \ref{discSec2}), and the precession cones' half opening angles 
range from $1-70\,^{\circ}$ \citep[see][]{luz05}.

\subsubsection{Smooth spheres}
The EELRs in J1351+0816 and J2050+0550 are featureless spheres in our data (J2050+0550 is accompanied 
by an ionized cloud, see above). All systems for which we observe outflow signatures are highly 
structured, suggesting that a different process has formed these spheres. J2050+0550 is in a 
cluster of galaxies and could be in an advanced merger state, engulfed in gas that is now ionized 
by the AGN. [\ion{O}{III}] is detected at least to a radius of 20\,kpc by our field redshift survey. 
J1351+0816, on the other hand, is isolated in the field. Our redshift survey detects [\ion{O}{III}] 
emission out to a radius of at least 48\,kpc. Some process must have transported the gas to these 
distances. Unfortunately, the depth and resolution of our spectral data are insufficient to obtain 
kinematics and further constraints.

\subsubsection{Peculiar systems}
Three GBs are set apart from the rest by their distinct nebular morphologies. First, J1455+0446 appears 
totally disrupted by a merger. Second, J1504+3409 is reminiscent of the 
\textit{Voorwerpjes}, ionization echoes found at low redshift \citep{kcb12}. It has several ionized 
clouds and bubbles superimposed on the body of a larger \textit{elliptical} galaxy, which distinguishes 
it from the Voorwerpjes (mostly spirals and irregulars).

Third, and most interesting, is J1155$-$0147. This is the brightest and also intrinsically most luminous 
object in our sample (both in terms of FUV/\Lya and [\ion{O}{III}]). It is also the largest object
in terms of area, and second largest in terms of linear diameter (second to J0113+0106). Curiously, it is
also located at the geometric centre of a relatively compact group. The ionized nebula is richly substructered,
fragmenting into smaller clouds. A detailed description is given in Appendix \ref{targetnotes}. Possibly, 
J1155$-$0147 has formed by accretion from the intracluster medium, and its \Lya emission is a mix of AGN 
photoionization and gravitational cooling radiation.

\section{Discussion -- AGN variability and LABs}\label{discussion}
The impact of variable AGN on the appearance of LABs has not yet been studied in detail. In Section 
\ref{discSec1} we review the literature, and in Section \ref{discSec2} we present
theoretical and observational evidence for significant episodic AGN phases. We discuss the effects 
of AGN variability on the \Lyan, MIR and optical properties in Sects. \ref{discSecLYA}, 
\ref{discSecMIR}, and \ref{discSecOPT}, respectively.

\subsection{\label{discSec1}Earlier considerations about variability}
AGN variability as an explanation for the ionization deficits in LABs has not been a serious 
contender in the light of cold accretion, shocks, starbursts, obscured AGN, and resonant 
scattering. Nonetheless, it has been mentioned early on: \citet{sas00} have emphasized the 
absence of strong radio and continuum sources in a luminous LAB and noted the possibility 
of a ``\textit{dead radio galaxy}'', albeit without elaborating the idea further. Later, 
\citet{kwc09} have stated in their summary that ``\textit{Among the proposed explanations 
for \Lya blobs ... [is] photoionization by active nuclei which may be obscured or transient}''. 
The discovery of \textit{Hanny's Voorwerp}, the prototypical quasar ionization echo, has been
published soon thereafter by \citet{lsk09}.


\citet{ond13} have found that LABs with $\llya\gtrsim5\times10^{43}$\,\ergs\; almost always harbour 
a luminous (obscured) quasar. Given that the AGN duty cycle is much shorter than that of cold 
accretion, they have argued that the high incidence of obscured quasars in these LABs implies a
substantial contribution to the ionization of the gas; the latter could still be provided 
by cold accretion streams. Furthermore, given the discovery of ionization echoes, they have stated 
that ``\textit{[...] episodic AGN activity may need to be considered as well when interpreting 
high-redshift LABs}.

\subsection{\label{discSec2}Evidence for AGN flickering}
Cosmological simulations require recurrent periods of rapid black hole growth, setting black hole 
scaling relations and unleashing strong feedback \citep{svg15}. Simulations resolving the gas dynamics 
on sub-kpc scales confirm these sharp intermittent bursts of AGN activity, followed by rapid shutdowns,
on time-scales of $\sim10^5$ years \citep{hoq10,noc11,gdg14}. Mergers, disk bar instabilities, and clumpy 
accretion may boost the quasar-modes further \citep[e.g.][]{bdt11,bjf12}. 

These predictions have been verified observationally by discoveries of ionization 
echoes at $z=0.05-0.35$ \citep{sev10,kls12,kcb12,sdh13,ssk13,kmb15}. AGN must undergo several $100-1000$ 
of these duty cycles (``\textit{flickering}'') to build up their mass \citep{skb15}. Independent 
evidence for flickering, albeit on longer time-scales of $\gtrsim1-10$\,Myrs, has been reported by 
\citet{kit08} and \citet{ful11} studying the transverse proximity effect in the hydrogen and 
helium \Lya forest of selected quasars, respectively \citep[see also][]{khm16}.

\subsection{\label{discSecLYA}AGN duty cycles and delayed \Lya response}
What does AGN flickering mean for LABs? The escape of \Lya photons is delayed because of 
resonant scattering, and the \Lya flux will lag behind the light curve of the ionizing 
source. The effect increases with the optical depth $\tau$, in particular if the ionizing 
source is a central AGN. The mean optical depth at the \Lya line centre is
\begin{equation}
\tau_0=\pi^{-1/2}\,1.04\times10^7\,\left(\frac{T}{10^4\,{\rm K}}\right)^{-1/2}\,\left(\frac{N_{\rm HI}}{10^{20}\,{\rm cm^2}}\right),
\end{equation}
\citep[e.g.][]{neu90,rsf10}, where $N_{\rm HI}$ is the column number density of neutral hydrogen.
Accordingly, LABs with typical temperatures of $\sim10^4$\,K can have a great range of optical 
depths, $\tau=10^{2-4}$ \citep[e.g.][]{dhs06}, or even higher. At much higher temperatures hydrogen 
is mostly ionized and optically thin to \Lya.

The spatial transfer of resonantly scattered \Lya photons does not follow a pure Brownian random 
walk because frequency scattering moves photons out of resonance, and therefore they propagate faster. 
For these purposes the photon frequency is commonly parametrized as $x=(\nu-\nu_0)/\Delta\nu_{\rm D}$, 
measuring the frequency deviation from the resonance frequency, $\nu_0$, in units of the Doppler 
broadening, $\Delta\nu_{\rm D}$. The two maxima of the double-peaked \Lya line profile occur 
at $|x|\gtrsim2$ \citep[e.g.][]{rsf10}.

\citet{rsf10} and \citet{xwf11} have studied the \Lya response of spherical Damped \Lya haloes
(DLAs; constant hydrogen density and temperature) to an ionizing flash of finite duration. The
effect of dust on the escape times is negligible \citep{yrs11}. We summarize their results, 
pertinent to our work, as follows:
\begin{enumerate}
\item{Typical \Lya escape times for $10-100$ kpc DLAs are $10-100$ times longer than in the 
  absence of resonant scattering; the delay scales roughly with $10\,\tau$. \textit{The \Lya response 
  is a very damped version of the light curve of the ionizing source, and the \Lya peak brightness 
  might be reached long after the central source has switched off.}}
\item{Photons with $|x|<4$ are \textit{effectively trapped in an optically thick halo and stored for a 
  long time}, approximately proportional to $\tau$.}
\item{Photons with $|x|<2$ are \textit{thermalised about 10 times sooner than their typical escape time, 
  meaning they have lost all memory about the location, spectrum and time variability of the source.}}
\end{enumerate}

The analyses of \citet{rsf10} and \citet{xwf11} were performed for spherical DLAs with high 
$\tau\geq10^{6-7}$. LABs are complex objects as witnessed by their vast range of morphologies, 
both for our low-z LABs as well as those at high redshift. AGN outflows may clear escape paths 
for the \Lya photons through the neutral hydrogen, whereas other regions maintain a high optical 
thickness. Regardless, the typical double peaked \Lya line profiles show that the effects of 
resonant scattering are ubiquitous and paramount in LABs. Therefore, the three main results
listed above still apply. A more differentiated analysis of these effects on LABs is desirable,
and will be presented in Malhotra et al. (2016; in prep.).

We conclude that the observed \Lya fluxes effectively decorrelate from typical AGN flickering.
LABs with high Ly$\alpha$ luminosities do not require currently powerful AGN. The LABs could 
simply be gradually releasing stored photons from earlier high states, while the AGN actually 
is in a low state. Conversely, a previously dormant AGN could experience several duty cycles, 
stocking up the halo with \Lya photons well before any Ly$\alpha$ manages to escapes. 
\citet{hwr15} report such \Lya deficient radio-quiet quasars, but do not consider variability 
as an explanation.

To give an order-of-magnitude calculation: \citet{noc11} find that AGN spend perhaps 1 per cent 
of their time in quasar mode. If the storage time of an optically thick LAB is 10 times longer than the 
typical burst duration of $\sim10^5$ years \citep{skb15}, then, statistically, in 90 per cent
of the cases we would \textit{not} detect a quasar in X-rays in a randomly selected sample of 
the most luminous LABs (if the AGN is below our detection threshold while being in the low state).
This explains at least some of the non-detections that have been attributed to
heavy obscuration \citep[e.g.][]{gal09,ond13}.

\subsection{\label{discSecMIR}AGN duty cycles and delayed MIR response (thermal echoes)}
\subsubsection{\label{argmirx}GBs must have faded recently to violate the MIR X-ray relation}
The X-ray data require the GBs to be intrinsically weak, violating the MIR X-ray relation 
(Fig. \ref{ichikawa12}). For a fixed column density of $N_{\rm H}=10^{23}$\,\cmss\;, 
the GB sample is a factor of $\sim30$ fainter than expected from the MIR X-ray relation. 
This discrepancy increases to a factor of $80$ for $N_{\rm H}=10^{22}$\,\cmss.

The violation is naturally explained by rapid fading of the AGN. Information about a change 
in nuclear luminosity will take $t_{\,\rm lag}=r_{\,\rm sub}/c$ years to reach the dusty torus at 
its sublimation radius, $r_{\,\rm sub}$. This time lag is between ten to a few hundred years for 
typical tori, and is the \textit{minimum} time for the torus to start a response in the MIR. 
The actual shape of the response, and the time needed by the torus to reach a new thermal 
equilibrium, depend on the torus' radial dust distribution. The directly illuminated surfaces 
at $r_{\rm sub}$ react quickly \citep{nsi08}, whereas shielded and indirectly illuminated parts 
further in the back are delayed. \citet{hok11} have analysed the MIR (and NIR) response of 
various dusty torus configurations to a discrete pulse from the nucleus with $0.5\,t_{\,\rm lag}$ 
length. They find that \textit{compact} tori reach their peak brightness quickly, coincident 
with the arrival time of the end of the pulse at the sublimation radius at $1.5\,t_{\,\rm lag}$. 
At $2.5\,t_{\,\rm lag}$ the MIR luminosity has already dropped again to 50 per cent of the peak 
flux. For \textit{thick} tori the peak brightness will be reached at a much later time,  
$\gtrsim10\,t_{\,\rm lag}$, and the full MIR response can easily be delayed by $10^3$ years 
or more (\textit{thermal echoes}).

The MIR X-ray relation is easily violated by a transient AGN, in particular for ``slow'' tori. 
An individual AGN with high MIR luminosity and low or absent X-ray flux is not necessarily 
deeply obscured; it might just be fading. 

We conclude that the GBs, as a group, have faded by several factors $10-100$ and quicker than
the response times of their dusty tori; more accurate evaluations of $N_{\rm H}$ require deeper
X-ray observations, or observations extending to higher X-ray energies, and have commenced for 
some GBs already. Depending on the response times, this change occurred over $10-1000$ years. The 
[\ion{O}{III}] excess with respect to the MIR emission reported in \citetalias{sdh13} implies 
another drop in luminosity by a factor of $5-50$ over the light crossing time of the optical 
EELR ($10^{4-5}$ years). Taking the thermal echoes and ionization echoes together, the AGN in GBs 
likely faded by $3-4$ orders of magnitude over the last $10^{4-5}$ years.

\subsubsection{Strong thermal echoes are underrepresented in the MIR X-ray relation}
The MIR X-ray relation is based on observational data, and as such individual geometric 
properties, anisotropic shielding as well as AGN variability contribute to its intrinsic 
scatter. The influence of strong variability, however, is small. \citet{hoq10} have shown 
in their sub-kpc simulations that even for active systems with large time-averaged accretion 
rates, the instantaneous inflow rates are modest most of the time. The black hole mass is 
typically built up during many short duty cycles with rapid switch on/off times. The AGN  
thus spend very little time in the transition phases. \citet{noc11} show in their 2D 
simulations that the AGN duty cycles (defined as the fraction of time above a certain 
Eddington ratio) are short, typically $10^{-3}-10^{-2}$ or less for an Eddington ratio of 
0.1. They do not, however, elaborate on the fraction of time the AGN spend on 
\textit{switching} from high to low accretion states (when -- and shortly thereafter -- 
we would observe them as echoes). \citet{skb15} use observations of low-z ionization echoes 
to argue that the time used to switch states is about ten times shorter than the time spent 
in the high state. 

Taking these results together, in a randomly selected sample of AGN, a fraction of 
$\lesssim10^{-4}-10^{-2}$ is expected to be in a significant transient or echo state. The  
small numbers of known ionization echoes suggest an even lower occurrence, both at redshifts 
$z<0.1$ \citep[\textit{Voorwerpjes,}][]{kcb12,kmb15} and at $z\sim0.3$ (this work). However, 
this is a lower limit as these objects need sufficient gas within the ionization cones to 
work as an echo screen in first place; otherwise we cannot recognize them as echoes.

We conclude that the intrinsic scatter observed in the MIR X-ray relation (constructed from 127 
sources) is caused mostly by intrinsic properties rather than AGN flickering. The MIR X-ray 
relation is not an adequate tool to infer properties for AGN that are suspected to be transient.

\subsection{\label{discSecOPT}AGN duty cycles and instantaneous optical response}
The hydrogen recombination time-scale is $\tau_r=(n_{\rm e}\alpha_{\rm B})^{-1}$, where $n_{\rm e}$ 
is the electron density and $\alpha_{\rm B}$ the recombination coefficient for ``Case B'' 
\citep{osf06}. For typical densities of $n_{\rm e}=50-200$\,cm$^{-3}$ \citetext{\citetalias{sdh13}, 
\citealp{dst15}} $\tau_r\sim500$ years in the denser parts of the GBs, and $\gtrsim2000$ years 
for the lower densities further out in the nebula. These are short compared to the typical 
light crossing times of the resolution elements in our data (a few kpc, Table \ref{genprops}), 
unless the density becomes very low $(n_{\rm e}\sim1$\,cm$^{-3})$. The recombination time-scale of 
O$^{++}$ is about one order of magnitude shorter than that of hydrogen \citep{bir87}. Therefore, 
the response of the GBs' [\ion{O}{III}] and \Ha lines to a sudden change of the ionization 
parameter can be considered instantaneous, which we do throughout this paper.

\subsubsection{Reconstructing historic X-ray light curves}
On a side note, their quick optical response makes ionization echoes suitable to reconstruct 
historic X-ray light curves. If the echoes' physical extent exceeds 10 kpc as in GBs, then
the reconstructed time line would reach $10^{4-5}$ years into the past, directly
testing AGN accretion models. 

This requires that the ionization parameter and the ionizing spectrum can be inferred locally 
and with good accuracy. The hardness of the ionizing spectrum can change locally e.g. due to 
anisotropic shielding of the nucleus (ionization cones) and local star bursts. In addition, 
the 3D cloud must be de-projected, translating angular separations into true physical distances 
(light travel times) to the nucleus. Such a de-projection is facilitated using Doppler mapping 
and extinction maps to break the foreground-background degeneracy. Additionally, differential 
decay times of various optical lines would help \citep{bir87}. Nonetheless, this task is 
formidable. Such a reconstruction of the X-ray light curve has yet to be demonstrated. GBs are 
ideal for this purpose as they are well resolved and offer high flux densities suitable 
for 3D spectroscopy.



\section{Discussion -- Evolution of LABs}
\subsection{\label{discussion1}The LAB size--luminosity function}
Figure \ref{labstats} shows the size--luminosity function for our low-z LABs and some high-z LABs. 
For this plot we assume that 75 per cent of the \textit{GALEX} FUV flux is caused by \Lya 
(Sects. \ref{starlight}$-$\ref{fuvJ1155}). The GBs' \Lya luminosities overlap well with those of 
the high-z LABs, whereas the GBs' ([\ion{O}{III}]) nebulae appear more compact than high-z LABs. 
We evaluate the validity of this comparison below.

\subsubsection{Size estimates}
\citet{myh04} and \citet{yze10} list isophotal surface areas which, for our comparison, we converted 
to physical diameters assuming circular shapes. These diameters are biased toward smaller values 
when compared to major elliptical diameters used by \citet{ebs11}, \citet{pdj13} and also by us.
No correction was made for this effect.

Can [\ion{O}{III}] be used to estimate the \Lya extents? \citet{dst15} show for one GB that 
[\ion{O}{III}] and H$\alpha$ occupy similar volumes, and thus the [\ion{O}{III}] emission will 
provide a lower limit to the \Lya extent because of resonant scattering. Long-slit observations 
near J1351+0816, J2050+0550, and J2202$+$2309 reveal [\ion{O}{III}] emission out to large radii, 
suggesting that the nebulae are up to 4 times larger than measured in our images (see Appendix 
\ref{targetnotes}). 

\subsubsection{Survey depths}
\citet{sso06} and \citet{yze10} find that different survey depths significantly affect size 
estimates for LABs. \citet{myh04,myh11}, \citet{yze10}, and \citet{ebs11} use $1\sigma$ surface 
brightness limits of 2.2, 5.5, 1.5 and $1.8\times10^{-18}$\,\ergscm\,arcsec$^{-2}$, targeting $z=3.1$, 
2.3, 2.3 and 3.1, respectively. We did not correct sizes for differential survey depths
\citep[see][for a discussion]{sbs11}.

How does our survey depth compare to theirs? Our limiting $r$-band isophotes are 
$2\sigma$ above the sky noise (Table \ref{targetlist}, $\sim2.3\times10^{-16}$\,\ergscm\,arcsec$^{-2}$). 
Redshifted to $z=3.1$, this becomes $2.3\times10^{-18}$\,\ergscm\,arcsec$^{-2}$ if all flux was caused 
by a single emission line, and thus our depth is comparable. However, this is built on the assumption 
that the [\ion{O}{III}] surface brightness is an unbiased estimator of the \Lya surface brightness. 
Line ratios of \Lyan/[\ion{O}{III}]$\,\sim0.1-8$ for other \Lya emitters \citep{kww02,mfr11,ond13,mrr14} 
show that this will not hold up in general.

We conclude that low-z LABs have similar \Lya extents as high-z LABs, yet direct \Lya imaging is 
required for an unbiased view. While the \Lya luminosities of low- and high-z LABs match well, the 
high-z Universe is capable of producing more powerful LABs. This could e.g. be caused by cooling 
flows, either alone or in addition to AGN ionization.

\begin{figure}
  \includegraphics[width=1.0\hsize]{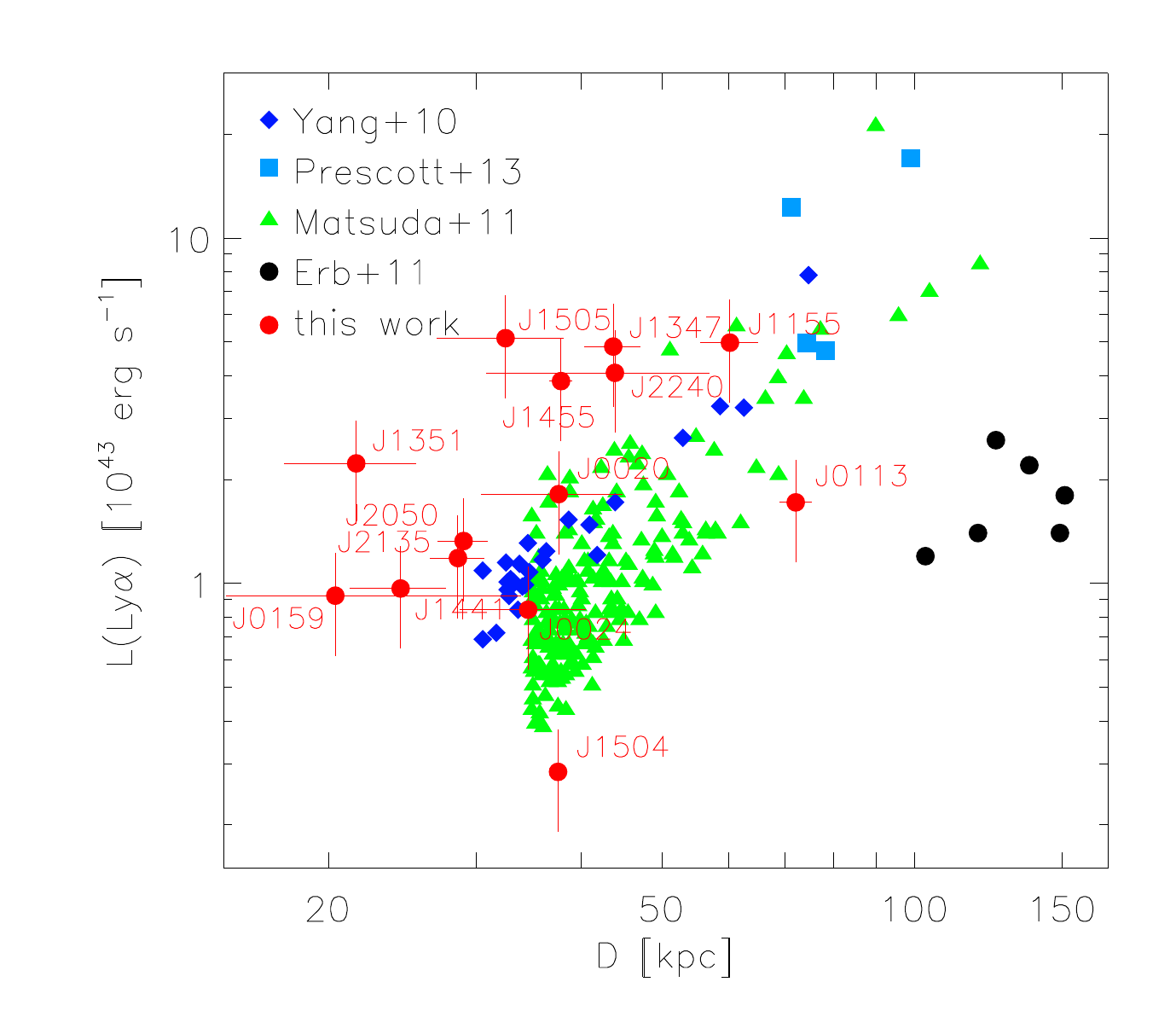}
  \caption{\label{labstats}{Qualitative comparison of LAB sizes and luminosities. The data 
      have different selection functions and size definitions (see text). Our sample is shown
      by the red dots. Therein, the lower and upper ends of the vertical error bars represent 
      fractions of 50 and 100 per cent of the FUV flux being caused by \Lyan, the data points 
      are drawn at 75 per cent.}}
\end{figure}

\begin{figure*}
  \includegraphics[width=1.0\hsize]{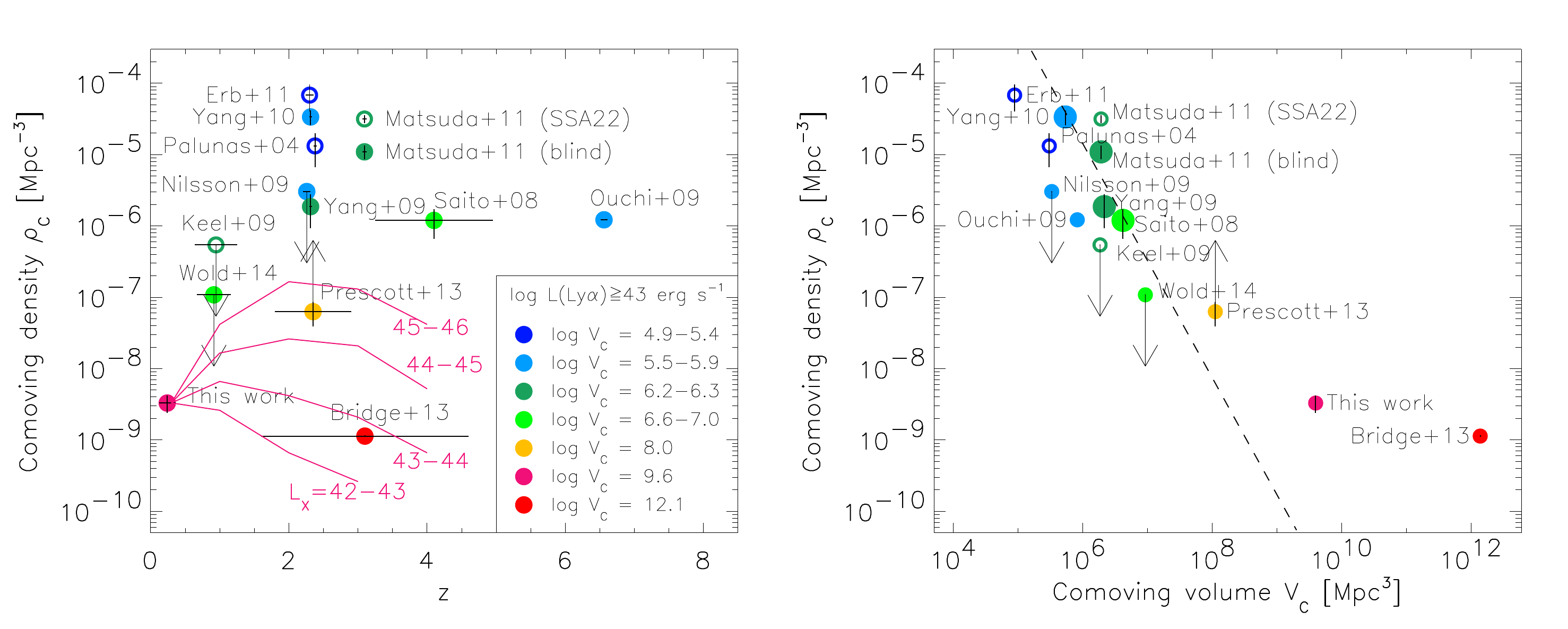}
  \caption{\label{lab_comoving_dens}{Left: Comoving density of LABs 
      ($L_{{\rm Ly}\alpha}\geq10^{43}$\,\ergs, $D\geq20$\,kpc) as a function of redshift. Solid dots 
      represent blind surveys and open dots cluster areas. Horizontal lines show the redshift range 
      covered, vertical lines are the Poisson errors. Arrows represent upper limits. Symbol colours 
      display the survey comoving volumes (see also the right panel). The pink lines show how GBs 
      evolve for different X-ray luminosities if linked to the AGN density evolution (Section \ref{diffevolution}). Right: Comoving 
      density as a function of the surveys' comoving volume. There should be no dependence on volume
      if the surveys are independent of the environment. The dashed line shows the best fit to the 
      blind surveys at $z=2-4$ (larger filled dots). We kept the colour coding for easier comparison 
      with the left panel.}}
\end{figure*}

\subsection{Evolution of the LAB comoving density}
With our complete sample of GBs, selected from a comoving volume of 3.9 Gpc$^3$ ($z=0.12-0.36$ over 
14500 deg$^2$ of SDSS), we can for the first time pin down the LAB comoving density, $\rho_c$, 
and its evolution at low redshift. Figure \ref{lab_comoving_dens} displays $\rho_c$ for the GBs and 
for some high-z LABs from the literature. The symbol colour encodes the comoving volumes 
probed by the surveys, spanning seven orders of magnitude between $10^{5-12}$ Mpc$^3$. The left panel 
in Fig. \ref{lab_comoving_dens} reveals a range of five orders of magnitude in density between surveys. 
Highest densities are found for (proto-)cluster structures at $z=2-3$ (open symbols), whereas blind 
surveys (solid dots) yield lower densities. We describe the evolution as a power law, 
$\rho_c\propto(1+z)^n$. Our main results are summarized as follows:
\begin{itemize}
\item{We confirm the previously reported strong evolution below $z=2$; LABs mostly
  disappear before $z=0.3$. The slope $n$ remains uncertain. It could be as low as $n=1.7$ 
  over $z=4$ to $z=0.3$, or as high as $n=6$ between $z=0.85$ and $z=0.3$.}
\item{High-z and low-z LAB populations are fundamentally different. Most likely, cold accretion 
  streams exhaust sometime between $z=2$ and $z=0.3$. At $z=0.3$, LABs are mostly powered 
  by AGN, and these follow a flatter evolution all the way from $z=4$.}
\item{There is an expected strong dependence of the density on the survey volume, 
  $\rho_c\propto V_c^{\,-1.6}$. At least Gpc volumes should be probed to appreciate the cosmic 
  large scale structure and cosmic variance, otherwise density measurements from different 
  surveys are difficult to compare.}
\end{itemize}

The following unpublished information has also been used for our analysis: \citet{bbb13} have 
confirmed that all spectroscopically verified LABs in their sample have $\llya\geq10^{43}$\,\ergs. 
\citet{wbc14} have confirmed that no other resolved \Lya emitters than the one presented in 
\citet{bcw12} are present in their survey. Finally, \citet{myh11} have provided the full list of 
LABs including those with diameters less than 100\,kpc. We combine their cluster field (SSA22Sb1) 
and the adjacent fields (SSA22Sb2-7) into one data point and consider it a cluster field 
(``Matsuda+11 (SSA22)''); we refer to their comparison field survey as ``Matsuda+11 (blind)''.

\subsubsection{Selection effects and biases}
There is currently no definition for LABs \citep[see e.g.][]{sbs11}. For this analysis we require 
a minimum diameter of $D\geq20$\,kpc and a minimum luminosity of $\llya\geq10^{43}$\,\ergs. The latter 
is a safeguard against completeness issues, biasing us against LABs with low \Lya escape fractions. 
13 of the 17 GBs are retained within these limits. Most surveys used for our comparison are sensitive 
to several $10^{42}$\,\ergs, and thus should have picked up most LABs with $\llya\geq10^{43}$\,\ergs\; 
in their search volumes. The broad-band survey of \citet{pdj12,pdj13} is an exception, as only the 
brightest and largest LABs are detected. We then count how many LABs in the comparison surveys fulfil 
our criteria, and we recompute the comoving volumes and densities according to our cosmology. For some 
surveys only a sub-sample of the LAB candidates was followed up spectroscopically, and the total number 
is extrapolated from the fraction of verified sources. We adopt these correction factors whenever 
available \citep[e.g.][]{pdj13,bbb13}.

Constraining the density evolution is complicated by heterogeneous surveys and selection 
functions: \citet{myh11}, \citet{sso08} and \citet{yzt09} use narrow-band imaging to select \Lya 
directly, and \citet{pdj12} use broad-band imaging. Spectroscopy is chosen by \citet{kwc09}, 
\citet{bcw12} and \citet{wbc14}, whereas \citet{bbb13} select dusty \Lya emitters using MIR 
data. Lastly, our LABs were selected based on their [\ion{O}{III}] strength in $r$-band.
We did not correct for any such selection effects. Further biases arise due to cosmic variance 
\citep{yzt09} and because of the original purpose of a survey (clusters, empty fields, etc). 
We analyse the density evolution separately for blind surveys and cluster surveys.

Lastly, LABs with different ionization sources (cold accretion, AGN, etc.) evolve differently and 
should be studied separately. However, determining the power source(s) of individual LABs is 
notoriously difficult (Section \ref{intro}). We ignore this issue for the moment and address it in 
Section \ref{diffevolution}.

\subsubsection{\label{clusterdensity}Representative surveys require $10^8$Mpc$^3$ volumes}
Ideally, measurements of the comoving density are independent of the survey volume. If objects 
such as LABs are rare, and preferentially found in certain environments such as clusters, then 
the search volume must be large to become representative. The colour coding in the left panel 
of Fig. \ref{lab_comoving_dens} shows that this is not the case for most LAB surveys. We illustrate this 
in the right panel of Fig. \ref{lab_comoving_dens}, plotting the density as a function of volume. Based 
on the blind surveys at $z=2-4$, excluding surveys with upper and lower limits and the work of \citet{bbb13}, 
we find $\rho_c\propto V_c^{\,-1.6\pm0.5}$ (dashed line). In particular narrow-band surveys are observationally 
expensive when it comes to probing large volumes, as they cut out thin redshift slices, only.

Most current search volumes do not appreciate the full cosmic environment required to produce LABs. How 
big should these volumes be? LABs with $\llya\geq10^{43}$\,\ergs\; and $D\geq20$\,kpc are rare objects at 
$z\sim2-4$. They are preferentially found in massive structures with masses of up to $10^{15}$ $M_\odot$ 
\citep{ptf04,myh11,ebs11,kyi15}, which will evolve into the richest clusters known today. What is 
the volume density of such massive clusters? \citet{eeh01} have compiled a statistically complete sample 
of the most massive X-ray selected clusters between $z=0.3-0.7$, for which we calculate an average weak 
lensing mass of $(1.3\pm0.4)\times10^{15}$ $M_\odot$ based on the data of \citet{alk14}. We derive the 
clusters' comoving density as $3.8\pm0.4$ Gpc$^{-3}$. To be \textit{fully} representative of the environment, 
an LAB search volume should therefore contain at least one Gpc$^3$. However, much smaller volumes of 0.1 
Gpc$^3$ should be sufficient when studying the LAB density in the field, e.g. when probing the fast 
evolution between $z=0.5-1.5$ (Section \ref{densityevolutionfield}).

Our subsequent analysis of the LAB density evolution is limited by this effect. We apply approximate 
corrections when meaningful.

\subsubsection{Previous evidence for rapid density evolution}
\citet{bcw12} have searched 0.7 deg$^2$ of the Chandra Deep Field South for \Lya emitters between 
$z=0.67-1.16$ using \textit{GALEX} grism images. They have found only one extended source at 
$z=0.977$ with $\llya=7.2\times10^{42}$\,\ergs\; and $D=120$\,kpc. This was the lowest redshift LAB 
known so far. For comparison, J1155$-$0147 ($z=0.306$) in our sample is nine times more luminous 
(Table \ref{targetUVprop}) while having $D=65$\,kpc (in [\ion{O}{III}]). \citet{wbc14} 
have extended the survey of \citet{bcw12} to four times the area, but have not found any other 
resolved \Lya emitters.

To study the evolution of LABs in clusters, \citet{kwc09} have conducted a \textit{GALEX} grism search 
between $z=0.64-1.25$ toward two massive clusters. Numerous LABs \citep[like in][]{myh04} should 
have been found in the absence of evolution, yet none was discovered. \citet{kwc09} have concluded 
that the LAB density in clusters must be evolving at least as fast as $n=3.0-3.5$ between 
$z\sim0.8$ and $z=2-3$.

\subsubsection{\label{densityevolutionfield}LAB density evolution in the field}
For the moment, we marginalize the nature and powering sources of LABs, and we neglect the work of 
\citet{bbb13} (see Section \ref{bridge}). Our survey is a blind survey at $z=0.3$. In comparison with 
the blind surveys of \citet{sso08}, \citet{yzt09,yze10} and \citet{myh11} we find $n=2.3\pm1.0$ for 
$z=0.3-4$, using an unweighted fit. Comparing against \citet{yzt09,yze10} only, 
the evolution accelerates to $n=3.6\pm1.2$ between $z=2$ and $z=0.3$. Comparing our result to the 
upper limit of \citet{wbc14}, only, the evolution could have been as fast as $n\sim4$ between 
$z=0.85$ and $z=0.3$.

Unfortunately, the sampling between $z=0.4-1.5$ is sparse, and it is this interval
of cosmic time when $99-99.9$ per cent of the LABs disappear.

\subsubsection{\label{densityevolutionclusters}LAB density evolution in clusters}
At high redshift, LABs are preferentially found in massive structures with masses of up to 
$10^{15}$\,$M_\odot$. These systems evolve into the most massive clusters known at low redshift. 
However, the majority of our low-z LABs are found in isolation or small groups. Only four 
of them (J1155$-$0147, J1505+1944, J2050+0550, J2202+2309) are in ``denser'' environments with 
masses of up to a few $10^{13}$\,$M_{\odot}$.

Have we possibly missed LABs in massive clusters due to selection effects? We do detect luminous 
EELRs if superimposed on brighter ellipticals (e.g. J1504+3439 and J2202+2309 with $M_i\sim-22.5$ 
mag). However, we would not detect them on top of BCGs in the most massive clusters, as 
the [\ion{O}{III}] EW would become too small to distinguish the galaxies in broad-band colour 
space (Fig. \ref{GB_colorspace}). Some BCGs in massive clusters host emission-line nebula, 
powered by AGN outflows, starbursts or cooling flows from the IGM. For a few of them \Lya 
luminosities are known. \citet{hue92} find $L_{\rm Ly\alpha}=(1-5)\times10^{41}$\,\ergs\; for 
clusters at $z=0.02-0.06$, including NGC 1275A, the BCG of the Perseus cluster. This is a 
factor of $\sim100$ below the typical \Lya luminosities of GBs. Going to $z=0.06-0.29$, 
\citet{obm04,oqo10} find $L_{\rm Ly\alpha}=(0.3-2.8)\times10^{43}$\,\ergs\; for 9 massive clusters 
with a median of $0.56\times10^{43}$\,\ergs\; that is at the bottom of the GB \Lya luminosity 
function. Abell 1835 ($z=0.253$) is the most luminous system in their sample, reaching 
about half the luminosity of the most luminous GBs. Such objects would not normally be considered 
as LABs because of their strong continuum.

Given our non-detection, we conclude that LABs are largely absent in clusters with 
$M_{200}\gtrsim10^{14}$\,$M_{\odot}$ (unless they are superimposed on the most massive BCGs). This 
corresponds to an upper limit of $\rho_c\sim2.5\times10^{-10}$\,Mpc$^{-3}$ for clusters at $z\sim0.3$,
integrated over our entire survey volume. However, we cannot compare this \textit{directly} with 
the densities for clusters at $z=2-3$ \citep{ptf04,myh11,ebs11} due to the environmental dependency 
of the latter studies caused by their small search volumes. Statistically, the density of clusters 
with masses $\gtrsim10^{15}$\,$M_{\odot}$ at $z=0.3-0.7$ is $3.8$ Gpc$^{-3}$ (Section \ref{clusterdensity}). 
We expect 15 such clusters in our search volume. If at most one of them contains a LAB, then we have 
a conservative upper limit of $\rho_c<3.5\times10^{-8}$ Mpc$^{-3}$ over the same search volume 
as \citet{myh11}. This yields a slope of $n>2.8\pm0.6$, compared to $n>3.0-3.5$ obtained by 
\citet{kwc09}.

We conclude that, like in empty fields, the conditions that support LABs in dense environments at 
$z\sim2$ do not hold anymore at low redshift. This implies that cold accretion streams are important 
at high redshift when pristine gas is still abundant in the IGM, whereas they must have been nearly 
depleted at low redshift. 

\subsubsection{\label{diffevolution}Differential evolution of LABs, or how many GBs do exist at 
high redshift?}
We have shown that (1) only LABs in low density environments survive redshift evolution, (2) that low-z
LABs are powered by AGN (Section \ref{agnoutflows}), and (3) that cold accretion streams have likely 
ceased at $z\sim0.3$. Consequently, the low-z LAB population is profoundly different from the high-z 
population, for which accretion streams abound. If the low-z population of AGN-powered LABs (i.e. the GBs)
exists independently of the high-z population, how many GBs would we expect at 
high redshift? In the following consideration we ignore the possibility that high-z LABs, powered 
by both AGN and cold accretion, could become pure AGN-powered LABs at low redshift. We also ignore 
AGN flickering which moves individual AGN across luminosity bins. And we assume that the occurrence 
of an AGN-related GB phase is independent of redshift.

If the GB phase is a (short-lived) phenomenon common amongst active galaxies, then the GB volume 
density is tied to the evolution of the AGN volume density. The latter is well known 
\citep[e.g.][]{rgw13,mhs15} for different AGN X-ray luminosity bins. We predict the GB volume 
density using Fig. 6 in \cite{mhs15} for redshifts $z=1-4$. The predictions are shown by the pink 
lines emerging from our data point in the left panel of Fig. \ref{lab_comoving_dens}. With the exception
of the most luminous bin, it is evident that this evolution is much flatter than the evolution of 
``typical'' high-z LABs.

If GBs adhere to the lowest luminosity bin, log$(L_x)=42-43$, then with respect to $z=0.3$ their 
density would be unchanged at $z=1$ and dropped by a factor of 10 at $z=3$. For the second lowest 
bin, log$(L_x)=43-44$, the density would remain constant within a factor of 2 out to $z=3$. For the 
two brightest bins, the density would increase significantly by factors $3-10$ at $z=1$. If GBs 
were represented by the brightest bin only, then their density would increase further by a factor 
50 at $z=2-3$.

Now we can estimate how many high-z GBs have been found by previous LAB surveys. For 
the volume probed by \citet{yzt09}, we expect 0.001, 0.01, 0.06 and 0.4 GBs for the four bins of 
increasing X-ray luminosity, respectively. For the blind survey part of \citet{myh11} 
we expect 0.0006, 0.005, 0.05, and 0.3 GBs. However, these surveys have found 4 and 21 LABs, respectively, 
that match our selection criteria ($D\geq20$\,kpc, $\llya\geq10^{43}$\,\ergs). For the broad-band survey 
of \citet{pdj13} we find 0.02, 0.1, 1, and 6 GBs, averaging over the survey's large redshift range, and 
based on a sub-sample of GBs with $\llya\geq5\times10^{43}$\,\ergs\; detectable by this survey. 
\citet{pdj13} extrapolate that 18 such LABs exist in their volume. We conclude that it is unlikely that 
GBs have been detected in any current high-z survey, unless the survey volume exceeds $10^7$\,Mpc$^3$ 
and GBs belong to the intrinsically most luminous AGN (currently they are in a low state).


\subsubsection{\label{bridge}Comparison with \citet{bbb13}}
Now we can also better interpret the comoving density of \cite{bbb13}, with $1.1\times10^{-9}$ Mpc$^{-3}$
$100-10000$ times below that of other measurements at $z=2-4$ (Fig. \ref{lab_comoving_dens}). The MIR 
approach chosen by \citet{bbb13} selects LABs with dusty AGN, most of them between $z=1.6-2.8$. X-ray 
observations of a representative large sub-sample are not yet available. \citet{sla14} have observed 3 of 
the sources (one of which an LAB). They argue that high MIR fluxes and low X-ray counts favour 
Compton-thick obscuration over intrinsically weak sources; variability is not considered (see also Section 
\ref{fadingagn}). Larger samples are required to characterize this population of LAB/AGN. 

The left panel of Fig. \ref{lab_comoving_dens} shows that GBs would evolve to similar volume densities 
at $z\sim2$ for log$(L_X)=42-44$. It is plausible that at least some of the LABs described by 
\citet{bbb13} and the GBs belong to the same physical population. Bridge, C.R. report that the \Lya 
emission of their LABs is rather filamentary based on the extent of the emission measured at 
various position angles (priv. comm.). This would also be the case for our GBs, a large fraction of which 
reveals elongated outflows. Another commonality is that both populations are unusually red in the MIR,
although for GBs this is based on the W3 and W4 passbands \citepalias{sdh13} compared to W1 and W2 for
\citet{bbb13}.

\subsubsection{\label{overzier}Meeting the postulation of \citet{ond13}}
\citet{ond13} argue that most luminous LABs with $\llya\gtrsim5\times10^{43}$\,\ergs\;harbour 
obscured AGN, and that these AGN are responsible for most of the ionization in the LABs. They consider 
that ``\textit{If the luminous LAB phenomenon is associated with powerful AGNs, we would naively expect 
that the population of LABs should extend toward much lower redshifts [...]}''. Since the colder gas-rich 
conditions of $z\sim2$ proto-clusters, in which the most luminous LABs are found, are replaced by 
a hot intracluster medium at $z\lesssim1$, \citet{ond13} suspect that 
``\textit{[...] luminous LABs at low redshift may still be found in the more typical, lower density, 
and gas-rich environments of actively accreting galaxies and AGNs.}''.

This is a spot-on prediction of the properties of our low-z LABs: 85 per cent have \Lya luminosities
in excess of $10^{43}$\,\ergs, 75 per cent are found in isolation or with a small number of companion
galaxies, the remaining 25 per cent are found in low mass groups with low concentration.
And lastly, all of our LABs harbour AGN.

\subsection{\label{incompleteness}Are there type-1 Green Beans, and Green Beans with currently active AGN?}
Our optical broad-band selection criteria (Section \ref{contamination}) favour absorbed (type-2)
AGN; unobscured AGN would reduce the observed EW of [\ion{O}{III}] by means of their high continuum 
fluxes, making the galaxies integrated colours inconspicuous. An unobscured AGN would also partially 
or fully outshine the EELRs, and the latter would be difficult to recognize without PSF subtraction.

The observed obscuration of an AGN strongly depends on the observer's viewing angle \citep{ant93}, and 
therefore unobscured (type-1) GBs must exist as well. 
It is also well-known that the fraction of obscured AGN decreases with increasing X-ray 
luminosity \citep[e.g.][]{mbb14}. Our work, and that of e.g. \citet{hoq10,noc11,gdg14,skb15} show 
that high amplitude X-ray flickering is an inherent property of AGN. This suggests that intrinsic 
obscuration is also variable and roughly contemporal with the burst duration. The details
are beyond the scope of our paper. We expect that unobscured low-z LABs (or type-1 GB analogs) 
must exist in our SDSS search volume, perhaps as many as described in this work.

The following question must be seen in the same context: How many GBs exist with their AGN still 
being in the high state, and how would they look like? \citet{bir87} have shown that [\ion{O}{III}] 
responds nearly instantaneously to a change of ionizing radiation, at least when considering volumes 
with kpc scales such as in GBs. That means that the life-time of an observable GB phase is tightly 
linked with the life-time of the AGN bursts; a strong ionization echo is only observed as long as 
the ionizing radiation from a previous high state is still escaping the galaxy. Therefore, the ratio 
of galaxies that appear as echoes (GBs) and their progenitors (with the AGN still in a high state) 
is similar to the ratio of the AGN fading time-scale and the duration of the AGN high state. Both 
occur on scales of thousands to hundreds of thousands of years 
\citep[this work, and][]{hoq10,noc11,gdg14,skb15}. It is plausible that there might be $1-100$ 
times as many GB ``progenitors'' as GBs themselves. As argued above, the optical appearance of the 
progenitors (and their obscuration) is likely to change \citep[see also the \textit{changing look 
quasars}][]{lcm15,rcr16}; we expect galaxies with brighter nuclei, making it more difficult to 
recognize their EELRs.

Imaging surveys with greater angular resolution than SDSS (e.g. LSST) would simplify the 
identification for unobscured GBs and their progeenitors. They are likely already contained 
in public quasar catalogues because of their high fluxes ($r\lesssim18$ mag).

\section{Summary and conclusions}
Green Bean galaxies (GBs, $z\sim0.3$) are spectacular and the most powerful emission-line objects 
known in the nearby Universe. Their extended emission-line regions (EELRs) measure between 
$20-70$\,kpc, and they are ionized by radio-weak type-2 quasars. GBs are extremely rare with a 
surface density 
of $1.1\times10^{-3}$ deg$^{-2}$; only 17 of them are known. They were selected photometrically from 
the 14500 deg$^2$ footprint of SDSS-DR8. In \citetalias{sdh13} we have 
suspected that the EELRs are ionization echoes, i.e. the AGN have faded recently and more quickly 
than the EELRs' typical light crossing times. GBs have not been investigated further apart from 
the single case study of \citet{dst15}, 

In this paper we have presented multi-wavelength observations to understand the unusual nature of 
GBs and to view them in the context of galaxy evolution. With \textit{Chandra} we have probed the current 
activity of the AGN and the intrinsic obscuration along the line of sight toward 10 GBs. With 
\textit{GALEX} archival images we have estimated the far-UV properties of 15 GBs, and we have obtained 
high resolution optical images in $gri$ filters for all objects. 

\subsection{Main results}

\begin{enumerate}
\item{\textit{Chandra} has revealed low counts, moderate hardness ratios and weak or absent 
K$\alpha$ lines. This implies that these AGN are intrinsically weak rather than Compton-thick. 
The EELRs' high [\ion{O}{III}] luminosities require recent and rapid fading of the AGN, confirming 
them as ionization echoes (Section \ref{fadingagn}).}
\item{Strongly variable AGN do not follow the MIR X-ray relation. The MIR response of a dusty torus 
can be delayed by up to $\sim10^3$ years \citep{hok11}, forming a thermal echo. Accordingly, the AGN 
in the GBs must have faded by several factors $10-100$ over the last $100-1000$ years (Section 
\ref{argmirx}). Combining the thermal and ionization echoes, the AGN must have faded by $3-4$ orders 
of magnitude within the last $10^{4-5}$ years. This rate is similar to that observed in the 
\textit{changing look} quasars \citep[][also attributed to change of accretion rates]{lcm15,rcr16};
however, in GBs it could be sustained over much longer periods of time.}
\item{\textit{GALEX} FUV images require that at least 85 per cent of the GBs have \Lya luminosities 
in excess of $10^{43}$\,\ergs. They form Lyman-$\alpha$ blobs or LABs (Section \ref{GALEXresults} and
\ref{discussion1}). We have proven that LABs still exist in the Universe $4-7$ billion 
years later than previously known. Ultimately, the \Lya emission has to be confirmed with 
FUV spectroscopy using \textit{HST} or \textit{Astrosat} \citep{hut14}.}
\item{We propose rapid duty cycles (AGN flickering) as a natural explanation for the mysterious 
ionization deficits observed in LABs. Resonant \Lya photons are efficiently stored
in LABs and only gradually released, \textit{decorrelating} from AGN variability on scales of 
up to $10^6$ years depending on the optical depth. An AGN may undergo several duty cycles before 
\Lya escapes; a luminous LAB does not require a \textit{currently} powerful AGN, independent of
obscuration (Section \ref{discSecLYA}). This does not mean that we can relinquish e.g. cold accretion, 
star formation and shocks as ionization sources. It means that multi-wavelength observations are 
required to identify the ionizing source(s) of individual LABs.}
\item{Low-z LABs live mostly in isolation or in low density environments (Section \ref{longslit}), 
whereas high-z LABs are preferentially found in massive structures. Sometime between $z=2$ and $z=0.3$ 
these structures must lose their ability to form and sustain LABs, probably because cold accretion 
streams have depleted. AGN survive, and may continue powering LABs at low redshift (Section 
\ref{agnoutflows}).}
\item{Our comoving volume is 3.9 Gpc$^3$, $100-1000$ times larger than other LAB surveys, second only 
to \cite{bbb13}. The density at $z\sim0.3$ is $\rho_c=3.3\pm0.9$\,Gpc$^{-3}$ for LABs with 
$\llya\geq10^{43}$\,\ergs\; and $D\geq20$\,kpc. The density evolves with $\rho_c\propto(1+z)^{3-4}$ for 
both clusters and in the field between $z=0.3$ and $z=2$. During this time, $99-99.9$ per cent of the 
LABs disappear. A more accurate determination of the evolution requires (1) better sampling between 
$z=0.5-1.5$, and (2) volumes of $0.1-1$\,Gpc$^3$ to overcome cosmic variance and 
to be independent of environment. Otherwise, densities are not directly comparable (Section 
\ref{clusterdensity}).}
\item{LABs with different ionization sources evolve differently. Gravitationally powered 
LABs do not survive redshift evolution as the cold accretion streams are depleting. We find only 
one LAB, J1155$-$0147 ($z=0.306$), that could still be powered by cold streams in addition to an 
AGN. The density of LABs powered solely by AGN is evolving much slower, if at all, depending on the
AGN's intrinsic X-ray luminosity (Sects. \ref{diffevolution} and \ref{bridge}).}
\end{enumerate}

\subsection{Conclusion and future observations}
LABs should be considered as efficient \Lya photon stores, albeit with a badly maintained inventory. 
They trap \Lya photons for a time much longer than their light crossing time. When the thermalised 
photons are eventually released in a gradual manner, all memory about the location, spectrum and 
time variability of their source has been lost. Typical storage times are short compared to the life 
times of cooling flows, star bursts, and shock ionization populating the store with photons. However, 
the storage times become long compared to episodic AGN bursts and their transition from high to 
low states (and vice versa). Even if optical and MIR observations point to the most powerful AGN, the
absence of the latter in X-rays does not necessarily mean Compton-thick obscuration. 

The AGN timescale 
is much shorter than the typical galaxy time-scale and the galaxies response time to AGN activity; 
care must be taken when seeking or applying relations between AGN and other ``galactic'' observables,
in particular if the AGN are suspected to be transient \citep[see also][]{hma14}.

The GBs and the associated LABs published in this paper are \textit{much} easier to study than their 
high-z counterparts, as low fluxes, redshift incompleteness, and physical resolution are not a problem. 
They are perfectly suited to study AGN feedback, large-scale outflows, quasar duty cycles, mode switching, 
and the \Lya escape fraction, the latter being controlled by an interplay of geometry and velocity field, 
metallicity, hydrogen density, dust obscuration and AGN variability.

Our own \textit{NUSTAR} observations have commenced in cycle 2 for the two X-ray brightest GBs, 
J1155$-$0147 and J0113+0106, to accurately determine the obscuration and the actual shut-down 
``depth'' of their AGN. Time has also been awarded for an initial survey with the 
\textit{Hubble Space Telescope}, to determine the FUV properties of the same two GBs, and 
J2240$-$0927 \citep{dst15}, using ACS/SBC imaging and spectroscopy.

\section*{Acknowledgements}
I (MS) thank my wife Karianne and my children Hendrik and Jakob for their patience over the last 
four years. Carrie Bridge, Gary Ferland, Pascale Hibon, Kohei Ichikawa, Lia Sartori, Mark Schartmann, Peter Schneider, 
Isak Wold, Yujin Yang, and the anonymous referee helped to improve this paper with discussions and 
comments. Carrie Bridge, Yuichi Matsuda and Isak Wold shared unpublished data for which I am very 
grateful. 

The authors wish to recognize and acknowledge the very significant cultural role and reverence 
that the summit of Mauna Kea has always had within the indigenous Hawaiian community. We are most 
fortunate to have the opportunity to conduct observations from this mountain. 

Author contributions: MS obtained all data, performed the scientific analysis and wrote the 
manuscript; SM pointed out the delayed \Lya escape times with respect to the hydrogen recombination 
time-scale; ultimately, that led to the solution of the LAB ionization deficit problem; NAL reduced 
the \textit{Chandra} data and extracted the X-ray fluxes; HF first suggested the possibility that 
GBs might be low-z LAB analogs; MS, RLD, HF, WCK, and PT discussed AGN ionization echoes in depth; 
VNB, AP and WCK provided missing redshifts for two GBs using the Lick observatory; JEHT double-checked 
the flux calibration of the IFU data.

Support for this work was provided by the National Aeronautics and Space Administration 
through \textit{Chandra} Award Number GO4-15110X (PI: M. Schirmer) issued by the 
\textit{Chandra} X-ray Observatory Center, which is operated by the Smithsonian Astrophysical 
Observatory for and on behalf of the National Aeronautics Space Administration under contract 
NAS8-03060. VNB gratefully acknowledges assistance from a National Science Foundation (NSF) 
Research at Undergraduate Institutions (RUI) grant AST-1312296. Note that findings and conclusions 
do not necessarily represent views of the NSF.

The scientific results reported in this article are based in part on observations made by 
the \textit{Chandra} X-ray Observatory, and on data obtained from the \textit{Chandra} Data 
Archive. We also made use of the software provided by the \textit{Chandra} X-ray Center (CXC) 
in the application packages CIAO. 

Based on observations obtained at the Gemini Observatory, which is operated by the Association 
of Universities for Research in Astronomy, Inc., under a cooperative agreement with the NSF on 
behalf of the Gemini partnership: the National Science Foundation (United States), the National 
Research Council (Canada), CONICYT (Chile), Ministerio de Ciencia, Tecnolog\'{i}a e Innovaci\'{o}n 
Productiva (Argentina), and Minist\'{e}rio da Ci\^{e}ncia, Tecnologia e Inova\c{c}\~{a}o (Brazil). 

Based on observations made with the NASA Galaxy Evolution Explorer.
\textit{GALEX} is a NASA Small Explorer launched in 2003 April. We gratefully acknowledge 
NASA's support for the construction, operation, and science analysis for the \textit{GALEX} 
mission, developed in cooperation with the Centre National d'Etudes Spatiales of France and 
the Korean Ministry of Science and Education.

Based on data products from the Wide-field Infrared Survey Explorer, 
which is a joint project of the University of California, Los Angeles, and the Jet Propulsion 
Laboratory/California Institute of Technology, funded by the National Aeronautics and Space 
Administration.

Based on observations obtained with MegaPrime/MegaCam, a joint project of CFHT and CEA/DAPNIA, 
at the Canada-France-Hawaii Telescope (CFHT) which is operated by the National Research Council 
(NRC) of Canada, the Institut National des Sciences de l'Univers of the Centre National de la 
Recherche Scientifique of France, and the University of Hawaii.

Based on observations made with the European Southern Observatory under Program 
287.B-5008, Chile.

Based on observations obtained at the Southern Astrophysical Research (SOAR) telescope, which 
is a joint project of the Minist\'{e}rio da Ci\^{e}ncia, Tecnologia, e Inova\c{c}\~{a}o (MCTI) 
da Rep\'{u}blica Federativa do Brasil, the U.S. National Optical Astronomy Observatory (NOAO), 
the University of North Carolina at Chapel Hill (UNC), and Michigan State University (MSU).

This research has also made use of NASA's Astrophysics Data System Bibliographic Services;
the NASA/IPAC Extragalactic Database (NED) which is operated by the Jet Propulsion Laboratory, 
California Institute of Technology, under contract with the National Aeronautics and Space 
Administration; the Python {\tt matplotlib} package \citep{hun07}.


\bibliographystyle{mnras}
\bibliography{mybib}

\begin{thebibliography}{}
\makeatletter
\relax
\def\mn@urlcharsother{\let\do\@makeother \do\$\do\&\do\#\do\^\do\_\do\%\do\~}
\def\mn@doi{\begingroup\mn@urlcharsother \@ifnextchar [ {\mn@doi@}
  {\mn@doi@[]}}
\def\mn@doi@[#1]#2{\def\@tempa{#1}\ifx\@tempa\@empty \href
  {http://dx.doi.org/#2} {doi:#2}\else \href {http://dx.doi.org/#2} {#1}\fi
  \endgroup}
\def\mn@eprint#1#2{\mn@eprint@#1:#2::\@nil}
\def\mn@eprint@arXiv#1{\href {http://arxiv.org/abs/#1} {{\tt arXiv:#1}}}
\def\mn@eprint@dblp#1{\href {http://dblp.uni-trier.de/rec/bibtex/#1.xml}
  {dblp:#1}}
\def\mn@eprint@#1:#2:#3:#4\@nil{\def\@tempa {#1}\def\@tempb {#2}\def\@tempc
  {#3}\ifx \@tempc \@empty \let \@tempc \@tempb \let \@tempb \@tempa \fi \ifx
  \@tempb \@empty \def\@tempb {arXiv}\fi \@ifundefined
  {mn@eprint@\@tempb}{\@tempb:\@tempc}{\expandafter \expandafter \csname
  mn@eprint@\@tempb\endcsname \expandafter{\@tempc}}}

\bibitem[\protect\citeauthoryear{{Allen} et~al.,}{{Allen} et~al.}{2015}]{ass15}
{Allen} J.~T.,  et~al., 2015, \mn@doi [\mnras] {10.1093/mnras/stv1121}, \href
  {http://adsabs.harvard.edu/abs/2015MNRAS.451.2780A} {451, 2780}

\bibitem[\protect\citeauthoryear{{Amor{\'{\i}}n}, {P{\'e}rez-Montero}  \&
  {V{\'{\i}}lchez}}{{Amor{\'{\i}}n} et~al.}{2010}]{apv10}
{Amor{\'{\i}}n} R.~O.,  {P{\'e}rez-Montero} E.,   {V{\'{\i}}lchez} J.~M.,
  2010, \mn@doi [\apjl] {10.1088/2041-8205/715/2/L128}, \href
  {http://adsabs.harvard.edu/abs/2010ApJ...715L.128A} {715, L128}

\bibitem[\protect\citeauthoryear{{Amor{\'{\i}}n}, {P{\'e}rez-Montero},
  {V{\'{\i}}lchez}  \& {Papaderos}}{{Amor{\'{\i}}n} et~al.}{2012}]{apv12}
{Amor{\'{\i}}n} R.,  {P{\'e}rez-Montero} E.,  {V{\'{\i}}lchez} J.~M.,
  {Papaderos} P.,  2012, \mn@doi [\apj] {10.1088/0004-637X/749/2/185}, \href
  {http://adsabs.harvard.edu/abs/2012ApJ...749..185A} {749, 185}

\bibitem[\protect\citeauthoryear{{Antonucci}}{{Antonucci}}{1993}]{ant93}
{Antonucci} R.,  1993, \mn@doi [\araa] {10.1146/annurev.aa.31.090193.002353},
  \href {http://adsabs.harvard.edu/abs/1993ARA%26A..31..473A} {31, 473}

\bibitem[\protect\citeauthoryear{{Applegate} et~al.,}{{Applegate}
  et~al.}{2014}]{alk14}
{Applegate} D.~E.,  et~al., 2014, \mn@doi [\mnras] {10.1093/mnras/stt2129},
  \href {http://adsabs.harvard.edu/abs/2014MNRAS.439...48A} {439, 48}

\bibitem[\protect\citeauthoryear{{Barger}, {Cowie}  \& {Wold}}{{Barger}
  et~al.}{2012}]{bcw12}
{Barger} A.~J.,  {Cowie} L.~L.,   {Wold} I.~G.~B.,  2012, \mn@doi [\apj]
  {10.1088/0004-637X/749/2/106}, \href
  {http://adsabs.harvard.edu/abs/2012ApJ...749..106B} {749, 106}

\bibitem[\protect\citeauthoryear{{Bassani}, {Dadina}, {Maiolino}, {Salvati},
  {Risaliti}, {Della Ceca}, {Matt}  \& {Zamorani}}{{Bassani}
  et~al.}{1999}]{bdm99}
{Bassani} L.,  {Dadina} M.,  {Maiolino} R.,  {Salvati} M.,  {Risaliti} G.,
  {Della Ceca} R.,  {Matt} G.,   {Zamorani} G.,  1999, \mn@doi [\apjs]
  {10.1086/313202}, \href {http://adsabs.harvard.edu/abs/1999ApJS..121..473B}
  {121, 473}

\bibitem[\protect\citeauthoryear{{Basu-Zych} \& {Scharf}}{{Basu-Zych} \&
  {Scharf}}{2004}]{bzs04}
{Basu-Zych} A.,  {Scharf} C.,  2004, \mn@doi [\apjl] {10.1086/426390}, \href
  {http://adsabs.harvard.edu/abs/2004ApJ...615L..85B} {615, L85}

\bibitem[\protect\citeauthoryear{{Bertin}}{{Bertin}}{2006}]{ber06}
{Bertin} E.,  2006, in {Gabriel} C.,  {Arviset} C.,  {Ponz} D.,   {Enrique} S.,
   eds,  ASP Conf. Ser. Vol. 351, ADASS XV. p.~112

\bibitem[\protect\citeauthoryear{{Binette} \& {Robinson}}{{Binette} \&
  {Robinson}}{1987}]{bir87}
{Binette} L.,  {Robinson} A.,  1987, \aap, \href
  {http://adsabs.harvard.edu/abs/1987A%26A...177...11B} {177, 11}

\bibitem[\protect\citeauthoryear{{Bournaud}, {Dekel}, {Teyssier}, {Cacciato},
  {Daddi}, {Juneau}  \& {Shankar}}{{Bournaud} et~al.}{2011}]{bdt11}
{Bournaud} F.,  {Dekel} A.,  {Teyssier} R.,  {Cacciato} M.,  {Daddi} E.,
  {Juneau} S.,   {Shankar} F.,  2011, \mn@doi [\apjl]
  {10.1088/2041-8205/741/2/L33}, \href
  {http://adsabs.harvard.edu/abs/2011ApJ...741L..33B} {741, L33}

\bibitem[\protect\citeauthoryear{{Bournaud} et~al.,}{{Bournaud}
  et~al.}{2012}]{bjf12}
{Bournaud} F.,  et~al., 2012, \mn@doi [\apj] {10.1088/0004-637X/757/1/81},
  \href {http://adsabs.harvard.edu/abs/2012ApJ...757...81B} {757, 81}

\bibitem[\protect\citeauthoryear{{Bridge} et~al.,}{{Bridge}
  et~al.}{2013}]{bbb13}
{Bridge} C.~R.,  et~al., 2013, \mn@doi [\apj] {10.1088/0004-637X/769/2/91},
  \href {http://adsabs.harvard.edu/abs/2013ApJ...769...91B} {769, 91}

\bibitem[\protect\citeauthoryear{{Bruzual} \& {Charlot}}{{Bruzual} \&
  {Charlot}}{2003}]{brc03}
{Bruzual} G.,  {Charlot} S.,  2003, \mn@doi [\mnras]
  {10.1046/j.1365-8711.2003.06897.x}, \href
  {http://adsabs.harvard.edu/abs/2003MNRAS.344.1000B} {344, 1000}

\bibitem[\protect\citeauthoryear{{Cardamone} et~al.,}{{Cardamone}
  et~al.}{2009}]{css09}
{Cardamone} C.,  et~al., 2009, \mn@doi [\mnras]
  {10.1111/j.1365-2966.2009.15383.x}, \href
  {http://adsabs.harvard.edu/abs/2009MNRAS.399.1191C} {399, 1191}

\bibitem[\protect\citeauthoryear{{Cen} \& {Zheng}}{{Cen} \&
  {Zheng}}{2013}]{cez13}
{Cen} R.,  {Zheng} Z.,  2013, \mn@doi [\apj] {10.1088/0004-637X/775/2/112},
  \href {http://adsabs.harvard.edu/abs/2013ApJ...775..112C} {775, 112}

\bibitem[\protect\citeauthoryear{{Chandrasekhar}}{{Chandrasekhar}}{1943}]{cha43}
{Chandrasekhar} S.,  1943, \mn@doi [\apj] {10.1086/144517}, \href
  {http://adsabs.harvard.edu/abs/1943ApJ....97..255C} {97, 255}

\bibitem[\protect\citeauthoryear{{Chapman}, {Lewis}, {Scott}, {Richards},
  {Borys}, {Steidel}, {Adelberger}  \& {Shapley}}{{Chapman}
  et~al.}{2001}]{cls2001}
{Chapman} S.~C.,  {Lewis} G.~F.,  {Scott} D.,  {Richards} E.,  {Borys} C.,
  {Steidel} C.~C.,  {Adelberger} K.~L.,   {Shapley} A.~E.,  2001, \mn@doi
  [\apjl] {10.1086/318919}, \href
  {http://adsabs.harvard.edu/abs/2001ApJ...548L..17C} {548, L17}

\bibitem[\protect\citeauthoryear{{Chapman}, {Scott}, {Windhorst}, {Frayer},
  {Borys}, {Lewis}  \& {Ivison}}{{Chapman} et~al.}{2004}]{csw04}
{Chapman} S.~C.,  {Scott} D.,  {Windhorst} R.~A.,  {Frayer} D.~T.,  {Borys} C.,
   {Lewis} G.~F.,   {Ivison} R.~J.,  2004, \mn@doi [\apj] {10.1086/382778},
  \href {http://adsabs.harvard.edu/abs/2004ApJ...606...85C} {606, 85}

\bibitem[\protect\citeauthoryear{{Chung}, {Yoon}  \& {Lee}}{{Chung}
  et~al.}{2011}]{cyl11}
{Chung} C.,  {Yoon} S.-J.,   {Lee} Y.-W.,  2011, \mn@doi [\apjl]
  {10.1088/2041-8205/740/2/L45}, \href
  {http://adsabs.harvard.edu/abs/2011ApJ...740L..45C} {740, L45}

\bibitem[\protect\citeauthoryear{{Comerford} \& {Greene}}{{Comerford} \&
  {Greene}}{2014}]{cog14}
{Comerford} J.~M.,  {Greene} J.~E.,  2014, \mn@doi [\apj]
  {10.1088/0004-637X/789/2/112}, \href
  {http://adsabs.harvard.edu/abs/2014ApJ...789..112C} {789, 112}

\bibitem[\protect\citeauthoryear{{Comerford}, {Gerke}, {Stern}, {Cooper},
  {Weiner}, {Newman}, {Madsen}  \& {Barrows}}{{Comerford} et~al.}{2012}]{cgs12}
{Comerford} J.~M.,  {Gerke} B.~F.,  {Stern} D.,  {Cooper} M.~C.,  {Weiner}
  B.~J.,  {Newman} J.~A.,  {Madsen} K.,   {Barrows} R.~S.,  2012, \mn@doi
  [\apj] {10.1088/0004-637X/753/1/42}, \href
  {http://adsabs.harvard.edu/abs/2012ApJ...753...42C} {753, 42}

\bibitem[\protect\citeauthoryear{{Conroy} \& {Gunn}}{{Conroy} \&
  {Gunn}}{2010}]{cog10}
{Conroy} C.,  {Gunn} J.~E.,  2010, \mn@doi [\apj]
  {10.1088/0004-637X/712/2/833}, \href
  {http://adsabs.harvard.edu/abs/2010ApJ...712..833C} {712, 833}

\bibitem[\protect\citeauthoryear{{Davies}, {Schirmer}  \& {Turner}}{{Davies}
  et~al.}{2015}]{dst15}
{Davies} R.~L.,  {Schirmer} M.,   {Turner} J.~E.~H.,  2015, \mn@doi [\mnras]
  {10.1093/mnras/stv343}, \href
  {http://adsabs.harvard.edu/abs/2015MNRAS.449.1731D} {449, 1731}

\bibitem[\protect\citeauthoryear{{DeGraf}, {Dekel}, {Gabor}  \&
  {Bournaud}}{{DeGraf} et~al.}{2014}]{gdg14}
{DeGraf} C.,  {Dekel} A.,  {Gabor} J.,   {Bournaud} F.,  2014, preprint, \href
  {http://adsabs.harvard.edu/abs/2014arXiv1412.3819D} {} (\mn@eprint {arXiv}
  {1412.3819})

\bibitem[\protect\citeauthoryear{{Dey} et~al.,}{{Dey} et~al.}{2005}]{dbs05}
{Dey} A.,  et~al., 2005, \mn@doi [\apj] {10.1086/430775}, \href
  {http://adsabs.harvard.edu/abs/2005ApJ...629..654D} {629, 654}

\bibitem[\protect\citeauthoryear{{Dijkstra} \& {Loeb}}{{Dijkstra} \&
  {Loeb}}{2009}]{dil09}
{Dijkstra} M.,  {Loeb} A.,  2009, \mn@doi [\mnras]
  {10.1111/j.1365-2966.2009.15533.x}, \href
  {http://adsabs.harvard.edu/abs/2009MNRAS.400.1109D} {400, 1109}

\bibitem[\protect\citeauthoryear{{Dijkstra}, {Haiman}  \& {Spaans}}{{Dijkstra}
  et~al.}{2006}]{dhs06}
{Dijkstra} M.,  {Haiman} Z.,   {Spaans} M.,  2006, \mn@doi [\apj]
  {10.1086/506243}, \href {http://adsabs.harvard.edu/abs/2006ApJ...649...14D}
  {649, 14}

\bibitem[\protect\citeauthoryear{{Ebeling}, {Edge}  \& {Henry}}{{Ebeling}
  et~al.}{2001}]{eeh01}
{Ebeling} H.,  {Edge} A.~C.,   {Henry} J.~P.,  2001, \mn@doi [\apj]
  {10.1086/320958}, \href {http://adsabs.harvard.edu/abs/2001ApJ...553..668E}
  {553, 668}

\bibitem[\protect\citeauthoryear{{Erb}, {Bogosavljevi{\'c}}  \&
  {Steidel}}{{Erb} et~al.}{2011}]{ebs11}
{Erb} D.~K.,  {Bogosavljevi{\'c}} M.,   {Steidel} C.~C.,  2011, \mn@doi [\apjl]
  {10.1088/2041-8205/740/1/L31}, \href
  {http://adsabs.harvard.edu/abs/2011ApJ...740L..31E} {740, L31}

\bibitem[\protect\citeauthoryear{Erben, Schirmer, Dietrich  et~al.}{Erben
  et~al.}{2005}]{esd05}
Erben T.,  Schirmer M.,  Dietrich J.,   et~al., 2005, AN, 326, 432

\bibitem[\protect\citeauthoryear{{Erben} et~al.,}{{Erben} et~al.}{2013}]{ehm13}
{Erben} T.,  et~al., 2013, \mn@doi [\mnras] {10.1093/mnras/stt928}, \href
  {http://adsabs.harvard.edu/abs/2013MNRAS.433.2545E} {433, 2545}

\bibitem[\protect\citeauthoryear{{Faucher-Gigu{\`e}re}, {Kere{\v s}},
  {Dijkstra}, {Hernquist}  \& {Zaldarriaga}}{{Faucher-Gigu{\`e}re}
  et~al.}{2010}]{fkd10}
{Faucher-Gigu{\`e}re} C.-A.,  {Kere{\v s}} D.,  {Dijkstra} M.,  {Hernquist} L.,
    {Zaldarriaga} M.,  2010, \mn@doi [\apj] {10.1088/0004-637X/725/1/633},
  \href {http://adsabs.harvard.edu/abs/2010ApJ...725..633F} {725, 633}

\bibitem[\protect\citeauthoryear{{Francis} et~al.,}{{Francis}
  et~al.}{2001}]{fwc01}
{Francis} P.~J.,  et~al., 2001, \mn@doi [\apj] {10.1086/321417}, \href
  {http://adsabs.harvard.edu/abs/2001ApJ...554.1001F} {554, 1001}

\bibitem[\protect\citeauthoryear{{Furlanetto} \& {Lidz}}{{Furlanetto} \&
  {Lidz}}{2011}]{ful11}
{Furlanetto} S.~R.,  {Lidz} A.,  2011, \mn@doi [\apj]
  {10.1088/0004-637X/735/2/117}, \href
  {http://adsabs.harvard.edu/abs/2011ApJ...735..117F} {735, 117}

\bibitem[\protect\citeauthoryear{{Gabor} \& {Bournaud}}{{Gabor} \&
  {Bournaud}}{2014}]{gab14}
{Gabor} J.~M.,  {Bournaud} F.,  2014, \mn@doi [\mnras] {10.1093/mnras/stu677},
  \href {http://adsabs.harvard.edu/abs/2014MNRAS.441.1615G} {441, 1615}

\bibitem[\protect\citeauthoryear{{Geach} et~al.,}{{Geach} et~al.}{2009}]{gal09}
{Geach} J.~E.,  et~al., 2009, \mn@doi [\apj] {10.1088/0004-637X/700/1/1}, \href
  {http://adsabs.harvard.edu/abs/2009ApJ...700....1G} {700, 1}

\bibitem[\protect\citeauthoryear{{Goerdt}, {Dekel}, {Sternberg}, {Ceverino},
  {Teyssier}  \& {Primack}}{{Goerdt} et~al.}{2010}]{gds10}
{Goerdt} T.,  {Dekel} A.,  {Sternberg} A.,  {Ceverino} D.,  {Teyssier} R.,
  {Primack} J.~R.,  2010, \mn@doi [\mnras] {10.1111/j.1365-2966.2010.16941.x},
  \href {http://adsabs.harvard.edu/abs/2010MNRAS.407..613G} {407, 613}

\bibitem[\protect\citeauthoryear{{Greene}, {Zakamska}  \& {Smith}}{{Greene}
  et~al.}{2012}]{gzs12}
{Greene} J.~E.,  {Zakamska} N.~L.,   {Smith} P.~S.,  2012, \mn@doi [\apj]
  {10.1088/0004-637X/746/1/86}, \href
  {http://adsabs.harvard.edu/abs/2012ApJ...746...86G} {746, 86}

\bibitem[\protect\citeauthoryear{{Guainazzi}, {Matt}  \& {Perola}}{{Guainazzi}
  et~al.}{2005}]{gmp05}
{Guainazzi} M.,  {Matt} G.,   {Perola} G.~C.,  2005, \mn@doi [\aap]
  {10.1051/0004-6361:20053643}, \href
  {http://adsabs.harvard.edu/abs/2005A%26A...444..119G} {444, 119}

\bibitem[\protect\citeauthoryear{{Haiman}, {Spaans}  \& {Quataert}}{{Haiman}
  et~al.}{2000}]{hsq00}
{Haiman} Z.,  {Spaans} M.,   {Quataert} E.,  2000, \mn@doi [\apjl]
  {10.1086/312754}, \href {http://adsabs.harvard.edu/abs/2000ApJ...537L...5H}
  {537, L5}

\bibitem[\protect\citeauthoryear{{Hainline}, {Hickox}, {Greene}, {Myers}  \&
  {Zakamska}}{{Hainline} et~al.}{2013}]{hhg13}
{Hainline} K.~N.,  {Hickox} R.,  {Greene} J.~E.,  {Myers} A.~D.,   {Zakamska}
  N.~L.,  2013, \mn@doi [\apj] {10.1088/0004-637X/774/2/145}, \href
  {http://adsabs.harvard.edu/abs/2013ApJ...774..145H} {774, 145}

\bibitem[\protect\citeauthoryear{{Han}, {Podsiadlowski}  \& {Lynas-Gray}}{{Han}
  et~al.}{2007}]{hpl07}
{Han} Z.,  {Podsiadlowski} P.,   {Lynas-Gray} A.~E.,  2007, \mn@doi [\mnras]
  {10.1111/j.1365-2966.2007.12151.x}, \href
  {http://adsabs.harvard.edu/abs/2007MNRAS.380.1098H} {380, 1098}

\bibitem[\protect\citeauthoryear{{Hansen}, {McKay}, {Wechsler}, {Annis},
  {Sheldon}  \& {Kimball}}{{Hansen} et~al.}{2005}]{hmw05}
{Hansen} S.~M.,  {McKay} T.~A.,  {Wechsler} R.~H.,  {Annis} J.,  {Sheldon}
  E.~S.,   {Kimball} A.,  2005, \mn@doi [\apj] {10.1086/444554}, \href
  {http://adsabs.harvard.edu/abs/2005ApJ...633..122H} {633, 122}

\bibitem[\protect\citeauthoryear{{Hansen}, {Sheldon}, {Wechsler}  \&
  {Koester}}{{Hansen} et~al.}{2009}]{hsw09}
{Hansen} S.~M.,  {Sheldon} E.~S.,  {Wechsler} R.~H.,   {Koester} B.~P.,  2009,
  \mn@doi [\apj] {10.1088/0004-637X/699/2/1333}, \href
  {http://adsabs.harvard.edu/abs/2009ApJ...699.1333H} {699, 1333}

\bibitem[\protect\citeauthoryear{{Harrison}, {Alexander}, {Mullaney}  \&
  {Swinbank}}{{Harrison} et~al.}{2014}]{ham14}
{Harrison} C.~M.,  {Alexander} D.~M.,  {Mullaney} J.~R.,   {Swinbank} A.~M.,
  2014, \mn@doi [\mnras] {10.1093/mnras/stu515}, \href
  {http://adsabs.harvard.edu/abs/2014MNRAS.441.3306H} {441, 3306}

\bibitem[\protect\citeauthoryear{{Hasinger}}{{Hasinger}}{2008}]{has08}
{Hasinger} G.,  2008, \mn@doi [\aap] {10.1051/0004-6361:200809839}, \href
  {http://adsabs.harvard.edu/abs/2008A%26A...490..905H} {490, 905}

\bibitem[\protect\citeauthoryear{{Hawley}}{{Hawley}}{2012}]{haw12}
{Hawley} S.~A.,  2012, \mn@doi [\pasp] {10.1086/663866}, \href
  {http://adsabs.harvard.edu/abs/2012PASP..124...21H} {124, 21}

\bibitem[\protect\citeauthoryear{{Hayes}, {Scarlata}  \& {Siana}}{{Hayes}
  et~al.}{2011}]{hss11}
{Hayes} M.,  {Scarlata} C.,   {Siana} B.,  2011, \mn@doi [\nat]
  {10.1038/nature10320}, \href
  {http://adsabs.harvard.edu/abs/2011Natur.476..304H} {476, 304}

\bibitem[\protect\citeauthoryear{{Heckman}, {Kauffmann}, {Brinchmann},
  {Charlot}, {Tremonti}  \& {White}}{{Heckman} et~al.}{2004}]{hkb04}
{Heckman} T.~M.,  {Kauffmann} G.,  {Brinchmann} J.,  {Charlot} S.,  {Tremonti}
  C.,   {White} S.~D.~M.,  2004, \mn@doi [\apj] {10.1086/422872}, \href
  {http://adsabs.harvard.edu/abs/2004ApJ...613..109H} {613, 109}

\bibitem[\protect\citeauthoryear{{Henry}, {Scarlata}, {Martin}  \&
  {Erb}}{{Henry} et~al.}{2015}]{hsm15}
{Henry} A.,  {Scarlata} C.,  {Martin} C.~L.,   {Erb} D.,  2015, \mn@doi [\apj]
  {10.1088/0004-637X/809/1/19}, \href
  {http://adsabs.harvard.edu/abs/2015ApJ...809...19H} {809, 19}

\bibitem[\protect\citeauthoryear{{Herenz}, {Wisotzki}, {Roth}  \&
  {Anders}}{{Herenz} et~al.}{2015}]{hwr15}
{Herenz} E.~C.,  {Wisotzki} L.,  {Roth} M.,   {Anders} F.,  2015, \mn@doi
  [\aap] {10.1051/0004-6361/201425580}, \href
  {http://adsabs.harvard.edu/abs/2015A%26A...576A.115H} {576, A115}

\bibitem[\protect\citeauthoryear{{Hickox}, {Mullaney}, {Alexander}, {Chen},
  {Civano}, {Goulding}  \& {Hainline}}{{Hickox} et~al.}{2014}]{hma14}
{Hickox} R.~C.,  {Mullaney} J.~R.,  {Alexander} D.~M.,  {Chen} C.-T.~J.,
  {Civano} F.~M.,  {Goulding} A.~D.,   {Hainline} K.~N.,  2014, \mn@doi [\apj]
  {10.1088/0004-637X/782/1/9}, \href
  {http://adsabs.harvard.edu/abs/2014ApJ...782....9H} {782, 9}

\bibitem[\protect\citeauthoryear{{Hildebrandt} et~al.,}{{Hildebrandt}
  et~al.}{2012}]{hek12}
{Hildebrandt} H.,  et~al., 2012, \mn@doi [\mnras]
  {10.1111/j.1365-2966.2012.20468.x}, \href
  {http://adsabs.harvard.edu/abs/2012MNRAS.421.2355H} {421, 2355}

\bibitem[\protect\citeauthoryear{{H{\"o}nig} \& {Kishimoto}}{{H{\"o}nig} \&
  {Kishimoto}}{2011}]{hok11}
{H{\"o}nig} S.~F.,  {Kishimoto} M.,  2011, \mn@doi [\aap]
  {10.1051/0004-6361/201117750}, \href
  {http://adsabs.harvard.edu/abs/2011A%26A...534A.121H} {534, A121}

\bibitem[\protect\citeauthoryear{{Hopkins} \& {Quataert}}{{Hopkins} \&
  {Quataert}}{2010}]{hoq10}
{Hopkins} P.~F.,  {Quataert} E.,  2010, \mn@doi [\mnras]
  {10.1111/j.1365-2966.2010.17064.x}, \href
  {http://adsabs.harvard.edu/abs/2010MNRAS.407.1529H} {407, 1529}

\bibitem[\protect\citeauthoryear{{Hu}}{{Hu}}{1992}]{hue92}
{Hu} E.~M.,  1992, \mn@doi [\apj] {10.1086/171374}, \href
  {http://adsabs.harvard.edu/abs/1992ApJ...391..608H} {391, 608}

\bibitem[\protect\citeauthoryear{{Humphrey}, {Vernet}, {Villar-Mart{\'{\i}}n},
  {di Serego Alighieri}, {Fosbury}  \& {Cimatti}}{{Humphrey}
  et~al.}{2013}]{hvv13}
{Humphrey} A.,  {Vernet} J.,  {Villar-Mart{\'{\i}}n} M.,  {di Serego Alighieri}
  S.,  {Fosbury} R.~A.~E.,   {Cimatti} A.,  2013, \mn@doi [\apjl]
  {10.1088/2041-8205/768/1/L3}, \href
  {http://adsabs.harvard.edu/abs/2013ApJ...768L...3H} {768, L3}

\bibitem[\protect\citeauthoryear{Hunter}{Hunter}{2007}]{hun07}
Hunter J.~D.,  2007, Computing In Science \& Engineering, 9, 90

\bibitem[\protect\citeauthoryear{{Hutchings}}{{Hutchings}}{2014}]{hut14}
{Hutchings} J.~B.,  2014, \mn@doi [\apss] {10.1007/s10509-014-1953-4}, \href
  {http://adsabs.harvard.edu/abs/2014Ap%26SS.354..143H} {354, 143}

\bibitem[\protect\citeauthoryear{{Ichikawa}, {Ueda}, {Terashima}, {Oyabu},
  {Gandhi}, {Matsuta}  \& {Nakagawa}}{{Ichikawa} et~al.}{2012}]{iut12}
{Ichikawa} K.,  {Ueda} Y.,  {Terashima} Y.,  {Oyabu} S.,  {Gandhi} P.,
  {Matsuta} K.,   {Nakagawa} T.,  2012, \mn@doi [\apj]
  {10.1088/0004-637X/754/1/45}, \href
  {http://adsabs.harvard.edu/abs/2012ApJ...754...45I} {754, 45}

\bibitem[\protect\citeauthoryear{{Izotov}, {Guseva}  \& {Thuan}}{{Izotov}
  et~al.}{2011}]{igt11}
{Izotov} Y.~I.,  {Guseva} N.~G.,   {Thuan} T.~X.,  2011, \mn@doi [\apj]
  {10.1088/0004-637X/728/2/161}, \href
  {http://adsabs.harvard.edu/abs/2011ApJ...728..161I} {728, 161}

\bibitem[\protect\citeauthoryear{{Jaskot} \& {Oey}}{{Jaskot} \&
  {Oey}}{2013}]{jao13}
{Jaskot} A.~E.,  {Oey} M.~S.,  2013, \mn@doi [\apj]
  {10.1088/0004-637X/766/2/91}, \href
  {http://adsabs.harvard.edu/abs/2013ApJ...766...91J} {766, 91}

\bibitem[\protect\citeauthoryear{{Johnston} et~al.,}{{Johnston}
  et~al.}{2007}]{jsw07}
{Johnston} D.~E.,  et~al., 2007, astro-ph/0709.1159, \href
  {http://adsabs.harvard.edu/abs/2007arXiv0709.1159J} {}

\bibitem[\protect\citeauthoryear{{Keel}, {Wu}, {Waddington}, {Windhorst}  \&
  {Pascarelle}}{{Keel} et~al.}{2002}]{kww02}
{Keel} W.~C.,  {Wu} W.,  {Waddington} I.,  {Windhorst} R.~A.,   {Pascarelle}
  S.~M.,  2002, \mn@doi [\aj] {10.1086/340696}, \href
  {http://adsabs.harvard.edu/abs/2002AJ....123.3041K} {123, 3041}

\bibitem[\protect\citeauthoryear{{Keel}, {White}, {Chapman}  \&
  {Windhorst}}{{Keel} et~al.}{2009}]{kwc09}
{Keel} W.~C.,  {White} III R.~E.,  {Chapman} S.,   {Windhorst} R.~A.,  2009,
  \mn@doi [\aj] {10.1088/0004-6256/138/3/986}, \href
  {http://adsabs.harvard.edu/abs/2009AJ....138..986K} {138, 986}

\bibitem[\protect\citeauthoryear{{Keel} et~al.,}{{Keel} et~al.}{2012a}]{kls12}
{Keel} W.~C.,  et~al., 2012a, \mn@doi [\aj] {10.1088/0004-6256/144/2/66}, \href
  {http://adsabs.harvard.edu/abs/2012AJ....144...66K} {144, 66}

\bibitem[\protect\citeauthoryear{{Keel} et~al.,}{{Keel} et~al.}{2012b}]{kcb12}
{Keel} W.~C.,  et~al., 2012b, \mn@doi [\mnras]
  {10.1111/j.1365-2966.2011.20101.x}, \href
  {http://adsabs.harvard.edu/abs/2012MNRAS.420..878K} {420, 878}

\bibitem[\protect\citeauthoryear{{Keel} et~al.,}{{Keel} et~al.}{2015}]{kmb15}
{Keel} W.~C.,  et~al., 2015, \mn@doi [\aj] {10.1088/0004-6256/149/5/155}, \href
  {http://adsabs.harvard.edu/abs/2015AJ....149..155K} {149, 155}

\bibitem[\protect\citeauthoryear{{Khrykin}, {Hennawi}, {McQuinn}  \&
  {Worseck}}{{Khrykin} et~al.}{2016}]{khm16}
{Khrykin} I.~S.,  {Hennawi} J.~F.,  {McQuinn} M.,   {Worseck} G.,  2016,
  \mn@doi [\apj] {10.3847/0004-637X/824/2/133}, \href
  {http://adsabs.harvard.edu/abs/2016ApJ...824..133K} {824, 133}

\bibitem[\protect\citeauthoryear{{Kirkman} \& {Tytler}}{{Kirkman} \&
  {Tytler}}{2008}]{kit08}
{Kirkman} D.,  {Tytler} D.,  2008, \mn@doi [\mnras]
  {10.1111/j.1365-2966.2008.13994.x}, \href
  {http://adsabs.harvard.edu/abs/2008MNRAS.391.1457K} {391, 1457}

\bibitem[\protect\citeauthoryear{{Kodaira} et~al.,}{{Kodaira}
  et~al.}{2003}]{ktk03}
{Kodaira} K.,  et~al., 2003, \mn@doi [\pasj] {10.1093/pasj/55.2.L17}, \href
  {http://adsabs.harvard.edu/abs/2003PASJ...55L..17K} {55, L17}

\bibitem[\protect\citeauthoryear{{Kollmeier}, {Zheng}, {Dav{\'e}}, {Gould},
  {Katz}, {Miralda-Escud{\'e}}  \& {Weinberg}}{{Kollmeier}
  et~al.}{2010}]{kzd10}
{Kollmeier} J.~A.,  {Zheng} Z.,  {Dav{\'e}} R.,  {Gould} A.,  {Katz} N.,
  {Miralda-Escud{\'e}} J.,   {Weinberg} D.~H.,  2010, \mn@doi [\apj]
  {10.1088/0004-637X/708/2/1048}, \href
  {http://adsabs.harvard.edu/abs/2010ApJ...708.1048K} {708, 1048}

\bibitem[\protect\citeauthoryear{{Krolik} \& {Kallman}}{{Krolik} \&
  {Kallman}}{1987}]{krk87}
{Krolik} J.~H.,  {Kallman} T.~R.,  1987, \mn@doi [\apjl] {10.1086/184966},
  \href {http://adsabs.harvard.edu/abs/1987ApJ...320L...5K} {320, L5}

\bibitem[\protect\citeauthoryear{{Kubo}, {Yamada}, {Ichikawa}, {Kajisawa},
  {Matsuda}  \& {Tanaka}}{{Kubo} et~al.}{2015}]{kyi15}
{Kubo} M.,  {Yamada} T.,  {Ichikawa} T.,  {Kajisawa} M.,  {Matsuda} Y.,
  {Tanaka} I.,  2015, \mn@doi [\apj] {10.1088/0004-637X/799/1/38}, \href
  {http://adsabs.harvard.edu/abs/2015ApJ...799...38K} {799, 38}

\bibitem[\protect\citeauthoryear{{LaMassa} et~al.,}{{LaMassa}
  et~al.}{2015}]{lcm15}
{LaMassa} S.~M.,  et~al., 2015, \mn@doi [\apj] {10.1088/0004-637X/800/2/144},
  \href {http://adsabs.harvard.edu/abs/2015ApJ...800..144L} {800, 144}

\bibitem[\protect\citeauthoryear{{Lamastra}, {Bianchi}, {Matt}, {Perola},
  {Barcons}  \& {Carrera}}{{Lamastra} et~al.}{2009}]{lbm09}
{Lamastra} A.,  {Bianchi} S.,  {Matt} G.,  {Perola} G.~C.,  {Barcons} X.,
  {Carrera} F.~J.,  2009, \mn@doi [\aap] {10.1051/0004-6361/200912023}, \href
  {http://adsabs.harvard.edu/abs/2009A%26A...504...73L} {504, 73}

\bibitem[\protect\citeauthoryear{{Laursen} \& {Sommer-Larsen}}{{Laursen} \&
  {Sommer-Larsen}}{2007}]{las07}
{Laursen} P.,  {Sommer-Larsen} J.,  2007, \mn@doi [\apjl] {10.1086/513191},
  \href {http://adsabs.harvard.edu/abs/2007ApJ...657L..69L} {657, L69}

\bibitem[\protect\citeauthoryear{{Lintott} et~al.,}{{Lintott}
  et~al.}{2009}]{lsk09}
{Lintott} C.~J.,  et~al., 2009, \mn@doi [\mnras]
  {10.1111/j.1365-2966.2009.15299.x}, \href
  {http://adsabs.harvard.edu/abs/2009MNRAS.399..129L} {399, 129}

\bibitem[\protect\citeauthoryear{{Liu}, {Shen}, {Strauss}  \& {Hao}}{{Liu}
  et~al.}{2011}]{lss11}
{Liu} X.,  {Shen} Y.,  {Strauss} M.~A.,   {Hao} L.,  2011, \mn@doi [\apj]
  {10.1088/0004-637X/737/2/101}, \href
  {http://adsabs.harvard.edu/abs/2011ApJ...737..101L} {737, 101}

\bibitem[\protect\citeauthoryear{{Liu}, {Shen}  \& {Strauss}}{{Liu}
  et~al.}{2012}]{lss12}
{Liu} X.,  {Shen} Y.,   {Strauss} M.~A.,  2012, \mn@doi [\apj]
  {10.1088/0004-637X/745/1/94}, \href
  {http://adsabs.harvard.edu/abs/2012ApJ...745...94L} {745, 94}

\bibitem[\protect\citeauthoryear{{Liu}, {Zakamska}, {Greene}, {Nesvadba}  \&
  {Liu}}{{Liu} et~al.}{2013}]{lzg13}
{Liu} G.,  {Zakamska} N.~L.,  {Greene} J.~E.,  {Nesvadba} N.~P.~H.,   {Liu} X.,
   2013, \mn@doi [\mnras] {10.1093/mnras/stt051}, \href
  {http://adsabs.harvard.edu/abs/2013MNRAS.430.2327L} {430, 2327}

\bibitem[\protect\citeauthoryear{{Lu} \& {Zhou}}{{Lu} \& {Zhou}}{2005}]{luz05}
{Lu} J.-F.,  {Zhou} B.-Y.,  2005, \mn@doi [\apjl] {10.1086/499333}, \href
  {http://adsabs.harvard.edu/abs/2005ApJ...635L..17L} {635, L17}

\bibitem[\protect\citeauthoryear{{Malhotra} \& {Rhoads}}{{Malhotra} \&
  {Rhoads}}{2002}]{mar02}
{Malhotra} S.,  {Rhoads} J.~E.,  2002, \mn@doi [\apjl] {10.1086/338980}, \href
  {http://adsabs.harvard.edu/abs/2002ApJ...565L..71M} {565, L71}

\bibitem[\protect\citeauthoryear{{Malizia}, {Stephen}, {Bassani}, {Bird},
  {Panessa}  \& {Ubertini}}{{Malizia} et~al.}{2009}]{msb09}
{Malizia} A.,  {Stephen} J.~B.,  {Bassani} L.,  {Bird} A.~J.,  {Panessa} F.,
  {Ubertini} P.,  2009, \mn@doi [\mnras] {10.1111/j.1365-2966.2009.15330.x},
  \href {http://adsabs.harvard.edu/abs/2009MNRAS.399..944M} {399, 944}

\bibitem[\protect\citeauthoryear{{Maraston}}{{Maraston}}{2005}]{mar05}
{Maraston} C.,  2005, \mn@doi [\mnras] {10.1111/j.1365-2966.2005.09270.x},
  \href {http://adsabs.harvard.edu/abs/2005MNRAS.362..799M} {362, 799}

\bibitem[\protect\citeauthoryear{{Maraston}, {Nieves Colmen{\'a}rez}, {Bender}
  \& {Thomas}}{{Maraston} et~al.}{2009}]{mnb09}
{Maraston} C.,  {Nieves Colmen{\'a}rez} L.,  {Bender} R.,   {Thomas} D.,  2009,
  \mn@doi [\aap] {10.1051/0004-6361:20066907}, \href
  {http://adsabs.harvard.edu/abs/2009A%26A...493..425M} {493, 425}

\bibitem[\protect\citeauthoryear{{Martin}, {Chang}, {Matuszewski}, {Morrissey},
  {Rahman}, {Moore}, {Steidel}  \& {Matsuda}}{{Martin} et~al.}{2014}]{mcm14}
{Martin} D.~C.,  {Chang} D.,  {Matuszewski} M.,  {Morrissey} P.,  {Rahman} S.,
  {Moore} A.,  {Steidel} C.~C.,   {Matsuda} Y.,  2014, \mn@doi [\apj]
  {10.1088/0004-637X/786/2/107}, \href
  {http://adsabs.harvard.edu/abs/2014ApJ...786..107M} {786, 107}

\bibitem[\protect\citeauthoryear{{Mateus}, {Sodr{\'e}}, {Cid Fernandes},
  {Stasi{\'n}ska}, {Schoenell}  \& {Gomes}}{{Mateus} et~al.}{2006}]{msc06}
{Mateus} A.,  {Sodr{\'e}} L.,  {Cid Fernandes} R.,  {Stasi{\'n}ska} G.,
  {Schoenell} W.,   {Gomes} J.~M.,  2006, \mn@doi [\mnras]
  {10.1111/j.1365-2966.2006.10565.x}, \href
  {http://adsabs.harvard.edu/abs/2006MNRAS.370..721M} {370, 721}

\bibitem[\protect\citeauthoryear{{Matsuda} et~al.,}{{Matsuda}
  et~al.}{2004}]{myh04}
{Matsuda} Y.,  et~al., 2004, \mn@doi [\aj] {10.1086/422020}, \href
  {http://adsabs.harvard.edu/abs/2004AJ....128..569M} {128, 569}

\bibitem[\protect\citeauthoryear{{Matsuda}, {Yamada}, {Hayashino}, {Yamauchi}
  \& {Nakamura}}{{Matsuda} et~al.}{2006}]{myh06}
{Matsuda} Y.,  {Yamada} T.,  {Hayashino} T.,  {Yamauchi} R.,   {Nakamura} Y.,
  2006, \mn@doi [\apjl] {10.1086/503362}, \href
  {http://adsabs.harvard.edu/abs/2006ApJ...640L.123M} {640, L123}

\bibitem[\protect\citeauthoryear{{Matsuda} et~al.,}{{Matsuda}
  et~al.}{2011}]{myh11}
{Matsuda} Y.,  et~al., 2011, \mn@doi [\mnras]
  {10.1111/j.1745-3933.2010.00969.x}, \href
  {http://adsabs.harvard.edu/abs/2011MNRAS.410L..13M} {410, L13}

\bibitem[\protect\citeauthoryear{{McDonald} et~al.,}{{McDonald}
  et~al.}{2012}]{mbb12}
{McDonald} M.,  et~al., 2012, \mn@doi [\nat] {10.1038/nature11379}, \href
  {http://adsabs.harvard.edu/abs/2012Natur.488..349M} {488, 349}

\bibitem[\protect\citeauthoryear{{McDonald} et~al.,}{{McDonald}
  et~al.}{2015}]{mmw15}
{McDonald} M.,  et~al., 2015, \mn@doi [\apj] {10.1088/0004-637X/811/2/111},
  \href {http://adsabs.harvard.edu/abs/2015ApJ...811..111M} {811, 111}

\bibitem[\protect\citeauthoryear{{McLinden} et~al.,}{{McLinden}
  et~al.}{2011}]{mfr11}
{McLinden} E.~M.,  et~al., 2011, \mn@doi [\apj] {10.1088/0004-637X/730/2/136},
  \href {http://adsabs.harvard.edu/abs/2011ApJ...730..136M} {730, 136}

\bibitem[\protect\citeauthoryear{{McLinden}, {Rhoads}, {Malhotra},
  {Finkelstein}, {Richardson}, {Smith}  \& {Tilvi}}{{McLinden}
  et~al.}{2014}]{mrr14}
{McLinden} E.~M.,  {Rhoads} J.~E.,  {Malhotra} S.,  {Finkelstein} S.~L.,
  {Richardson} M.~L.~A.,  {Smith} B.,   {Tilvi} V.~S.,  2014, \mn@doi [\mnras]
  {10.1093/mnras/stu023}, \href
  {http://adsabs.harvard.edu/abs/2014MNRAS.439..446M} {439, 446}

\bibitem[\protect\citeauthoryear{{Meink{\"o}hn} \& {Richling}}{{Meink{\"o}hn}
  \& {Richling}}{2002}]{mer02}
{Meink{\"o}hn} E.,  {Richling} S.,  2002, \mn@doi [\aap]
  {10.1051/0004-6361:20020951}, \href
  {http://adsabs.harvard.edu/abs/2002A%26A...392..827M} {392, 827}

\bibitem[\protect\citeauthoryear{{Merloni} et~al.,}{{Merloni}
  et~al.}{2014}]{mbb14}
{Merloni} A.,  et~al., 2014, \mn@doi [\mnras] {10.1093/mnras/stt2149}, \href
  {http://adsabs.harvard.edu/abs/2014MNRAS.437.3550M} {437, 3550}

\bibitem[\protect\citeauthoryear{{Miyaji} et~al.,}{{Miyaji}
  et~al.}{2015}]{mhs15}
{Miyaji} T.,  et~al., 2015, \mn@doi [\apj] {10.1088/0004-637X/804/2/104}, \href
  {http://adsabs.harvard.edu/abs/2015ApJ...804..104M} {804, 104}

\bibitem[\protect\citeauthoryear{{Mullaney}, {Alexander}, {Fine}, {Goulding},
  {Harrison}  \& {Hickox}}{{Mullaney} et~al.}{2013}]{maf13}
{Mullaney} J.~R.,  {Alexander} D.~M.,  {Fine} S.,  {Goulding} A.~D.,
  {Harrison} C.~M.,   {Hickox} R.~C.,  2013, \mn@doi [\mnras]
  {10.1093/mnras/stt751}, \href
  {http://adsabs.harvard.edu/abs/2013MNRAS.433..622M} {433, 622}

\bibitem[\protect\citeauthoryear{{Nenkova}, {Sirocky}, {Ivezi{\'c}}  \&
  {Elitzur}}{{Nenkova} et~al.}{2008}]{nsi08}
{Nenkova} M.,  {Sirocky} M.~M.,  {Ivezi{\'c}} {\v Z}.,   {Elitzur} M.,  2008,
  \mn@doi [\apj] {10.1086/590482}, \href
  {http://adsabs.harvard.edu/abs/2008ApJ...685..147N} {685, 147}

\bibitem[\protect\citeauthoryear{{Neufeld}}{{Neufeld}}{1990}]{neu90}
{Neufeld} D.~A.,  1990, \mn@doi [\apj] {10.1086/168375}, \href
  {http://adsabs.harvard.edu/abs/1990ApJ...350..216N} {350, 216}

\bibitem[\protect\citeauthoryear{{Nilsson}, {Fynbo}, {M{\o}ller},
  {Sommer-Larsen}  \& {Ledoux}}{{Nilsson} et~al.}{2006}]{nfm06}
{Nilsson} K.~K.,  {Fynbo} J.~P.~U.,  {M{\o}ller} P.,  {Sommer-Larsen} J.,
  {Ledoux} C.,  2006, \mn@doi [\aap] {10.1051/0004-6361:200600025}, \href
  {http://adsabs.harvard.edu/abs/2006A%26A...452L..23N} {452, L23}

\bibitem[\protect\citeauthoryear{{Novak}, {Ostriker}  \& {Ciotti}}{{Novak}
  et~al.}{2011}]{noc11}
{Novak} G.~S.,  {Ostriker} J.~P.,   {Ciotti} L.,  2011, \mn@doi [\apj]
  {10.1088/0004-637X/737/1/26}, \href
  {http://adsabs.harvard.edu/abs/2011ApJ...737...26N} {737, 26}

\bibitem[\protect\citeauthoryear{{Nussbaumer} \& {Schmutz}}{{Nussbaumer} \&
  {Schmutz}}{1984}]{nus84}
{Nussbaumer} H.,  {Schmutz} W.,  1984, \aap, \href
  {http://adsabs.harvard.edu/abs/1984A%26A...138..495N} {138, 495}

\bibitem[\protect\citeauthoryear{{Nusser} \& {Sheth}}{{Nusser} \&
  {Sheth}}{1999}]{nus99}
{Nusser} A.,  {Sheth} R.~K.,  1999, \mn@doi [\mnras]
  {10.1046/j.1365-8711.1999.02197.x}, \href
  {http://adsabs.harvard.edu/abs/1999MNRAS.303..685N} {303, 685}

\bibitem[\protect\citeauthoryear{{O'Connell}}{{O'Connell}}{1999}]{crw99}
{O'Connell} R.~W.,  1999, \mn@doi [\araa] {10.1146/annurev.astro.37.1.603},
  \href {http://adsabs.harvard.edu/abs/1999ARA%26A..37..603O} {37, 603}

\bibitem[\protect\citeauthoryear{{O'Dea}, {Baum}, {Mack}, {Koekemoer}  \&
  {Laor}}{{O'Dea} et~al.}{2004}]{obm04}
{O'Dea} C.~P.,  {Baum} S.~A.,  {Mack} J.,  {Koekemoer} A.~M.,   {Laor} A.,
  2004, \mn@doi [\apj] {10.1086/422402}, \href
  {http://adsabs.harvard.edu/abs/2004ApJ...612..131O} {612, 131}

\bibitem[\protect\citeauthoryear{{O'Dea} et~al.,}{{O'Dea} et~al.}{2010}]{oqo10}
{O'Dea} K.~P.,  et~al., 2010, \mn@doi [\apj] {10.1088/0004-637X/719/2/1619},
  \href {http://adsabs.harvard.edu/abs/2010ApJ...719.1619O} {719, 1619}

\bibitem[\protect\citeauthoryear{{Osterbrock} \& {Ferland}}{{Osterbrock} \&
  {Ferland}}{2006}]{osf06}
{Osterbrock} D.~E.,  {Ferland} G.~J.,  2006, {Astrophysics of gaseous nebulae
  and active galactic nuclei}.
University Science Books

\bibitem[\protect\citeauthoryear{{Ouchi} et~al.,}{{Ouchi} et~al.}{2009}]{ooe09}
{Ouchi} M.,  et~al., 2009, \mn@doi [\apj] {10.1088/0004-637X/696/2/1164}, \href
  {http://adsabs.harvard.edu/abs/2009ApJ...696.1164O} {696, 1164}

\bibitem[\protect\citeauthoryear{{Overzier}, {Nesvadba}, {Dijkstra}, {Hatch},
  {Lehnert}, {Villar-Mart{\'{\i}}n}, {Wilman}  \& {Zirm}}{{Overzier}
  et~al.}{2013}]{ond13}
{Overzier} R.~A.,  {Nesvadba} N.~P.~H.,  {Dijkstra} M.,  {Hatch} N.~A.,
  {Lehnert} M.~D.,  {Villar-Mart{\'{\i}}n} M.,  {Wilman} R.~J.,   {Zirm} A.~W.,
   2013, \mn@doi [\apj] {10.1088/0004-637X/771/2/89}, \href
  {http://adsabs.harvard.edu/abs/2013ApJ...771...89O} {771, 89}

\bibitem[\protect\citeauthoryear{{Palunas}, {Teplitz}, {Francis}, {Williger}
  \& {Woodgate}}{{Palunas} et~al.}{2004}]{ptf04}
{Palunas} P.,  {Teplitz} H.~I.,  {Francis} P.~J.,  {Williger} G.~M.,
  {Woodgate} B.~E.,  2004, \mn@doi [\apj] {10.1086/381145}, \href
  {http://adsabs.harvard.edu/abs/2004ApJ...602..545P} {602, 545}

\bibitem[\protect\citeauthoryear{{Pengelly} \& {Seaton}}{{Pengelly} \&
  {Seaton}}{1964}]{pes64}
{Pengelly} R.~M.,  {Seaton} M.~J.,  1964, \mn@doi [\mnras]
  {10.1093/mnras/127.2.165}, \href
  {http://adsabs.harvard.edu/abs/1964MNRAS.127..165P} {127, 165}

\bibitem[\protect\citeauthoryear{{Pilyugin}, {V{\'{\i}}lchez}, {Mattsson}  \&
  {Thuan}}{{Pilyugin} et~al.}{2012}]{pvm12}
{Pilyugin} L.~S.,  {V{\'{\i}}lchez} J.~M.,  {Mattsson} L.,   {Thuan} T.~X.,
  2012, \mn@doi [\mnras] {10.1111/j.1365-2966.2012.20420.x}, \href
  {http://adsabs.harvard.edu/abs/2012MNRAS.421.1624P} {421, 1624}

\bibitem[\protect\citeauthoryear{{Pogge} \& {De Robertis}}{{Pogge} \& {De
  Robertis}}{1993}]{pdr93}
{Pogge} R.~W.,  {De Robertis} M.~M.,  1993, \mn@doi [\apj] {10.1086/172308},
  \href {http://adsabs.harvard.edu/abs/1993ApJ...404..563P} {404, 563}

\bibitem[\protect\citeauthoryear{{Prescott}, {Kashikawa}, {Dey}  \&
  {Matsuda}}{{Prescott} et~al.}{2008}]{pkd08}
{Prescott} M.~K.~M.,  {Kashikawa} N.,  {Dey} A.,   {Matsuda} Y.,  2008, \mn@doi
  [\apjl] {10.1086/588606}, \href
  {http://adsabs.harvard.edu/abs/2008ApJ...678L..77P} {678, L77}

\bibitem[\protect\citeauthoryear{{Prescott}, {Dey}  \& {Jannuzi}}{{Prescott}
  et~al.}{2012}]{pdj12}
{Prescott} M.~K.~M.,  {Dey} A.,   {Jannuzi} B.~T.,  2012, \mn@doi [\apj]
  {10.1088/0004-637X/748/2/125}, \href
  {http://adsabs.harvard.edu/abs/2012ApJ...748..125P} {748, 125}

\bibitem[\protect\citeauthoryear{{Prescott}, {Dey}  \& {Jannuzi}}{{Prescott}
  et~al.}{2013}]{pdj13}
{Prescott} M.~K.~M.,  {Dey} A.,   {Jannuzi} B.~T.,  2013, \mn@doi [\apj]
  {10.1088/0004-637X/762/1/38}, \href
  {http://adsabs.harvard.edu/abs/2013ApJ...762...38P} {762, 38}

\bibitem[\protect\citeauthoryear{{Prescott}, {Momcheva}, {Brammer}, {Fynbo}  \&
  {M{\o}ller}}{{Prescott} et~al.}{2015}]{pmb15}
{Prescott} M.~K.~M.,  {Momcheva} I.,  {Brammer} G.~B.,  {Fynbo} J.~P.~U.,
  {M{\o}ller} P.,  2015, \mn@doi [\apj] {10.1088/0004-637X/802/1/32}, \href
  {http://adsabs.harvard.edu/abs/2015ApJ...802...32P} {802, 32}

\bibitem[\protect\citeauthoryear{{Rampadarath} et~al.,}{{Rampadarath}
  et~al.}{2010}]{rgj10}
{Rampadarath} H.,  et~al., 2010, \mn@doi [\aap] {10.1051/0004-6361/201014782},
  \href {http://adsabs.harvard.edu/abs/2010A%26A...517L...8R} {517, L8}

\bibitem[\protect\citeauthoryear{{Ree}, {Jeong}, {Oh}, {Chung}, {Lee}, {Kim}
  \& {Kyeong}}{{Ree} et~al.}{2012}]{rjo12}
{Ree} C.~H.,  {Jeong} H.,  {Oh} K.,  {Chung} C.,  {Lee} J.~H.,  {Kim} S.~C.,
  {Kyeong} J.,  2012, \mn@doi [\apjl] {10.1088/2041-8205/744/1/L10}, \href
  {http://adsabs.harvard.edu/abs/2012ApJ...744L..10R} {744, L10}

\bibitem[\protect\citeauthoryear{{Rengstorf} et~al.,}{{Rengstorf}
  et~al.}{2004}]{rma04}
{Rengstorf} A.~W.,  et~al., 2004, \mn@doi [\apj] {10.1086/425246}, \href
  {http://adsabs.harvard.edu/abs/2004ApJ...617..184R} {617, 184}

\bibitem[\protect\citeauthoryear{{Reyes} et~al.,}{{Reyes} et~al.}{2008}]{rzs08}
{Reyes} R.,  et~al., 2008, \mn@doi [\aj] {10.1088/0004-6256/136/6/2373}, \href
  {http://adsabs.harvard.edu/abs/2008AJ....136.2373R} {136, 2373}

\bibitem[\protect\citeauthoryear{{Risaliti}, {Maiolino}  \&
  {Salvati}}{{Risaliti} et~al.}{1999}]{rms99}
{Risaliti} G.,  {Maiolino} R.,   {Salvati} M.,  1999, \mn@doi [\apj]
  {10.1086/307623}, \href {http://adsabs.harvard.edu/abs/1999ApJ...522..157R}
  {522, 157}

\bibitem[\protect\citeauthoryear{{Rosdahl} \& {Blaizot}}{{Rosdahl} \&
  {Blaizot}}{2012}]{rob12}
{Rosdahl} J.,  {Blaizot} J.,  2012, \mn@doi [\mnras]
  {10.1111/j.1365-2966.2012.20883.x}, \href
  {http://adsabs.harvard.edu/abs/2012MNRAS.423..344R} {423, 344}

\bibitem[\protect\citeauthoryear{{Ross} et~al.,}{{Ross} et~al.}{2013}]{rgw13}
{Ross} N.~P.,  et~al., 2013, \mn@doi [\apj] {10.1088/0004-637X/773/1/14}, \href
  {http://adsabs.harvard.edu/abs/2013ApJ...773...14R} {773, 14}

\bibitem[\protect\citeauthoryear{{Roy}, {Shu}  \& {Fang}}{{Roy}
  et~al.}{2010}]{rsf10}
{Roy} I.,  {Shu} C.-W.,   {Fang} L.-Z.,  2010, \mn@doi [\apj]
  {10.1088/0004-637X/716/1/604}, \href
  {http://adsabs.harvard.edu/abs/2010ApJ...716..604R} {716, 604}

\bibitem[\protect\citeauthoryear{{Runnoe} et~al.,}{{Runnoe}
  et~al.}{2016}]{rcr16}
{Runnoe} J.~C.,  et~al., 2016, \mn@doi [\mnras] {10.1093/mnras/stv2385}, \href
  {http://adsabs.harvard.edu/abs/2016MNRAS.455.1691R} {455, 1691}

\bibitem[\protect\citeauthoryear{{Saito}, {Shimasaku}, {Okamura}, {Ouchi},
  {Akiyama}  \& {Yoshida}}{{Saito} et~al.}{2006}]{sso06}
{Saito} T.,  {Shimasaku} K.,  {Okamura} S.,  {Ouchi} M.,  {Akiyama} M.,
  {Yoshida} M.,  2006, \mn@doi [\apj] {10.1086/505678}, \href
  {http://adsabs.harvard.edu/abs/2006ApJ...648...54S} {648, 54}

\bibitem[\protect\citeauthoryear{{Saito}, {Shimasaku}, {Okamura}, {Ouchi},
  {Akiyama}, {Yoshida}  \& {Ueda}}{{Saito} et~al.}{2008}]{sso08}
{Saito} T.,  {Shimasaku} K.,  {Okamura} S.,  {Ouchi} M.,  {Akiyama} M.,
  {Yoshida} M.,   {Ueda} Y.,  2008, \mn@doi [\apj] {10.1086/527282}, \href
  {http://adsabs.harvard.edu/abs/2008ApJ...675.1076S} {675, 1076}

\bibitem[\protect\citeauthoryear{{Schawinski} et~al.,}{{Schawinski}
  et~al.}{2010}]{sev10}
{Schawinski} K.,  et~al., 2010, \mn@doi [\apjl] {10.1088/2041-8205/724/1/L30},
  \href {http://adsabs.harvard.edu/abs/2010ApJ...724L..30S} {724, L30}

\bibitem[\protect\citeauthoryear{{Schawinski}, {Koss}, {Berney}  \&
  {Sartori}}{{Schawinski} et~al.}{2015}]{skb15}
{Schawinski} K.,  {Koss} M.,  {Berney} S.,   {Sartori} L.~F.,  2015, \mn@doi
  [\mnras] {10.1093/mnras/stv1136}, \href
  {http://adsabs.harvard.edu/abs/2015MNRAS.451.2517S} {451, 2517}

\bibitem[\protect\citeauthoryear{{Schirmer}}{{Schirmer}}{2013}]{sch13}
{Schirmer} M.,  2013, \mn@doi [\apjs] {10.1088/0067-0049/209/2/21}, \href
  {http://adsabs.harvard.edu/abs/2013ApJS..209...21S} {209, 21}

\bibitem[\protect\citeauthoryear{{Schirmer}, {Hildebrandt}, {Kuijken}  \&
  {Erben}}{{Schirmer} et~al.}{2011}]{shk11}
{Schirmer} M.,  {Hildebrandt} H.,  {Kuijken} K.,   {Erben} T.,  2011, \mn@doi
  [\aap] {10.1051/0004-6361/201016348}, \href
  {http://adsabs.harvard.edu/abs/2011A%26A...532A..57S} {532, 57}

\bibitem[\protect\citeauthoryear{{Schirmer}, {Diaz}, {Holhjem}, {Levenson}  \&
  {Winge}}{{Schirmer} et~al.}{2013}]{sdh13}
{Schirmer} M.,  {Diaz} R.,  {Holhjem} K.,  {Levenson} N.~A.,   {Winge} C.,
  2013, \mn@doi [\apj] {10.1088/0004-637X/763/1/60}, \href
  {http://adsabs.harvard.edu/abs/2013ApJ...763...60S} {763, 60}

\bibitem[\protect\citeauthoryear{{Schlafly} \& {Finkbeiner}}{{Schlafly} \&
  {Finkbeiner}}{2011}]{scf11}
{Schlafly} E.~F.,  {Finkbeiner} D.~P.,  2011, \mn@doi [\apj]
  {10.1088/0004-637X/737/2/103}, \href
  {http://adsabs.harvard.edu/abs/2011ApJ...737..103S} {737, 103}

\bibitem[\protect\citeauthoryear{{Schweizer}, {Seitzer}, {Kelson}, {Villanueva}
   \& {Walth}}{{Schweizer} et~al.}{2013}]{ssk13}
{Schweizer} F.,  {Seitzer} P.,  {Kelson} D.~D.,  {Villanueva} E.~V.,   {Walth}
  G.~L.,  2013, \mn@doi [\apj] {10.1088/0004-637X/773/2/148}, \href
  {http://adsabs.harvard.edu/abs/2013ApJ...773..148S} {773, 148}

\bibitem[\protect\citeauthoryear{{Shen}, {Liu}, {Greene}  \& {Strauss}}{{Shen}
  et~al.}{2011}]{slg11}
{Shen} Y.,  {Liu} X.,  {Greene} J.~E.,   {Strauss} M.~A.,  2011, \mn@doi [\apj]
  {10.1088/0004-637X/735/1/48}, \href
  {http://adsabs.harvard.edu/abs/2011ApJ...735...48S} {735, 48}

\bibitem[\protect\citeauthoryear{{Sijacki}, {Vogelsberger}, {Genel},
  {Springel}, {Torrey}, {Snyder}, {Nelson}  \& {Hernquist}}{{Sijacki}
  et~al.}{2015}]{svg15}
{Sijacki} D.,  {Vogelsberger} M.,  {Genel} S.,  {Springel} V.,  {Torrey} P.,
  {Snyder} G.~F.,  {Nelson} D.,   {Hernquist} L.,  2015, \mn@doi [\mnras]
  {10.1093/mnras/stv1340}, \href
  {http://adsabs.harvard.edu/abs/2015MNRAS.452..575S} {452, 575}

\bibitem[\protect\citeauthoryear{{Silk}}{{Silk}}{2013}]{sil13}
{Silk} J.,  2013, \mn@doi [\apj] {10.1088/0004-637X/772/2/112}, \href
  {http://adsabs.harvard.edu/abs/2013ApJ...772..112S} {772, 112}

\bibitem[\protect\citeauthoryear{{Silk}, {Di Cintio}  \& {Dvorkin}}{{Silk}
  et~al.}{2013}]{sdd13}
{Silk} J.,  {Di Cintio} A.,   {Dvorkin} I.,  2013, preprint, \href
  {http://adsabs.harvard.edu/abs/2013arXiv1312.0107S} {} (\mn@eprint {arXiv}
  {1312.0107})

\bibitem[\protect\citeauthoryear{{Snyder}}{{Snyder}}{1998}]{sny98}
{Snyder} J.~A.,  1998, in {D'Odorico} S.,  ed.,  Society of Photo-Optical
  Instrumentation Engineers (SPIE) Conference Series Vol. 3355, Optical
  Astronomical Instrumentation. pp 635--645

\bibitem[\protect\citeauthoryear{{Steidel}, {Pettini}  \& {Hamilton}}{{Steidel}
  et~al.}{1995}]{sph95}
{Steidel} C.~C.,  {Pettini} M.,   {Hamilton} D.,  1995, \mn@doi [\aj]
  {10.1086/117709}, \href {http://adsabs.harvard.edu/abs/1995AJ....110.2519S}
  {110, 2519}

\bibitem[\protect\citeauthoryear{{Steidel}, {Adelberger}, {Shapley}, {Pettini},
  {Dickinson}  \& {Giavalisco}}{{Steidel} et~al.}{2000}]{sas00}
{Steidel} C.~C.,  {Adelberger} K.~L.,  {Shapley} A.~E.,  {Pettini} M.,
  {Dickinson} M.,   {Giavalisco} M.,  2000, \mn@doi [\apj] {10.1086/308568},
  \href {http://adsabs.harvard.edu/abs/2000ApJ...532..170S} {532, 170}

\bibitem[\protect\citeauthoryear{{Steidel}, {Bogosavljevi{\'c}}, {Shapley},
  {Kollmeier}, {Reddy}, {Erb}  \& {Pettini}}{{Steidel} et~al.}{2011}]{sbs11}
{Steidel} C.~C.,  {Bogosavljevi{\'c}} M.,  {Shapley} A.~E.,  {Kollmeier} J.~A.,
   {Reddy} N.~A.,  {Erb} D.~K.,   {Pettini} M.,  2011, \mn@doi [\apj]
  {10.1088/0004-637X/736/2/160}, \href
  {http://adsabs.harvard.edu/abs/2011ApJ...736..160S} {736, 160}

\bibitem[\protect\citeauthoryear{{Stern} et~al.,}{{Stern} et~al.}{2014}]{sla14}
{Stern} D.,  et~al., 2014, \mn@doi [\apj] {10.1088/0004-637X/794/2/102}, \href
  {http://adsabs.harvard.edu/abs/2014ApJ...794..102S} {794, 102}

\bibitem[\protect\citeauthoryear{{Swinbank} et~al.,}{{Swinbank}
  et~al.}{2015}]{svs15}
{Swinbank} A.~M.,  et~al., 2015, \mn@doi [\mnras] {10.1093/mnras/stv366}, \href
  {http://adsabs.harvard.edu/abs/2015MNRAS.449.1298S} {449, 1298}

\bibitem[\protect\citeauthoryear{{Tamura} et~al.,}{{Tamura}
  et~al.}{2013}]{tmi13}
{Tamura} Y.,  et~al., 2013, \mn@doi [\mnras] {10.1093/mnras/stt077}, \href
  {http://adsabs.harvard.edu/abs/2013MNRAS.430.2768T} {430, 2768}

\bibitem[\protect\citeauthoryear{{Taniguchi} \& {Shioya}}{{Taniguchi} \&
  {Shioya}}{2000}]{tas00}
{Taniguchi} Y.,  {Shioya} Y.,  2000, \mn@doi [\apjl] {10.1086/312557}, \href
  {http://adsabs.harvard.edu/abs/2000ApJ...532L..13T} {532, L13}

\bibitem[\protect\citeauthoryear{{Tokovinin} et~al.,}{{Tokovinin}
  et~al.}{2010}]{tts10}
{Tokovinin} A.,  et~al., 2010, in Society of Photo-Optical Instrumentation
  Engineers (SPIE) Conference Series. p.~3, \mn@doi{10.1117/12.856407}

\bibitem[\protect\citeauthoryear{{Tokovinin}, {Tighe}, {Schurter},
  {Cantarutti}, {van der Bliek}, {Martinez}, {Mondaca}  \&
  {Heathcote}}{{Tokovinin} et~al.}{2012}]{tts12}
{Tokovinin} A.,  {Tighe} R.,  {Schurter} P.,  {Cantarutti} R.,  {van der Bliek}
  N.,  {Martinez} M.,  {Mondaca} E.,   {Heathcote} S.,  2012, in Society of
  Photo-Optical Instrumentation Engineers (SPIE) Conference Series. p.~4,
  \mn@doi{10.1117/12.925705}

\bibitem[\protect\citeauthoryear{{Trebitsch}, {Verhamme}, {Blaizot}  \&
  {Rosdahl}}{{Trebitsch} et~al.}{2014}]{tvb14}
{Trebitsch} M.,  {Verhamme} A.,  {Blaizot} J.,   {Rosdahl} J.,  2014, in
  {Ballet} J.,  {Martins} F.,  {Bournaud} F.,  {Monier} R.,   {Reyl{\'e}} C.,
  eds, SF2A-2014: Proceedings of the Annual meeting of the French Society of
  Astronomy and Astrophysics. pp 375--377

\bibitem[\protect\citeauthoryear{{Ueda}, {Akiyama}, {Hasinger}, {Miyaji}  \&
  {Watson}}{{Ueda} et~al.}{2014}]{uah14}
{Ueda} Y.,  {Akiyama} M.,  {Hasinger} G.,  {Miyaji} T.,   {Watson} M.~G.,
  2014, \mn@doi [\apj] {10.1088/0004-637X/786/2/104}, \href
  {http://adsabs.harvard.edu/abs/2014ApJ...786..104U} {786, 104}

\bibitem[\protect\citeauthoryear{{Verhamme}, {Schaerer}  \&
  {Maselli}}{{Verhamme} et~al.}{2006}]{vsm06}
{Verhamme} A.,  {Schaerer} D.,   {Maselli} A.,  2006, \mn@doi [\aap]
  {10.1051/0004-6361:20065554}, \href
  {http://adsabs.harvard.edu/abs/2006A%26A...460..397V} {460, 397}

\bibitem[\protect\citeauthoryear{{Webb}, {Yamada}, {Huang}, {Ashby}, {Matsuda},
  {Egami}, {Gonzalez}  \& {Hayashimo}}{{Webb} et~al.}{2009}]{wyh09}
{Webb} T.~M.~A.,  {Yamada} T.,  {Huang} J.-S.,  {Ashby} M.~L.~N.,  {Matsuda}
  Y.,  {Egami} E.,  {Gonzalez} M.,   {Hayashimo} T.,  2009, \mn@doi [\apj]
  {10.1088/0004-637X/692/2/1561}, \href
  {http://adsabs.harvard.edu/abs/2009ApJ...692.1561W} {692, 1561}

\bibitem[\protect\citeauthoryear{{Weijmans}, {Bower}, {Geach}, {Swinbank},
  {Wilman}, {de Zeeuw}  \& {Morris}}{{Weijmans} et~al.}{2010}]{wbg10}
{Weijmans} A.-M.,  {Bower} R.~G.,  {Geach} J.~E.,  {Swinbank} A.~M.,  {Wilman}
  R.~J.,  {de Zeeuw} P.~T.,   {Morris} S.~L.,  2010, \mn@doi [\mnras]
  {10.1111/j.1365-2966.2009.16055.x}, \href
  {http://adsabs.harvard.edu/abs/2010MNRAS.402.2245W} {402, 2245}

\bibitem[\protect\citeauthoryear{White, Becker, Helfand  \& Gregg}{White
  et~al.}{1997}]{wbh97}
White R.~L.,  Becker R.~H.,  Helfand D.~J.,   Gregg M.~D.,  1997, ApJ, 475, 479

\bibitem[\protect\citeauthoryear{{Williamson} et~al.,}{{Williamson}
  et~al.}{2011}]{wbh11}
{Williamson} R.,  et~al., 2011, \mn@doi [\apj] {10.1088/0004-637X/738/2/139},
  \href {http://adsabs.harvard.edu/abs/2011ApJ...738..139W} {738, 139}

\bibitem[\protect\citeauthoryear{{Wold}, {Barger}  \& {Cowie}}{{Wold}
  et~al.}{2014}]{wbc14}
{Wold} I.~G.~B.,  {Barger} A.~J.,   {Cowie} L.~L.,  2014, \mn@doi [\apj]
  {10.1088/0004-637X/783/2/119}, \href
  {http://adsabs.harvard.edu/abs/2014ApJ...783..119W} {783, 119}

\bibitem[\protect\citeauthoryear{{Xu}, {Wu}  \& {Fang}}{{Xu}
  et~al.}{2011}]{xwf11}
{Xu} W.,  {Wu} X.-P.,   {Fang} L.-Z.,  2011, \mn@doi [\mnras]
  {10.1111/j.1365-2966.2011.19539.x}, \href
  {http://adsabs.harvard.edu/abs/2011MNRAS.418..853X} {418, 853}

\bibitem[\protect\citeauthoryear{{Yajima}, {Li}  \& {Zhu}}{{Yajima}
  et~al.}{2013}]{ylz13}
{Yajima} H.,  {Li} Y.,   {Zhu} Q.,  2013, \mn@doi [\apj]
  {10.1088/0004-637X/773/2/151}, \href
  {http://adsabs.harvard.edu/abs/2013ApJ...773..151Y} {773, 151}

\bibitem[\protect\citeauthoryear{{Yang}, {Zabludoff}, {Tremonti}, {Eisenstein}
  \& {Dav{\'e}}}{{Yang} et~al.}{2009}]{yzt09}
{Yang} Y.,  {Zabludoff} A.,  {Tremonti} C.,  {Eisenstein} D.,   {Dav{\'e}} R.,
  2009, \mn@doi [\apj] {10.1088/0004-637X/693/2/1579}, \href
  {http://adsabs.harvard.edu/abs/2009ApJ...693.1579Y} {693, 1579}

\bibitem[\protect\citeauthoryear{{Yang}, {Zabludoff}, {Eisenstein}  \&
  {Dav{\'e}}}{{Yang} et~al.}{2010}]{yze10}
{Yang} Y.,  {Zabludoff} A.,  {Eisenstein} D.,   {Dav{\'e}} R.,  2010, \mn@doi
  [\apj] {10.1088/0004-637X/719/2/1654}, \href
  {http://adsabs.harvard.edu/abs/2010ApJ...719.1654Y} {719, 1654}

\bibitem[\protect\citeauthoryear{{Yang}, {Zabludoff}, {Jahnke}, {Eisenstein},
  {Dav{\'e}}, {Shectman}  \& {Kelson}}{{Yang} et~al.}{2011a}]{yzj11}
{Yang} Y.,  {Zabludoff} A.,  {Jahnke} K.,  {Eisenstein} D.,  {Dav{\'e}} R.,
  {Shectman} S.~A.,   {Kelson} D.~D.,  2011a, \mn@doi [\apj]
  {10.1088/0004-637X/735/2/87}, \href
  {http://adsabs.harvard.edu/abs/2011ApJ...735...87Y} {735, 87}

\bibitem[\protect\citeauthoryear{{Yang}, {Roy}, {Shu}  \& {Fang}}{{Yang}
  et~al.}{2011b}]{yrs11}
{Yang} Y.,  {Roy} I.,  {Shu} C.-W.,   {Fang} L.-Z.,  2011b, \mn@doi [\apj]
  {10.1088/0004-637X/739/2/91}, \href
  {http://adsabs.harvard.edu/abs/2011ApJ...739...91Y} {739, 91}

\bibitem[\protect\citeauthoryear{{Yang}, {Zabludoff}, {Jahnke}  \&
  {Dav{\'e}}}{{Yang} et~al.}{2014}]{yzj14}
{Yang} Y.,  {Zabludoff} A.,  {Jahnke} K.,   {Dav{\'e}} R.,  2014, \mn@doi
  [\apj] {10.1088/0004-637X/793/2/114}, \href
  {http://adsabs.harvard.edu/abs/2014ApJ...793..114Y} {793, 114}

\bibitem[\protect\citeauthoryear{{Yang}, {Malhotra}, {Gronke}, {Rhoads},
  {Jaskot}, {Zheng}  \& {Dijkstra}}{{Yang} et~al.}{2015}]{ymg15}
{Yang} H.,  {Malhotra} S.,  {Gronke} M.,  {Rhoads} J.~E.,  {Jaskot} A.,
  {Zheng} Z.,   {Dijkstra} M.,  2015, preprint, \href
  {http://adsabs.harvard.edu/abs/2015arXiv150602885Y} {} (\mn@eprint {arXiv}
  {1506.02885})

\bibitem[\protect\citeauthoryear{{Zabl}, {N{\o}rgaard-Nielsen}, {Fynbo},
  {Laursen}, {Ouchi}  \& {Kj{\ae}rgaard}}{{Zabl} et~al.}{2015}]{znf15}
{Zabl} J.,  {N{\o}rgaard-Nielsen} H.~U.,  {Fynbo} J.~P.~U.,  {Laursen} P.,
  {Ouchi} M.,   {Kj{\ae}rgaard} P.,  2015, preprint, \href
  {http://adsabs.harvard.edu/abs/2015arXiv150501859Z} {} (\mn@eprint {arXiv}
  {1505.01859})

\bibitem[\protect\citeauthoryear{{Zakamska} et~al.,}{{Zakamska}
  et~al.}{2005}]{zss05}
{Zakamska} N.~L.,  et~al., 2005, \mn@doi [\aj] {10.1086/427543}, \href
  {http://adsabs.harvard.edu/abs/2005AJ....129.1212Z} {129, 1212}

\makeatother
\end{thebibliography}

\appendix

\section{Notes about individual targets}\label{targetnotes}
\subsection{J0020--0531 (z=0.334)}
The strong [\ion{O}{III}] emission mimics the widely opened arms of a SBc galaxy.
The gas has high surface brightness in the centre, and appears tidally warped
by two nearby elliptical galaxies at the same redshift. It is not clear whether the 
gas in the two 'arms' is outflowing due to AGN activity, infalling, or pulled out 
by tidal forces. It could be forced into the spiral shape by differential rotation. 
The projected distances to the two elliptical galaxies are 11 and 24\,kpc, respectively. 
Fan-shaped tidal debris, consisting of stars and/or gas, extends southwest over 35\,kpc, 
embedding one of the ellipticals. The host galaxy is compact.

\begin{figure}
  \includegraphics[width=1.0\hsize]{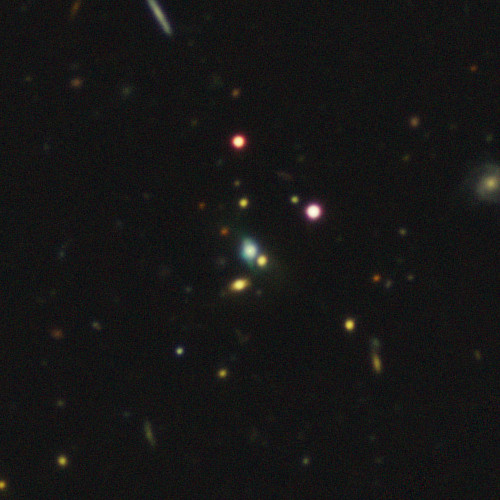}
  \caption{\label{J0020-0531}{J0020--0531. The images of the GBs in this Appendix are 
      true-colour renditions of the $gri$ data. The powerful [\ion{O}{III}] line dominates 
      the green channel. A selective non-linear stretch was applied to the GBs to avoid 
      saturation; they are significantly brighter with respect to the other field galaxies 
      than these images suggest.}}
\end{figure}

\subsection{J0024+3258 (z=0.293)}
This system has a complex, disrupted morphology. The [\ion{O}{III}] emission is high
in the centre. A collimated bipolar outflow extends over a total of 31\,kpc in North-South 
direction. It gets wider with distance from the nucleus, and shows a twist beginning about
halfway from the centre. It is probably ionized by a double ionization cone. Tidal debris of 
stellar origin (judged by its broad-band colours) intersects the outflow at an angle of about 
45 degrees. It extends over at least 54\,kpc and increases in surface brightness toward the 
nucleus.

The $gri$ colour image reveals several ellipticals and red spirals with similar colours as
the tidal debris. However, our long-slit spectra reveal that they belong to an unrelated 
foreground structure at $z=0.22$. J0024$+$3258 is an isolated, advanced merger system in a 
low density environment.

\begin{figure}
  \includegraphics[width=1.0\hsize]{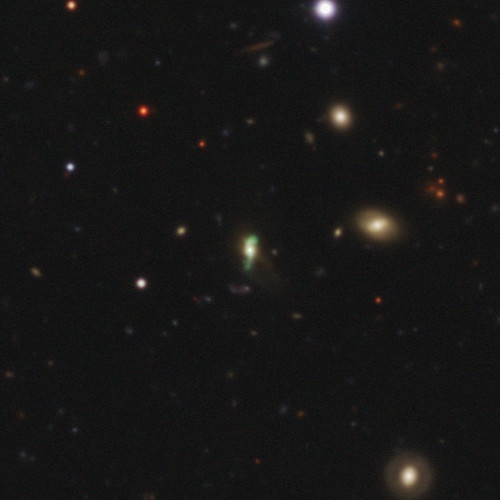}
  \caption{\label{J0024+3258}{J0024+3258}}
\end{figure}

\subsection{J0111+2253 (z=0.319)}
This galaxy is dominated by a 22\,kpc long luminous unipolar outflow to the Southeast. A 
fainter outflow of similar extent is visible to the South. The host galaxy is compact. 
35\,kpc to the East of J0111+2253 is an elliptical galaxy at $z=0.321$. Their relative radial 
velocities are $450\pm220$\,\kms. Both galaxies are symmetrically embedded in a diffuse halo 
stretching in a slightly curved manner over 103\,kpc from East to West, and over 23\,kpc 
North to South. The colours of this halo and of the elliptical galaxy are the same. These 
galaxies must have experienced a close encounter in the past.

Another elliptical galaxy 92 kpc south of J0111+2253 is at the same redshift. The field 
shows several other galaxies with unknown redshifts. The overall impression is that of 
a low density environment.

\begin{figure}
  \includegraphics[width=1.0\hsize]{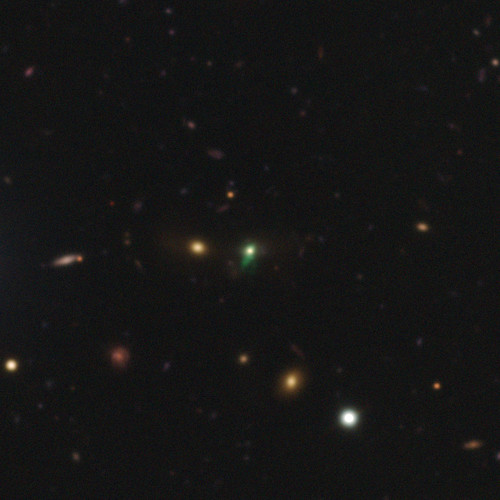}
  \caption{\label{J0111+2253}{J0111+2253}}
\end{figure}

\subsection{J0113+0106 (z=0.281)}
A spectacular system with two symmetric luminous [\ion{O}{III}] superbubbles on the 
northwestern and southeastern side of the nucleus, at a projected distance of 5.5\,kpc 
each. The northwestern bubble is somewhat fainter than the one in the Southeast, and 
their diameters are $5-8$\,kpc. There is also a 14\,kpc long curved arc in [\ion{O}{III}] 
at 24\,kpc separation from the nucleus. It is connected with the main system through a 
fainter [\ion{O}{III}] bridge. This could be an old ejecta from the AGN, or gas pulled out 
by tidal interaction. There is no galaxy visible in this arc, suggesting an AGN-driven 
outflow. The curvature of this arc and the misalignment with the two superbubbles is 
remarkable. In the absence of a shaping radio-jet this could perhaps indicate a nuclear 
spinflip, e.g. caused by an SMBH merger. It is evident that this AGN underwent at least 
two powerful events in the past. The youngest one, perhaps $2-5$ million years ago, 
ejected the two superbubbles. An older event $20-30$ million years ago sent out the arc,
and possibly some very faint [\ion{O}{III}] emission out to 43\,kpc south of the 
nucleus, roughly aligned with the two superbubbles. At the moment this is qualitative at 
best, as we have no radial velocities nor information about the true orientation in space.

The entire system, 16\,arcsec wide, is embedded in a low surface brightness halo whose 
colours suggest a stellar origin. The nucleus of J0113$+$0106 is also very luminous in 
[\ion{O}{III}] and makes it difficult 
to discern the host galaxy. The latter could be a compact elliptical, given the somewhat 
redder colours of the centre. A second compact elliptical with a diameter of 6\,kpc is found 
at a projected distance of 8\,kpc to the West of the nucleus. Its redshift is still unknown 
and this could be a chance projection. Regardless, J0113$+$0106 is a rather isolated system.
There is a sparse foreground galaxy cluster at $z=0.186$ to the North. 

Overall, J0113+0106 is reminiscent of SDSS J1356+1026 at $z=0.123$, where a pair of 10 
kpc outflows is found by \cite{gzs12}. Both systems are also radio-quiet.

\begin{figure}
  \includegraphics[width=1.0\hsize]{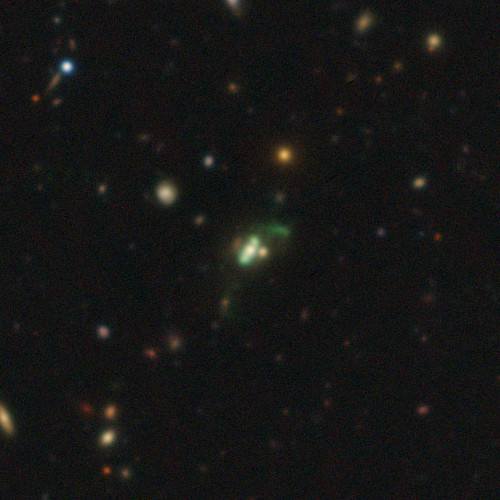}
  \caption{\label{J0113+0106}{J0113+0106}}
\end{figure}

\subsection{J0159+2703 (z=0.278)}
This is a rather untypical GB because the emission-line region is hosted in a 46\,kpc size, 
bright spiral galaxy with a pronounced bar. In addition, or because of this, the [\ion{O}{III}] 
EW is lower than in other GBs. A long-slit observation for our field spectroscopy survey 
intersects the outer areas of this galaxy and we detect [\ion{O}{III}] out to 25\,kpc
from the nucleus. The barred spiral is not visible in the SDSS images in which this 
object was discovered.

Field spectroscopy shows that J0159$+$2703 is relatively isolated. Only one elliptical at 
590\,kpc separation has so far been confirmed at the same redshift. Two more, smaller 
ellipticals of the same colour, and a few spirals are still pending redshift confirmation.

J0159$+$2703 has a mid-IR flux (18.1 mJy) close to the median of our sample (Table 
\ref{targetlist2}), yet its X-ray count rate is by far the lowest. This could indicate
an exceptionally strong fading of the AGN in recent times.

\begin{figure}
  \includegraphics[width=1.0\hsize]{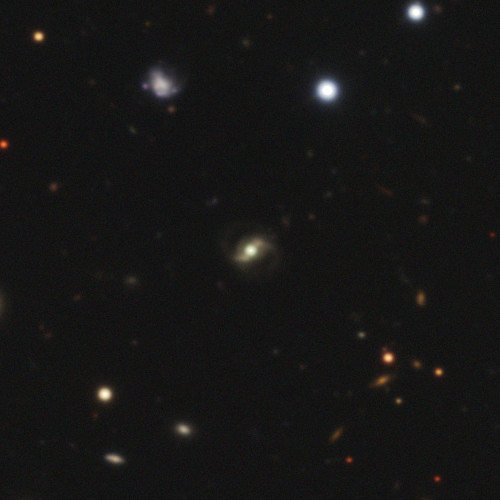}
  \caption{\label{J0159+2703}{J0159+2703}}
\end{figure}

\subsection{\label{j11550147target}J1155--0147 (z=0.306)}
This galaxy is extraordinary. First, its [\ion{O}{III}] emission has a diameter of 64\,kpc 
and is the largest nebula in our sample. It also has the highest \textit{GALEX} FUV luminosity, 
the highest X-ray counts and relatively low column density (see Section \ref{j1155chandra}). 
The host galaxy is likely an elliptical.

The [\ion{O}{III}] emission roughly decomposes into three main parts: a luminous centre
with a diameter of 12\,kpc, a medium-bright and somewhat elliptical zone of 30\,kpc, and a fainter 
envelope extending over 64\,kpc. The latter has a characteristic 'X'-like shape, which could either
be caused by a strong bipolar and older ionization cone, preferentially oriented in the plane of 
the sky, or it is simply a direct consequence of the spatial distribution of the gas. Ionizing 
radiation appears to escape in all directions as evidenced by the central surface brightness 
distribution.

A fragmentation into smaller clouds lends an unusual, mottled appearance to J1155$-$0147. The typical 
diameter of these clouds is about 0\myarcsec8 to 1\myarcsec6 or $3.5-7.0$\,kpc. We count $12-16$ 
of them. Possibly, the AGN has sputtered outflows in various directions during several duty cycles. 
Precession of a single or binary SMBH could also be responsible for the clouds and the `X'-like shape.

Another possibility is that the smaller clouds are shock-compressed regions in infalling gas.
Interestingly, J1155$-$0147 is located at the geometric centre of a low mass galaxy group.
The group has at least a dozen members, mostly ellipticals and red spirals, distributed over 
$\sim220$\,kpc. Using the mass-richness relation for red sequence galaxies \citep{hmw05,hsw09,jsw07}, 
we estimate the total mass to $M_{200}=(1.3\pm0.9)\times10^{13}\;M_{\odot}$. There is no indication 
of a soft X-ray group halo in the \textit{Chandra} data. None of the other member galaxies are 
seen in X-rays. The major axis of the [\ion{O}{III}] emission is aligned to within 20 degrees with 
the major axis of the galaxy distribution. Possibly, some cold accretion is still happening in 
J1155$-$0147, similar to the LABs of \citet{ebs11} which are preferentially aligned with filaments. 
The smaller clouds are of similar size and reminiscent of the structures seen in the \Lya 
fluorescence maps of \cite{kzd10} that arise in cold accretion streams. It is unclear, though, 
whether these \Lya structures would also reflect in [\ion{O}{III}].

Another odd feature is a broad tidal stream, $15x35$\,kpc in size, extending from the elliptical
galaxy J115545$-$014722 in direction to J1155$-$0147 (90\,kpc away, clearly visible in Fig. 
\ref{J1155-0147}). Our central long-slit spectrum of this galaxy does not reveal any emission 
lines. A possible dynamic encounter with J1155$-$0147 would have occurred $\gtrsim200$ million 
years ago if $200$\,\kms is the typical velocity of the cluster members. Likely, the two phenomena 
are not directly linked.

If the [\ion{O}{III}] nebula is outflowing, then this object would contribute significantly to the 
enrichment of the intergalactic medium in this group. The contrast between the essentially
gas depleted ellipticals and the gas-rich GB in the group centre is remarkable.

A curious comparison that comes to mind is that of the \textit{Phoenix cluster} \citep{wbh11,mbb12}
at $z=0.596$, where a massive cooling flow on to the BCG feeds a substantial star burst 
\citep{mmw15}. The authors also argue that the type-2 quasar present in the BCG is currently 
transition from a high state into a low state (like the GBs), given the simultaneous presence of 
radiative and kinematic feedback. A more detailed comparison of these two systems is worthwhile, 
given their similarities despite their vastly different environments.

\begin{figure}
  \includegraphics[width=1.0\hsize]{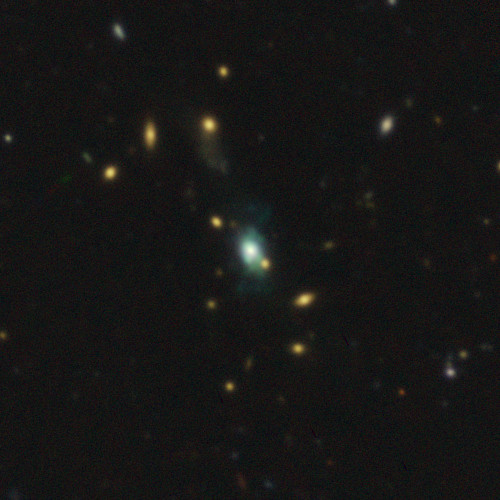}
  \caption{\label{J1155-0147}{J1155--0147}}
\end{figure}

\begin{figure}
  \includegraphics[width=1.0\hsize]{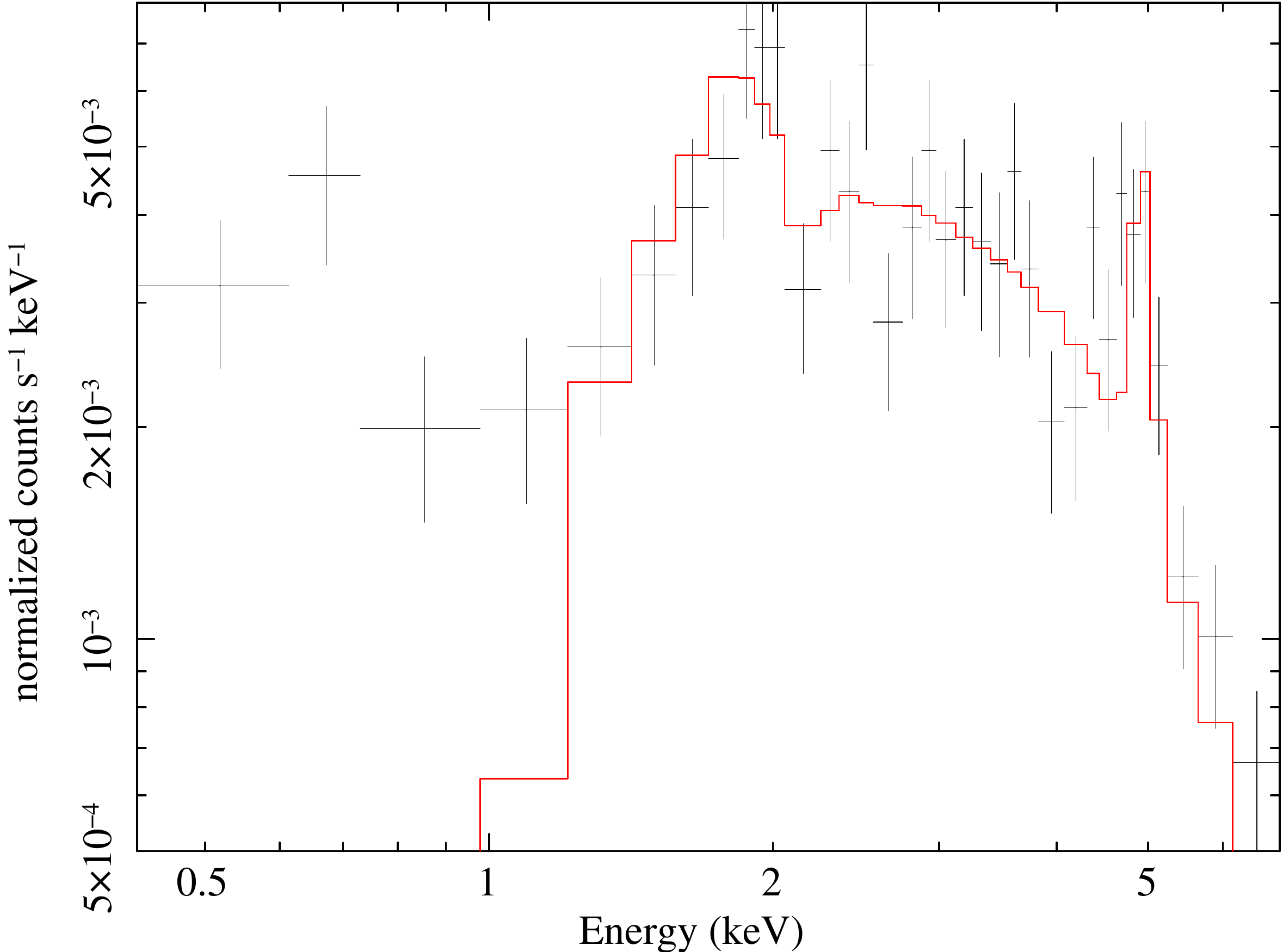}
  \caption{\label{j1155-0147_chandra}{\textit{Chandra} spectrum of J1155-0147. The 
      background-subtracted spectrum binned into groups of a minimum of 20 counts (black crosses) 
      is fit with a power law and the spectrally unresolved Fe K$\alpha$ line at 6.4\,keV in the 
      galaxy's rest frame (red solid line). The low-energy spectrum is fit poorly, suggesting 
      that an additional physical component is producing the soft X-rays.}}
\end{figure}

\subsubsection{\label{j1155chandra}Archival Chandra data of J1155-0147}
J1155$-$0147 is by far the X-ray brightest of the GBs. The 30\,ks \textit{Chandra} exposure 
provides a total of 550 counts in the useful energy range, revealing a weakly extended 
source with a relatively soft spectrum. We fit the full ($0.3-8.0$\,keV) spectrum as the sum 
of the AGN power law with photon index $\Gamma=1.9$, plus the unresolved Fe K$\alpha$ line 
(Fig. \ref{j1155-0147_chandra}). The very soft spectrum below 1\,keV is fit poorly, 
indicating the presence of photoionized emission corresponding to the strong optical 
[\ion{O}{III}] line. Thermal emission from a starburst is unlikely because of the high value 
of ${\rm log([\ion{O}{III}]/{\rm H}\beta)=1.163\pm0.006}$ \citepalias{sdh13}, requiring a 
hard spectrum. We recover the AGN in this fit with an intrinsic flux 
$F_{0.3-8}^{\rm intr}=1.18\times 10^{-13}$\,\ergscm\; and $N_{\rm H}=5.1\times10^{22}$\,\cmss.
The rest frame EW of the Fe line is 500\,eV. 


\subsection{J1347+5453 (z=0.332)}
This is a text-book example of an AGN jet and a double ionization cone lighting up a bipolar 
outflow. As such it is reminiscent of J0024+3258. The host galaxy is a red edge-on spiral with 
21\,kpc diameter, unresolved along its minor axis. The outflow is launched from the nucleus 
and nearly perpendicular to the disk. 

The southeastern outflow has a smaller opening angle, and after $\sim5$\,kpc it bends by 90\,deg 
and runs parallel to the disk. The northwestern outflow has irregular surface brightness distribution 
and a wide opening angle of $70-80$\,deg. It stretches over at least 12\,kpc. At a distance of 
18\,kpc we find a 4\,kpc wide (0\myarcsec8), faint blob of gas that is mostly visible in the $r$-band 
image. A weak detection in $g$-band could be caused by [\ion{O}{II}] and [\ion{Ne}{III}] emission. 
It is also weakly detected in $i$-band. Since \Ha is redshifted beyond $i$-band, the most plausible 
emission lines that could cause the $i$-band signal are [\ion{O}{I}]$\lambda$6302,66, possibly
indicating shock ionization. A high contrast stretch of the $r$-band image reveals that the 
northwestern outflow reaches further to least 32\,kpc from the nucleus, embedding this seemingly
isolated blob of gas.

Spectra of field galaxies have not yet been taken. Two ellipticals nearby have similar colours as 
the edge-on disk in J1347$+$5453, and could be at the same redshift. Overall, this is a low density 
area.

\begin{figure}
  \includegraphics[width=1.0\hsize]{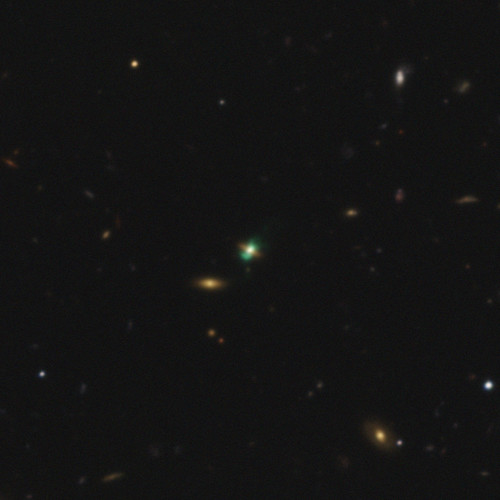}
  \caption{\label{J1347+5453}{J1347+5453}}
\end{figure}

\subsection{J1351+0816 (z=0.306)}
Imaging data reveal a featureless, spherically round nebula with symmetric brightness profile
peaking at the centre. The sky noise is met at a diameter of 28\,kpc. A compact blue star-forming galaxy
is superimposed at a distance of 9.4\,kpc of the nucleus. We took a long slit spectrum of it and detect 
a weak continuum and nebular emission lines. The latter place it either at a slightly lower redshift 
than the host galaxy $(\Delta z=-0.0007)$, or they are due to superimposed gas from the host. No other 
emission or absorption lines are seen. There is a small chance that this is a high-z galaxy in the 
redshift desert, or a foreground star. Otherwise, J1351$+$0816 appears to be an isolated galaxy in the 
field.

Noteworthy is a very faint halo in the $r$-band image (not visible in the images reproduced in this 
paper; check the public available coadded FITS images) extending 48\,kpc to the South. It coincides 
with our long-slit orientation and we detect [\ion{O}{III}] over its entire extent.

\begin{figure}
  \includegraphics[width=1.0\hsize]{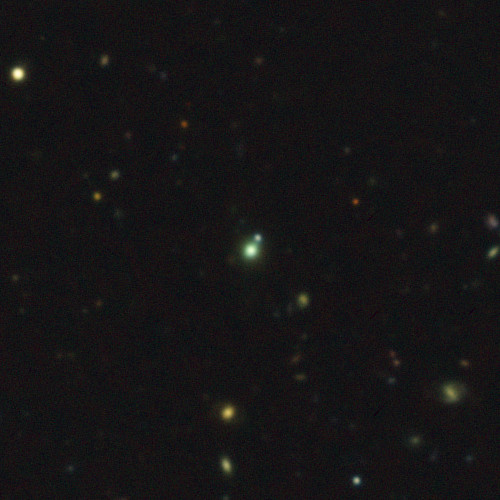}
  \caption{\label{J1351+0816}{J1351+0816}}
\end{figure}

\subsection{J1441+2517 (z=0.192)}
This GB has the lowest redshift in our sample. The [\ion{O}{III}] nebula consists of a 7\,kpc long
bright core elongated North-South. A host galaxy is likely compact. A fainter area of irregular surface 
brightness stretches over 17\,kpc to the North, with an opening angle of approximately 90 degrees. The 
southern part of the emission-line nebula is weaker and more diffuse than the northern part. Two red 
compact galaxies of unknown redshift are embedded in the southern nebula. If at the same redshift as 
J1441+2517, then their physical sizes are 6\,kpc and 4\,kpc, respectively. Other than that this galaxy
would live in isolation. A bridge in the [\ion{O}{III}] nebula appears to connect the larger of the two 
possible companions with the nucleus of J1441+2517. Judging from its colours, a mix of ionized gas and 
stellar tidal debris extends 22\,kpc to the South of the nucleus, in a broad fan-shaped fashion.

\begin{figure}
  \includegraphics[width=1.0\hsize]{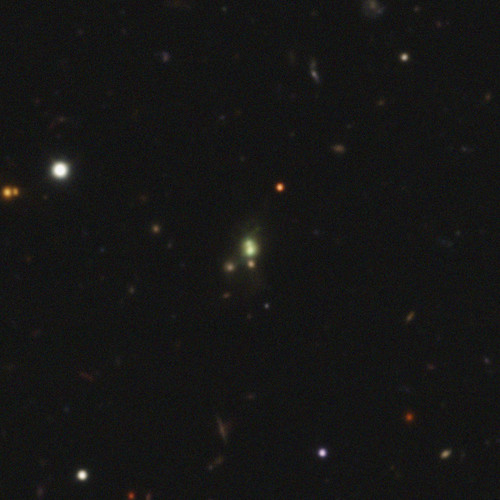}
  \caption{\label{J1441+2517}{J1441+2517}}
\end{figure}

\subsection{J1455+0446 (z=0.334)}
This is the most a bizarre system in our sample, extending over 40\,kpc. The [\ion{O}{III}] emission 
is brightest in the nucleus and has two wings that extend in East-West direction. The nucleus appears
at the edge of the galaxy. A condensation (probably a star cluster) is seen North of the nucleus. 
Significant colour gradients suggest a wild mix of tidally disturbed gas and stars, consequence of a 
violent merger. Spectroscopy of 7 field galaxies yield only one galaxy with similar redshift 
$(z=0.329)$, making it unlikely that these two galaxies interacted in the past. Otherwise, this
redshift difference would imply an encounter with a radial velocity of $\sim1100$\,\kms, inconceivable 
in the absence of a massive cluster. The large spiral to the South is a foreground object at $z=0.087$.

\begin{figure}
  \includegraphics[width=1.0\hsize]{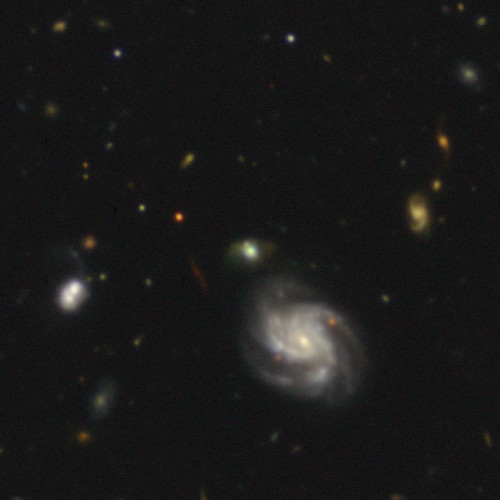}
  \caption{\label{J1455+0446}{J1455+0446}}
\end{figure}

\subsection{J1504+3439 (z=0.294)}
The [\ion{O}{III}] nebula in this galaxy is brightest at the nucleus, and fragments into several 
clouds of medium surface brightness. We find a bubble-like structure extending 12\,kpc to the East 
of the nucleus. Fainter parts reach radii of 18\,kpc in different directions. The ionized gas is 
superimposed over an elliptical galaxy measuring approximately $37\times24$\,kpc. This is the 
only GB where the gas distribution resembles that of the \textit{Voorwerpjes} (ionization echoes 
at $z\lesssim0.1$) described by \citet{kcb12,kmb15}, with the notable difference that 
\textit{Voorwerpjes} are not observed in ellipticals.

The field is populated by about 20 brighter spirals, ellipticals and interacting galaxies loosely 
scattered across the area. Most of them have SDSS photometric redshifts of $z=0.22-0.40$ 
($\Delta z=0.06-0.08$). Spectroscopic redshifts place two galaxies at different distances than 
J1504+3439. Spectra for three more galaxies have too low S/N for redshift determination,
and they are likely at higher redshift.

\begin{figure}
  \includegraphics[width=1.0\hsize]{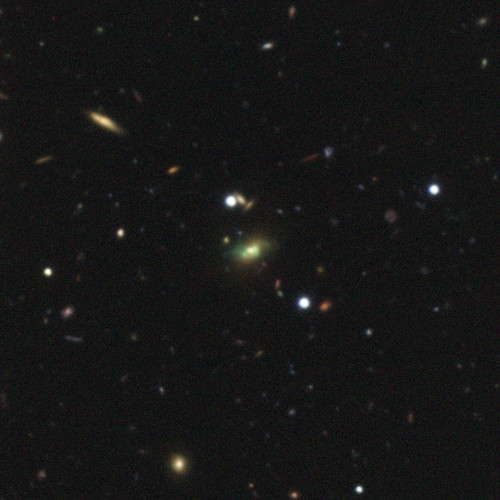}
  \caption{\label{J1504+3439}{J1504+3439}}
\end{figure}

\subsection{J1505+1944 (z=0.341)}
This is the highest redshift GB in our sample. Its nebula extends East-West over 28\,kpc with 
high central surface brightness. The eastern wing is brighter than the western wing, in 
which a small elliptical galaxy is embedded and confirmed at the same redshift. It appears to 
be tidally interacting with the ionized gas. J1505+1944 is the only GB where the [\ion{O}{III}]
and X-ray peaks are significantly offset from each other (see Section \ref{xrayoffsets}).

J1505$+$1944 is located in a small group of at least 8 spectroscopically verified members. The red 
sequence has $10-15$ members, mostly small ellipticals within $1.6$\,Mpc. The concentration 
of this group is low. Analogous to Section \ref{j11550147target} we estimate 
$M_{200}=(4\pm1)\times10^{13}\; M_{\odot}$. The host galaxy is compact, and small in comparison to the 
other group members.

J1505$+$1944 is superimposed on a small group of red ellipticals, likely at $z\sim0.8-0.9$. 

\begin{figure}
  \includegraphics[width=1.0\hsize]{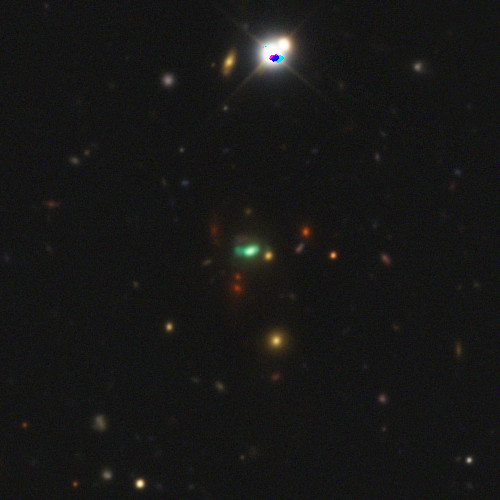}
  \caption{\label{J1505+1944}{J1505+1944}}
\end{figure}

\subsection{J2050+0550 (z=0.301)}
J2050+0550 has a smooth EELR with high central surface brightness. A 7\,kpc cloud is seen 4.5\,kpc 
to the Northwest, less pronounced yet similar to J2240$-$0927. One of our field long-slit spectra 
was taken southwest of the nucleus. We still detect [\ion{O}{III}] emission at a radius of $20$\,kpc. 
This shows that the [\ion{O}{III}] sizes listed in Table \ref{targetlist}, based on $r$-band surface 
brightness, may significantly underestimate the true extent of the ionized nebulae. The host galaxy 
is compact.

$10-12$ red sequence galaxies are found within a radius of about 540\,kpc of J2050$+$0550. We 
obtained spectra for three of them and they have the same redshift as J2050$+$0550. The 
cluster is sparse, with J2050$+$0550 located near the densest area (similar to J1155$-$0147). 
None of the other cluster members is detected in X-rays, nor do we see diffuse emission from a 
cluster X-ray halo. Analogous to Section \ref{j11550147target}) we estimate 
$M_{200}=(2.6\pm0.8)\times10^{13}\;M_{\odot}$.

\begin{figure}
  \includegraphics[width=1.0\hsize]{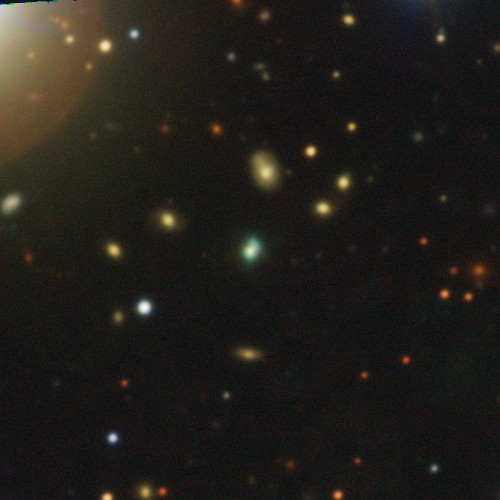}
  \caption{\label{J2050+0550}{J2050+0550}}
\end{figure}

\subsection{J2135--0314 (z=0.246)}
The host galaxy of J2135$-$0314 is compact, overpowered by the high nebular central surface brightness. 
Three fringes lend a tattered look to the nebula. The northeastern fringe curves to the South over 
$\sim 9$\,kpc, the southeastern fringe is straight an can be traced over 13\,kpc before disappearing 
in the sky noise. The galaxy is located in a remarkably empty area. Plausibly interacting companions 
are not seen.

\begin{figure}
  \includegraphics[width=1.0\hsize]{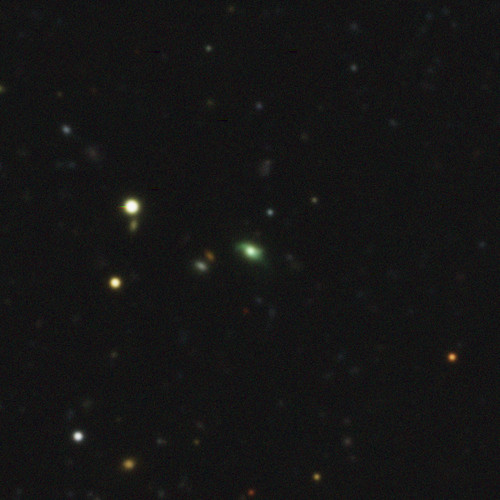}
  \caption{\label{J2135-0314}{J2135--0314}}
\end{figure}

\subsection{J2202+2309 (z=0.258)}
J2202$+$2309 is located in a structure with several red sequence galaxies confirmed at $z=0.258$.
Our data reveal about 40 red sequence galaxies with matching colours and within 600\,kpc radius. 
However, an unknown fraction of them belongs to a galaxy cluster projected about 330\,kpc to the West 
and with $z\sim0.230$ (about 80 Mpc in the foreground of J2202$+$2309, neglecting possible bulk motions 
along the line of sight). Like in case of J1155$-$0147 and J2050$+$0550, the GB is located near the centre 
of the galaxies at $z=0.258$. More field spectroscopy is required to clarify the structures along the 
line of sight, and to obtain reliable mass estimates. Assuming a contamination of 50 per cent, this
cluster could be as massive as $(5-8)\times10^{13}\; M_{\odot}$ (see Section \ref{j11550147target}), and 
would be the most massive structure in our survey. This field was not part of our X-ray observations.

J2202$+$2309 differs from the other GBs in the sense that its host galaxy is a luminous ($M_i\sim-22.5$
mag) elliptical. The [\ion{O}{III}] emission does not dominate the broad-band colours as much. Tidal 
disturbances are visible in the halo of J2202$+$2309. The physical extent of the elliptical host is at 
least $40\times25$\,kpc. A contrast-enhanced $r$-band image shows that this galaxy shares a common faint 
halo with J220215.5+230859, another elliptical of comparable size and luminosity, and possibly a third 
elliptical as well. This triple system could form the future BCG of this cluster.

One of our long slit positions has an impact parameter of 12\,arcsec (48\,kpc) with respect to the 
nucleus. There is a weak detection of [\ion{O}{III}] over 3\,arcsec length at the point of highest
proximity to J2202$+$2309. Our 20 minute $r$-band image (containing the redshifted [\ion{O}{III}] 
line) is not deep enough to reveal a signal at this location. This part of the sky has not been 
surveyed in the FUV with \textit{GALEX}.

\begin{figure}
  \includegraphics[width=1.0\hsize]{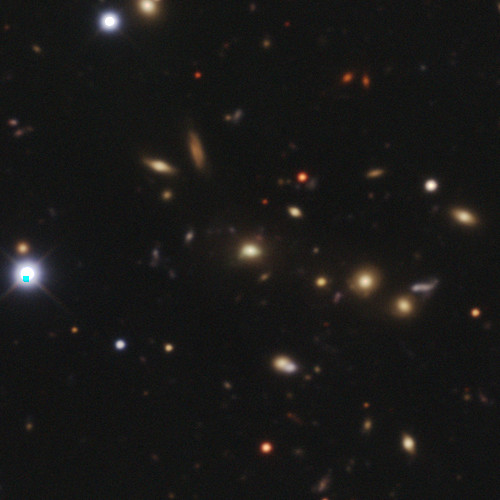}
  \caption{\label{J2202+2309}{J2202+2309}}
\end{figure}

\subsection{J2240--0927 (z=0.326)}
This is an interacting system in an otherwise empty area. A detailed study using GMOS-S 3D spectroscopy 
has been presented in \citet{dst15}, showing a complex and turbulent merger system with different gas 
phases and an ionization cone.

\begin{figure}
  \includegraphics[width=1.0\hsize]{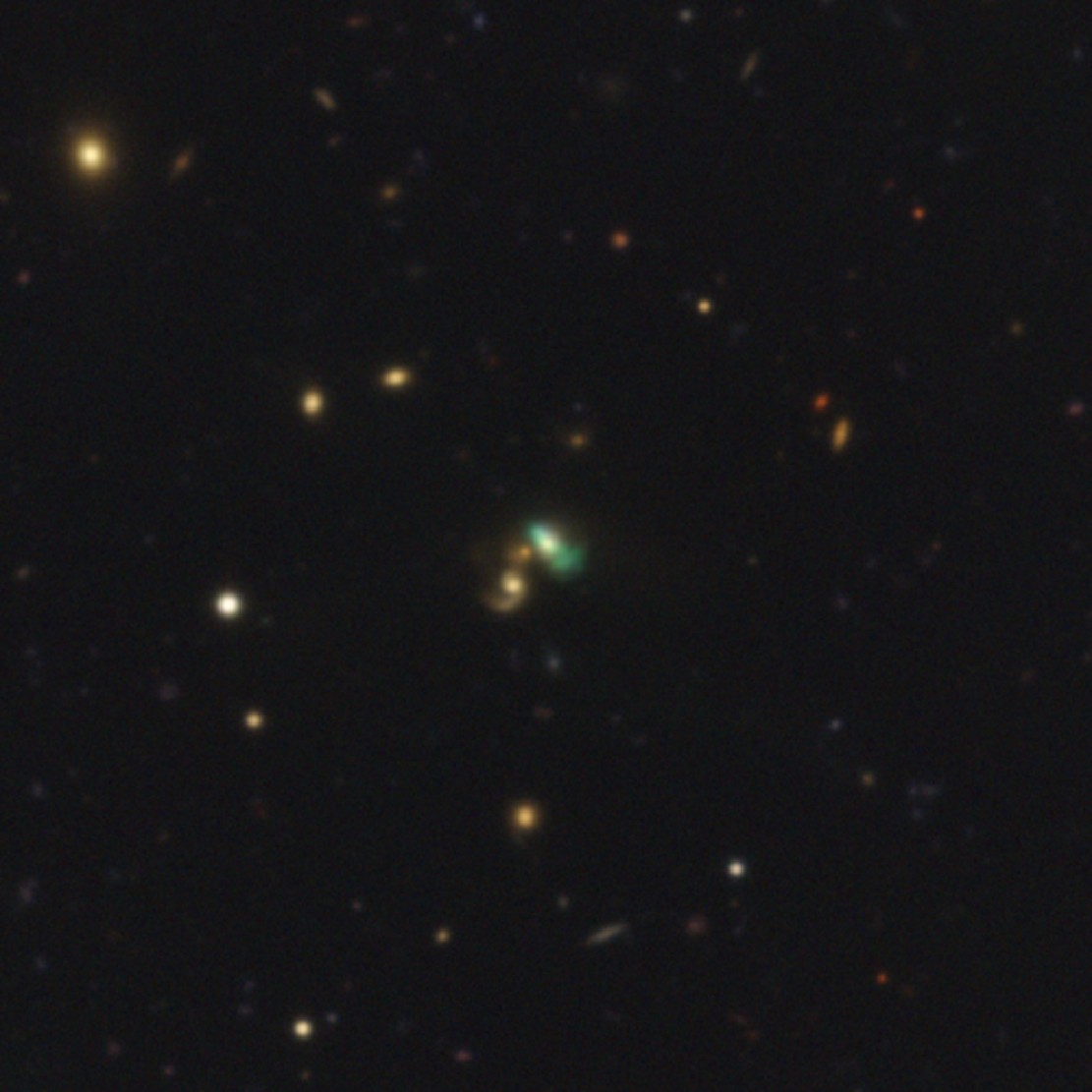}
  \caption{\label{J2240-0927}{J2240--0927}}
\end{figure}

\subsection{J2308+3303 (z=0.284)}
This system appears rather normal in our broad-band data, like a face-on spiral with insufficient 
resolution to discern the spiral arms. The nucleus is bright and barely resolved. Either the 
nebular flux is concentrated in the nucleus (and therefore within a physical diameter of 6\,kpc), 
or it is more extended but with low EW and of relatively uniform surface brightness. The surrounding 
featureless disk has a diameter of $22\times20$\,kpc. There is one spectroscopically confirmed neighbour, 
a distorted spiral, 29\,kpc to the North.

\begin{figure}
  \includegraphics[width=1.0\hsize]{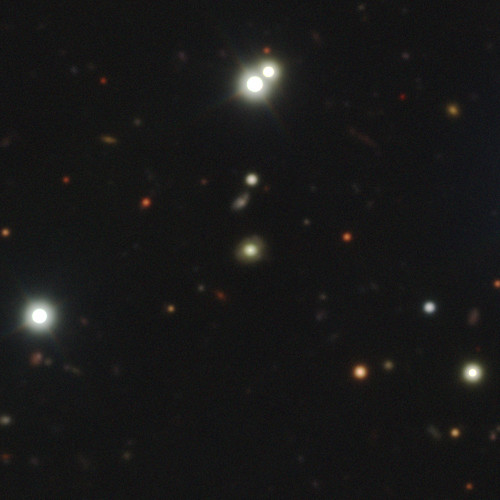}
  \caption{\label{J2308+3303}{J2308+3303}}
\end{figure}

\begin{landscape}
\begin{table}
\caption{\label{targetlist}
  Properties of the optical nebulae and host galaxies. Two estimates of the EELRs' physical diameters are 
  given in columns 2 and 3: $D_{\rm [\ion{o}{III}]}^{\,\rm min}$ is a lower limit based on the green appearance 
  in $gri$ colour images where [\ion{O}{III}] \textit{dominates} the $r$-band. $D_{\rm [\ion{o}{III}]}^{\,\rm max}$ 
  is the largest diameter of a contour $2\sigma$ above the sky noise (column 3). Column 4 lists the mean 
  surface brightness along this contour. Column 5 states whether spectra of field galaxies have been taken. 
  Column 6 lists the local environment of the GBs. We consider it a group if at least 2 other galaxies are 
  at the same redshift. Column 7 lists the size estimates of the host galaxies. These are approximate due 
  to the contamination of the broad-band filters by nebular emission. Column 8 is a morphological 
  classification of the hosts, often troubled by the limited spatial resolution in which case we classify 
  them as \textit{compact}. These could be small ellipticals as well as star-forming galaxies. Column 9 
  contains our best guess whether these systems are interacting or merging. Column 10 states whether the 
  gas distribution resembles AGN driven outflows. In column 11 we describe the gas distribution and other 
  system characteristics.
}
\begin{tabular}{lcccclclccl}
  \noalign{\smallskip}
  \hline 
  \noalign{\smallskip}
  Name & $D_{\rm [\ion{o}{III}]}^{\,\rm min}$ & $D_{\rm [\ion{o}{III}]}^{\,\rm max}$ & ${\rm SB}_{S/N=2}$ & Spec. & Environment 
  & Host size & Host type & Merger & Outflow & Comment\\
  & [kpc] & [kpc] & [\ergscm] & survey & & [kpc] & & & &\\
  &       &       &  arcsec$^{-2}$] & & & & & & &\\
  \noalign{\smallskip}
  \hline 
  \noalign{\smallskip}
SDSS J002016.44$-$053126.6 & 30 & 44 & $2.90\times10^{-16}$ & Yes   & Group            & $18\times14$ & compact    & Yes   & Yes   & Tidally warped outflow; compact group\\
SDSS J002434.90$+$325842.7 & 28 & 40 & $1.94\times10^{-16}$ & Yes   & Isolated         & $10\times7$  & irregular  & Yes   & Yes   & Bipolar outflow\\
SDSS J011133.31$+$225359.1 & 21 & 28 & $1.78\times10^{-16}$ & Yes   & Group            & $18\times14$ & compact    & Yes   & Yes   & Unipolar outflow; sparse group\\
SDSS J011341.11$+$010608.5 & 68 & 75 & $2.40\times10^{-16}$ & Yes   & Isolated         & $8\times6$   & compact    & Yes   & Yes   & Bipolar outflow and signatures for\\
                           &    &    &                     &     &                  &              &            &     &     & recurrent outflows\\
SDSS J015930.84$+$270302.2 &  6 & 33 & $1.80\times10^{-16}$ & Yes   & Isolated         & $46\times46$ & spiral     & $-$ & $-$ & Prominent spiral, one confirmed distant\\
                           &    &    &                     &     &                  &              &            &     &     & neighbour\\
SDSS J115544.59$-$014739.9 & 55 & 65 & $2.24\times10^{-16}$ & Yes   & Group            & $16\times12$ & elliptical & ?   & ?   & Perhaps a cold accretion nebula at the\\
                           &    &    &                     &     &                  &              &            &     &     & centre of a cluster\\
SDSS J134709.12$+$545310.9 & 40 & 47 & $2.77\times10^{-16}$ & $-$ & Isolated / Group & $21\times4$  & spiral     & $-$ & Yes & Very small group, perhaps $2-3$ members\\
SDSS J135155.48$+$081608.4 & 17 & 25 & $2.31\times10^{-16}$ & $-$ & Isolated / Group & $14\times14$ & spherical  & $-$ & $-$ & Smooth spherical nebula; Isolated or\\
                           &    &    &                     &     &                  &              &            &     &     & in very sparse group\\
SDSS J144110.95$+$251700.1 & 21 & 27 & $2.81\times10^{-16}$ & $-$ & Isolated / Group & $\lesssim6$  & compact    & Yes   & Yes   & Multiple faint outflows; two small\\
                           &    &    &                     &     &                  &              &            &     &     & companions (unconfirmed)\\
SDSS J145533.69$+$044643.2 & 36 & 39 & $2.24\times10^{-16}$ & Yes   & Isolated         & $9\times7$   & irregular  & Yes   & ?   & Totally disrupted system\\
SDSS J150420.68$+$343958.2 & 36 & 38 & $2.33\times10^{-16}$ & $-$ & Isolated         & $37\times24$ & elliptical & $-$ & Yes   & Compact clouds superimposed on\\
                           &    &    &                     &     &                  &              &            &     &     & undisturbed elliptical\\
SDSS J150517.63$+$194444.8 & 26 & 38 & $3.14\times10^{-16}$ & Yes   & Group            & $10\times7$  & compact    & ?   & Yes   & Recurrent outflows; red sequence\\
                           &    &    &                     &     &                  &              &            &     &     & galaxies, low concentration\\
SDSS J205058.08$+$055012.8 & 26 & 31 & $2.23\times10^{-16}$ & Yes   & Group            & $17\times16$ & compact    & $-$ & Yes   & Smooth with ejected (?) cloud\\
SDSS J213542.85$-$031408.8 & 26 & 30 & $2.70\times10^{-16}$ & Yes   & Isolated         & $8\times7$   & compact    & ?   & Yes   & Multiple outflows\\
SDSS J220216.71$+$230903.1 & 14 & 31 & $1.66\times10^{-16}$ & Yes   & Group / Cluster  & $40\times25$ & elliptical & Yes   & $-$ & Bright elliptical, gas morphology\\
                           &    &    &                     &     &                  &              &            &     &     & not well visible\\
SDSS J224024.11$-$092748.1 & 30 & 56 & $1.53\times10^{-16}$ & Yes   & Isolated / Group & $9\times9$   & spherical  & Yes   & Yes   & Case study in \citet{dst15}.\\
                           &    &    &                     &     &                  &              &            &     &     & Two companions, one confirmed.\\
SDSS J230829.37$+$330310.5 & 16 & 25 & $1.69\times10^{-16}$ & Yes   & Isolated pair    & $22\times20$ & spiral     & $-$ & $-$ & Inconspicuous, one nearby member\\
  \hline
\end{tabular}
\end{table}
\end{landscape}

\begin{table*}
\caption{\label{redshiftlist}
Redshift survey of field galaxies near GBs. The \textit{Membership} is in parentheses if the galaxy is a 
foreground/background object.}
\begin{tabular}{rrrrl}
  \noalign{\smallskip}
  \hline 
  \noalign{\smallskip}
  $\alpha_{2000.0}$ & $\delta_{2000.0}$ & z & Membership & Spectral features and morphology\\
  \noalign{\smallskip}
  \hline 
  \noalign{\smallskip}
  00:20:16.333 & $-$05:31:28.02 & 0.334 & J0020$-$0531 & CaH+K; elliptical\\
  00:20:16.550 & $-$05:31:31.54 & 0.334 & J0020$-$0531 & CaH+K; elliptical\\
  00:20:19.705 & $-$05:32:21.89 & 0.281 & (J0020$-$0531) & CaH+K; foreground elliptical\\
  \hline
  \noalign{\smallskip}
  00:24:24.557 & +33:00:24.74 & 0.204 & (J0024+3258) & [\ion{O}{II}], \Hbn, \Hgn, [\ion{O}{III}]; foreground spiral\\
  00:24:25.363 & +33:00:07.55 & 0.224 & (J0024+3258) & CaH+K, G-band, Mg$\lambda$5177; foreground elliptical\\
  00:24:29.161 & +32:58:54.05 & 0.224 & (J0024+3258) & CaH+K, G-band, Mg$\lambda$5177; foreground elliptical\\
  00:24:29.883 & +32:59:54.13 & 0.230 & (J0024+3258) & [\ion{O}{II}], \Hbn, [\ion{O}{III}], CaH+K; foreground spiral\\
  00:24:33.432 & +32:58:46.29 & 0.226 & (J0024+3258) & CaH+K, G-band; foreground elliptical\\
  \hline
  \noalign{\smallskip}
  01:11:32.875 & 22:53:40.15 & 0.319 & J0111$+$2253 & CaH+K, G-band; elliptical\\
  01:11:33.860 & 22:53:59.64 & 0.321 & J0111$+$2253 & CaH+K, G-band; elliptical\\
  \hline
  \noalign{\smallskip}
  01:13:41.285 & +01:06:45.47 & 0.218 & (J0113$+$0106) & CaK, \Hbn, [\ion{O}{III}]; foreground spiral\\
  01:13:41.930 & +01:06:16.91 & 0.433 & (J0113$+$0106) & [\ion{O}{II}], CaH; background spiral\\
  \hline
  \noalign{\smallskip}
  01:59:23.315 & +27:03:31.76 & 0.611? & (J0159$+$2703) & [\ion{O}{II}]; distant faint background, single line\\
  01:59:25.820 & +27:03:22.53 & 0.583 & (J0159$+$2703) & [\ion{O}{II}], CaH+K; background spiral\\
  01:59:30.467 & +27:03:06.48 & 0.710? & (J0159$+$2703) & [\ion{O}{II}]; distant faint background, single line\\
  01:59:33.741 & +27:02:54.31 & 0.118 & (J0159$+$2703) & [\ion{Ne}{V}]$\lambda3427$, [\ion{O}{II}], [\ion{Ne}{III}], \Hgn, HeII$\lambda$4687, \Hbn, [\ion{O}{III}], CaH; foreground spiral\\
  01:59:37.488 & +27:02:38.49 & 0.351 & (J0159$+$2703) & [\ion{O}{II}]; well resolved background spiral\\
  01:59:40.972 & +27:02:29.32 & 0.277 & J0159$+$2703   & CaH+K, G-band; elliptical\\
  01:59:43.630 & +27:02:18.01 & 0.220 & (J0159$+$2703) & [\ion{Ne}{V}]$\lambda3427$, [\ion{O}{II}], \Hbn, [\ion{O}{III}], CaH; foreground spiral\\
  \hline
  \noalign{\smallskip}
  11:55:43.272 & $-$01:47:21.76 & 0.304 & J1155$-$0147 & [\ion{O}{II}], CaH+K; spiral\\
  11:55:44.994 & $-$01:47:21.77 & 0.305 & J1155$-$0147 & CaH+K; elliptical, tidal tail\\
  11:55:45.565 & $-$01:47:23.20 & 0.306 & J1155$-$0147 & weak [\ion{O}{II}], CaH+K; lenticular\\
  \hline
  \noalign{\smallskip}
  14:55:28.676 & +04:49:04.95 & 0.231 & (J1455+0446) & [\ion{O}{II}], \Hbn; foreground spiral\\
  14:55:31.562 & +04:47:08.87 & 0.162 & (J1455+0446) & [\ion{O}{II}], \Hbn; foreground spiral\\
  14:55:31.812 & +04:46:59.10 & 0.571 & (J1455+0446) & [\ion{O}{II}], CaH+K (all features weak); background spiral\\
  14:55:32.054 & +04:46:48.65 & 0.368 & (J1455+0446) & [\ion{O}{II}], \Hbn, [\ion{O}{III}], CaH+K; background spiral\\
  14:55:35.007 & +04:45:46.29 & 0.329 & (J1455+0446) & [\ion{O}{II}], \Hbn, [\ion{O}{III}], CaH+K; foreground spiral\\
  14:55:35.452 & +04:46:36.35 & 0.162 & (J1455+0446) & [\ion{O}{II}], \Hbn, \Hgn, [\ion{O}{III}], CaH+K; foreground irregular\\
  14:55:36.831 & +04:49:07.70 & 0.473 & (J1455+0446) & [\ion{O}{II}]; background spiral\\
  \hline
  \noalign{\smallskip}
  15:04:25.586 & 34:39:16.91 & 0.461 & (J1504+3439) & [\ion{O}{II}], CaH+K; background irregular\\
  15:04:26.737 & 34:39:20.91 & 0.208 & (J1504+3439) & CaH+K, G-band, Mg$\lambda$5177; foreground spiral\\
  \hline
  \noalign{\smallskip}
  15:05:09.999 & +19:44:08.80 & 0.344 & J1505$+$1944 & CaH+K; elliptical\\
  15:05:13.744 & +19:44:48.03 & 0.341 & J1505$+$1944 & CaH+K; elliptical\\
  15:05:15.906 & +19:45:11.63 & 0.344 & J1505$+$1944 & [\ion{O}{II}], \Hbn, [\ion{O}{III}]; irregular\\
  15:05:16.031 & +19:42:30.65 & 0.433 & (J1505$+$1944) & [\ion{O}{III}], CaK; background red spiral\\
  15:05:16.113 & +19:41:54.87 & 0.342 & J1505$+$1944 & elliptical (SDSS spectrum)\\
  15:05:17.363 & +19:44:31.81 & 0.342 & J1505$+$1944 & CaH+K; elliptical\\
  15:05:17.446 & +19:44:44.18 & 0.343 & J1505$+$1944 & CaK; elliptical\\
  15:05:17.847 & +19:45:12.36 & 0.341 & J1505$+$1944 & CaH+K, [\ion{O}{III}]; red spiral (also SDSS)\\
  15:05:18.040 & +19:45:36.48 & 0.345 & J1505$+$1944 & weak [\ion{O}{II}], CaH+K; elliptical\\
  \hline
  \noalign{\smallskip}
  20:50:57.167 & +05:50:22.84 & 0.302 & J2050+0550 & CaH+K; elliptical\\
  20:50:57.336 & +05:50:19.10 & 0.301 & J2050+0550 & weak [\ion{Ne}{V}]$\lambda3427$ and [\ion{S}{II}]$\lambda$4069,76, CaH+K; elliptical\\
  20:51:00.906 & +05:49:15.77 & 0.300 & J2050+0550 & CaH, 4000\AA$\;$break; red spiral\\
  \hline
  \noalign{\smallskip}
  21:35:40.804 & $-$03:12:58.14 & 0.573 & (J2050+0550) & [\ion{O}{II}], CaH+K; background spiral\\
  \hline
  \noalign{\smallskip}
  22:02:15.498 & +23:08:58.67 & 0.258 & J2202+2309 & CaH+K, G-band; elliptical\\
  22:02:16.100 & +23:08:03.98 & 0.258 & J2202+2309 & CaH+K, G-band; red spiral\\
  22:02:17.666 & +23:09:15.42 & 0.258 & J2202+2309 & weak [\ion{O}{II}] and [\ion{O}{III}], CaH+K; red spiral\\
  22:02:17.795 & +23:09:38.82 & 0.228 &(J2202+2309) & CaH+K, G-band, Mg$\lambda$5177; foreground elliptical\\
  \hline
  \noalign{\smallskip}
  23:08:24.439 & +33:03:06.22 & 0.299 & (J2308+3303) & CaH+K; background elliptical\\
  23:08:29.482 & +33:03:17.49 & 0.283 & J2308+3303 & \Hbn, [\ion{O}{III}]; spiral\\
  23:08:32.958 & +33:03:38.28 & 0.249 & (J2308+3303) & [\ion{O}{II}], \Hbn, [\ion{O}{III}]; foreground spiral\\
  23:08:34.248 & +33:03:42.02 & 0.299 & (J2308+3303) & [\ion{O}{II}], \Hbn, [\ion{O}{III}], CaH+K; background spiral\\
\end{tabular}
\end{table*}

\begin{figure*}
  \includegraphics[width=1.0\hsize]{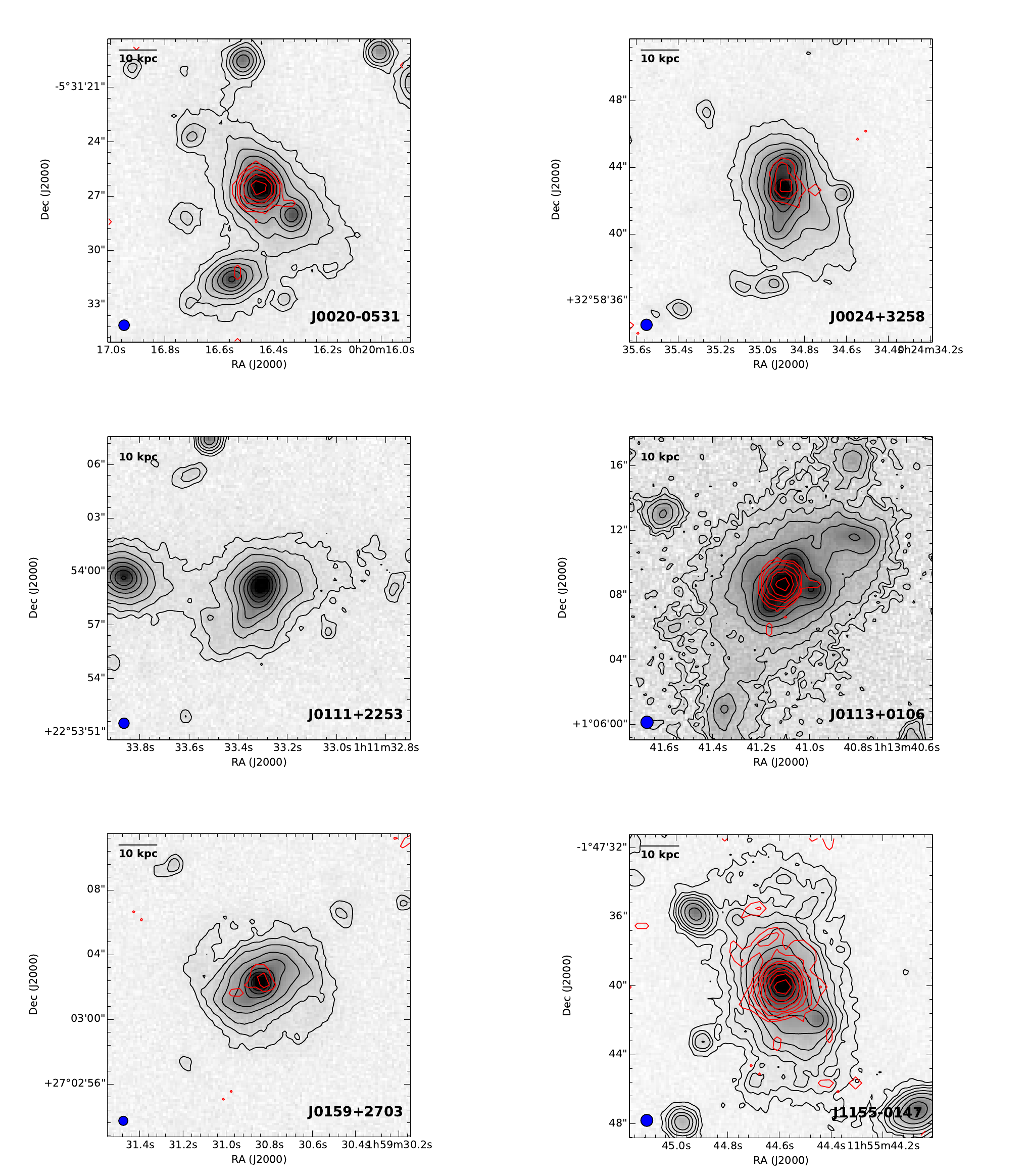}
  \caption{\label{chandramosaic1}Shown are the $r$-band images of the GBs and the $r$-band contours.
    A strong non-linear stretch using an asinh() function was applied to the grey-scale image to reveal 
    detail at all brightness levels. The small blue disk at the lower left represents the FWHM of the
    optical seeing disk. The \textit{Chandra} X-ray data are shown as red contours if available.}
\end{figure*}

\begin{figure*}
  \includegraphics[width=1.0\hsize]{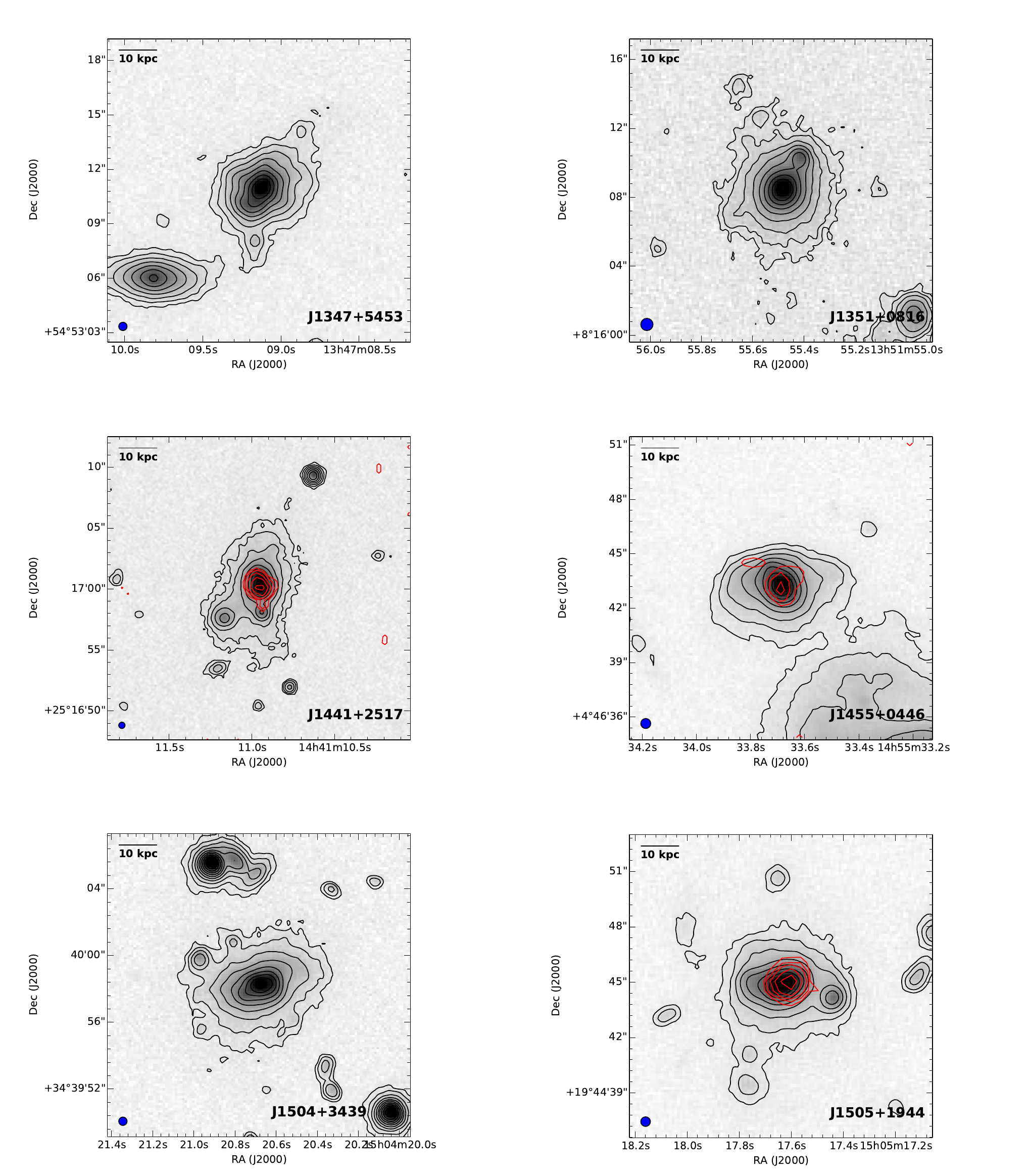}
  \caption{\label{chandramosaic2} Same as Fig. \ref{chandramosaic1}.}
\end{figure*}

\begin{figure*}
  \includegraphics[width=1.0\hsize]{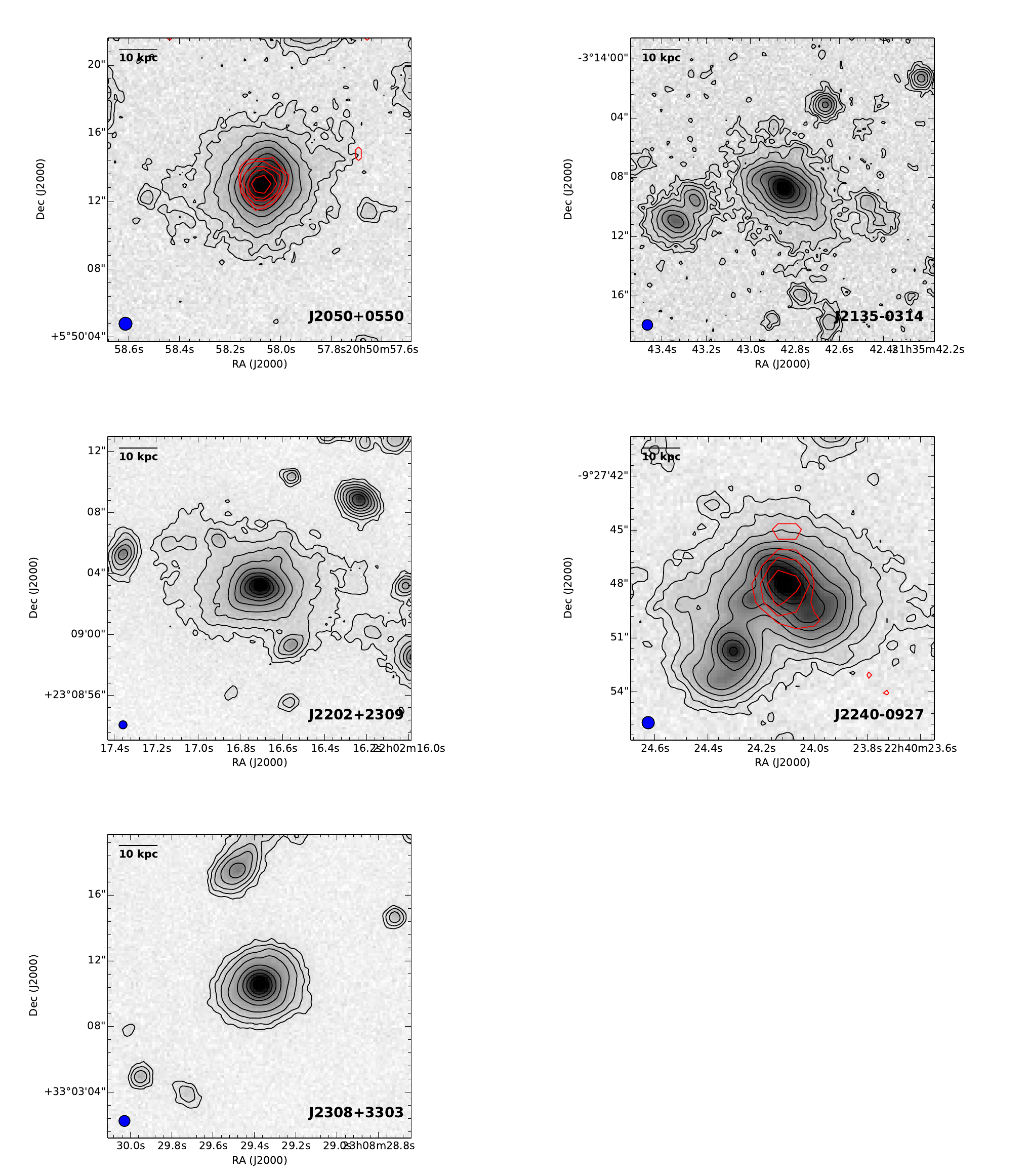}
  \caption{\label{chandramosaic3} Same as Fig. \ref{chandramosaic1}.}
\end{figure*}

\bsp	
\label{lastpage}
\end{document}